\documentclass[11pt]{amsart} 
\usepackage{amscd,amssymb,amsxtra}
\usepackage{mathrsfs}  
\DeclareSymbolFontAlphabet{\mathrsfs}{rsfs}
\usepackage[mathscr]{eucal} 
\usepackage{comment}
\usepackage{color}
\usepackage{enumitem} 
\usepackage{bbm}
\usepackage[colorlinks,backref=page,citecolor=refkey]{hyperref}
\usepackage{aliascnt}
\usepackage{tikz-cd}
\usetikzlibrary{arrows}
\colorlet{bgcolor}{black}
\usetikzlibrary{decorations.pathmorphing}

\usepackage{marginnote}

\makeatletter
\let\@secnumfont\bfseries
\def\section{\@startsection{section}{1}%
  \z@{4\linespacing\@plus\linespacing}{\linespacing}%
  {\bfseries\centering}}
\def\introsection{\@startsection{section}{1}%
  \z@{3\linespacing\@plus\linespacing}{\linespacing}%
  {\bfseries\centering}}
\def\subsection{\@startsection{subsection}{2}%
   \z@{1.25\linespacing\@plus.7\linespacing}{.5\linespacing}%
   {\normalfont\bfseries}}
\def\subsectionsinline{\def\subsection{\@startsection{subsection}{2}%
  \z@{1\linespacing\@plus.7\linespacing}{-.5em}%
  {\normalfont\bfseries}}}

\makeatother

\numberwithin{equation}{section}

\newcommand{\mynewtheorem}[2]{
  \newaliascnt{#1}{equation}
  \newtheorem{#1}[#1]{#2}
  \aliascntresetthe{#1}
  \expandafter\def\csname #1autorefname\endcsname{#2}
}

\theoremstyle{definition}
\mynewtheorem{definition}{Definition}
\mynewtheorem{example}{Example}
\mynewtheorem{problem}{\color{blue}Problem}
\mynewtheorem{probsec}{\color{blue}Problem}
\mynewtheorem{exercise}{Exercise}
\mynewtheorem{question}{\color{blue}Question}
\mynewtheorem{project}{\color{blue}Project}
\mynewtheorem{construction}{Construction}
\mynewtheorem{task}{\color{blue}Task}
\mynewtheorem{otask}{\color{green}Obsolete Task}
\mynewtheorem{notation}{Notation}

\newtheorem*{definition*}{Definition}
\newtheorem*{example*}{Example}
\newtheorem*{problem*}{\color{blue}Problem}
\newtheorem*{probsec*}{\color{blue}Problem}
\newtheorem*{exercise*}{Exercise}
\newtheorem*{question*}{\color{blue}Question}
\newtheorem*{project*}{\color{blue}Project}
\newtheorem*{construction*}{Construction}
\newtheorem*{notation*}{Notation}

\theoremstyle{remark}
\mynewtheorem{note}{Note}
\mynewtheorem{remark}{Remark}
\mynewtheorem{data}{Data}

\newtheorem*{note*}{Note}
\newtheorem*{remark*}{Remark}
\newtheorem*{data*}{Data}

\theoremstyle{plain}
\mynewtheorem{theorem}{Theorem}
\mynewtheorem{corollary}{Corollary}
\mynewtheorem{lemma}{Lemma}
\mynewtheorem{proposition}{Proposition}
\mynewtheorem{conjecture}{Conjecture}
\mynewtheorem{claim}{Claim}
\mynewtheorem{proposal}{Proposal}
\mynewtheorem{conclusion}{Conclusion}
\mynewtheorem{hypothesis}{Hypothesis}
\mynewtheorem{assumption}{Assumption}

\newtheorem*{theorem*}{Theorem}
\newtheorem*{corollary*}{Corollary}
\newtheorem*{lemma*}{Lemma}
\newtheorem*{proposition*}{Proposition}
\newtheorem*{conjecture*}{Conjecture}
\newtheorem*{claim*}{Claim}
\newtheorem*{proposal*}{Proposal}
\newtheorem*{conclusion*}{Conclusion}
\newtheorem*{hypothesis*}{Hypothesis}
\newtheorem*{assumption*}{Assumption}

\newenvironment{proof*}[1][\proofname]{
  \begin{proof}[#1]}{  
\end{proof}}

\definecolor{refkey}{rgb}{0,.6,.4}

\renewcommand{\:}{\colon}

\DeclareMathOperator{\Aut}{Aut}
\newcommand{\CC}{{\mathbb C}}

\DeclareMathOperator{\End}{End}

\DeclareMathOperator{\Hom}{Hom}
\DeclareMathOperator{\id}{id}

\DeclareMathOperator{\Map}{Map}

\DeclareMathOperator{\pt}{pt}
\newcommand{\QQ}{{\mathbb Q}}
\newcommand{\RP}{{\mathbb R\mathbb P}}
\newcommand{\RR}{{\mathbb R}}
\newcommand{\TT}{\mathbb T}
\DeclareMathOperator{\Spin}{Spin}

\DeclareMathOperator{\Tr}{Tr}

\newcommand{\ZZ}{{\mathbb Z}}

\newcommand{\chiup}{\raise.5ex\hbox{$\chi$}}
\newcommand{\cir}{S^1}

\newcommand{\inv}{^{-1}}
\DeclareRobustCommand{\mstrut}{^{\vphantom{1*\prime y\vee M}}}

\newcommand{\res}[1]{\negmedspace\bigm|\mstrut_{#1}}

\newcommand{\temsquare}{\raise3.5pt\hbox{\boxed{ }}}

\newcommand{\zmod}[1]{\ZZ/#1\ZZ}

\newcommand{\zt}{\zmod2}

\newcommand{\longhookrightarrow}{\lhook\joinrel\longrightarrow}
\newcommand{\hneg}{\mkern-.5\thinmuskip}
 
\DeclareFontFamily{U}{mathx}{}
\DeclareFontShape{U}{mathx}{m}{n}{<-> mathx10}{}
\DeclareSymbolFont{mathx}{U}{mathx}{m}{n}
\DeclareMathAccent{\widehat}{0}{mathx}{"70}
\DeclareMathAccent{\widecheck}{0}{mathx}{"71}
 
\DeclareMathSymbol{\bigtimes}{1}{mathx}{"91}

\DeclareMathOperator{\SO}{SO}

\let\O\relax
\DeclareMathOperator{\O}{O}
\DeclareMathOperator{\U}{U}
\DeclareMathOperator{\SU}{SU}
\DeclareMathOperator{\GL}{GL}

\usepackage[all,2cell]{xy}\renewcommand{\cir}{\ensuremath{S^1}}
\usepackage{graphicx}\usepackage{epsf}

\let\O\relax\DeclareMathOperator{\O}{O}

\definecolor{refkey}{rgb}{0,.8,.2}\definecolor{labelkey}{rgb}{1,0,0} 

\newcommand{\bmuu}{\mbox{$\raisebox{-.07em}{\rotatebox{9.9}
  {\tiny {\bf /}
  }}\hspace{-0.53em}\mu\hspace{-0.88em}\raisebox{-0.98ex}{\scalebox{2} 
  {$\color{white}\phantom{.}$}}\hspace{-0.416em}\raisebox{+0.88ex}
  {$\color{white}\phantom{.}$}\hspace{0.46em}$}} 
\newcommand{\bmut}{\bmu 2}
\newcommand{\bmu}[1]{\bmuu _{#1}}

\DeclareMathOperator{\Alg}{Alg}
\DeclareMathOperator{\Bord}{Bord}
\DeclareMathOperator{\Bun}{Bun}
\DeclareMathOperator{\Cat}{Cat}
\DeclareMathOperator{\Fam}{Fam}
\DeclareMathOperator{\Fun}{Fun}

\DeclareMathOperator{\Man}{Man}
\DeclareMathOperator{\Rep}{Rep}
\DeclareMathOperator{\Set}{Set}
\DeclareMathOperator{\Sq}{Sq}
\DeclareMathOperator{\Sum}{Sum}
\DeclareMathOperator{\Vect}{Vect}
\DeclareMathOperator{\codim}{codim}
\DeclareMathOperator{\germ}{germ}
\newcommand{\Ar}{A\mstrut _{\! A}}
\newcommand{\BO}{B\!\O}
\newcommand{\BSO}{B\!\SO}
\newcommand{\BnA}[1]{B^{#1}\hneg A}

\newcommand{\BtA}{B^2\!A}
\newcommand{\Cx}{\CC^{\times}}
\newcommand{\GA}{\CC[G]}

\newcommand{\LlX}{\sX^{S^{\ell -1}}}
\newcommand{\MTSO}{MT\!\SO}

\newcommand{\OX}{\Omega \sX}
\newcommand{\OlX}{\Omega ^\ell \sX}
\newcommand{\RepcA}{\Rep_c(A)}
\newcommand{\STd}{\Sigma ^n\sTd}
\newcommand{\XTd}{\sX\mstrut _{\Sigma ^n\sTd}}
\newcommand{\XT}{\sX\mstrut _{\!\sT}}
\newcommand{\Xln}[1]{(\sX_{#1},\lambda _{#1})}
\newcommand{\Xl}{(\sX,\lambda )}
\newcommand{\Xm}[1]{\sX^{#1}}
\newcommand{\bG}{\overline{G}}

\newcommand{\bX}{\boldsymbol{X}}
\newcommand{\bn}{\bar{\nu }}
\newcommand{\bone}{\mathbbm{1}}
\newcommand{\bs}{\!\hneg\bigm/_{\!\hneg\epsilon}\!\sigma }
\newcommand{\dual}{^\vee}
\newcommand{\eA}{\epsilon \mstrut _{\hneg A}}
\newcommand{\eP}{\epsilon \mstrut _{\Phi }}
\newcommand{\ed}{\epsilon \dual}
\newcommand{\gpd}{/\!/} 
\newcommand{\hB}{\widehat{B}}
\newcommand{\hD}{\widehat{D}}
\newcommand{\hF}{\widehat{F}}
\newcommand{\hdD}{\hat{\delta }_D}
\newcommand{\hth}{\widehat{\theta }}
\newcommand{\op}{^{\textnormal{op}}}
\newcommand{\rd}{\rho ^{\vee}}
\newcommand{\sA}{\mathscr{A}}
\newcommand{\sCt}{\sC^\textnormal{fd}}
\newcommand{\sC}{\mathscr{C}}
\newcommand{\sE}{\mathscr{E}}
\newcommand{\sF}{\mathcal{F}}
\newcommand{\sHl}{{}\mstrut _{\GA}\sH}
\newcommand{\sH}{\mathscr{H}}
\newcommand{\sL}{\mathcal{L}}
\newcommand{\sD}{\mathcal{D}}
\newcommand{\sM}{\mathscr{M}}
\newcommand{\sP}{\sigma\mstrut _{\!\Phi }}
\newcommand{\sTd}{\sT\dual}
\newcommand{\sT}{\boldsymbol{A}}
\newcommand{\sXd}[2]{\sigma _{#2}^{(#1)}}
\newcommand{\sXn}{\sigma _{\sX}^{(n+1)}}
\newcommand{\sX}{\mathscr{X}}
\newcommand{\sY}{\mathscr{Y}}
\newcommand{\sZ}{\mathscr{Z}}
\newcommand{\scP}{\mathscr{P}}
\newcommand{\scrT}{\mathscr{T}}
\newcommand{\sdrd}{(\sd ,\rd )}
\newcommand{\sd}{\sigma ^{\vee}}
\newcommand{\sqmo}{\sqrt{-1}}
\newcommand{\srtF}{(\sigma ,\rho ,\tF)}
\newcommand{\sr}{(\sigma ,\rho )}
\newcommand{\tFt}{(\tF,\theta )}
\newcommand{\tF}{\widetilde{F}}
\newcommand{\tZ}{\widetilde{Z}}
\newcommand{\tll}{\tau ^{\ell -1}\lambda }
\newcommand{\tsF}{\widetilde{\sF}}
\newcommand{\ts}{\widetilde{\sigma} }
\newcommand{\xreg}{x_{\text{reg}}}
\newcommand{\zoh}{[0,1)}
\newcommand{\zo}{[0,1]}
\newcommand{\tW}{\widetilde{W}}

\begin{document}

\abovedisplayskip18pt plus4.5pt minus9pt
\belowdisplayskip \abovedisplayskip
\abovedisplayshortskip0pt plus4.5pt
\belowdisplayshortskip10.5pt plus4.5pt minus6pt
\baselineskip=15 truept
\marginparwidth=55pt

\theoremstyle{definition}
\newtheorem{axiom}[equation]{Axiom System}
\newtheorem{iaxiom}[equation]{Informal Axiom System}
\newtheorem{ansatz}[equation]{Ansatz}
\newtheorem{thesis}[equation]{Thesis}

\makeatletter
\renewcommand{\tocsection}[3]{%
  \indentlabel{\@ifempty{#2}{\hskip1.5em}{\ignorespaces#1 #2.\;\;}}#3}
\renewcommand{\tocsubsection}[3]{%
  \indentlabel{\@ifempty{#2}{\hskip 2.5em}{\hskip 2.5em\ignorespaces#1%
    #2.\;\;}}#3} 
\renewcommand{\tocsubsubsection}[3]{%
  \indentlabel{\@ifempty{#2}{\hskip 5.5em}{\hskip 5.5em\ignorespaces#1%
    #2.\;\;}}#3} 
\def\@makefnmark{%
  \leavevmode
  \raise.9ex\hbox{\fontsize\sf@size\z@\normalfont\tiny\@thefnmark}} 
\def\multfoot{\textsuperscript{\tiny\color{red},}}
\def\footref#1{$\tiny\ref{#1}$}
\makeatother

\setcounter{tocdepth}{3}


 \title[Topological symmetry in QFT]{Topological Symmetry in Quantum Field
Theory} 
 \author[D. S. Freed]{Daniel S.~Freed}
 \address{Harvard University \\ Department of Mathematics \\ Science Center
Room 325 \\ 1 Oxford Street \\ Cambridge, MA 02138}
 \email{dafr@math.harvard.edu}

 \author[G. W. Moore]{Gregory W.~Moore}
 \address{NHETC and Department of Physics and Astronomy \\
Rutgers University \\ Piscataway, NJ 08855--0849}
 \email{gmoore@physics.rutgers.edu}

 \author[C. Teleman]{Constantin Teleman} 
  \address{Department of Mathematics \\ University of California \\ 970 Evans
 Hall \#3840 \\ Berkeley, CA 94720-3840}  
  \email{teleman@berkeley.edu}

 \thanks{This material is based upon work supported by the National Science
Foundation under Grant Number DMS-2005286, the Department of Energy under
Grant Number DOE-SC0010008, and by the Simons Foundation Award 888988 as part
of the Simons Collaboration on Global Categorical Symmetries.  Large parts of
this work were performed at the Aspen Center for Physics, which is supported
by National Science Foundation grant PHY-1607611.}

 \dedicatory{In memory of Vaughan Jones}
 \date{July 30, 2024}
 \begin{abstract} 
 We introduce a definition and framework for internal topological symmetries in
quantum field theory, including ``noninvertible symmetries'' and ``categorical
symmetries''.  We outline a calculus of topological defects which takes
advantage of well-developed theorems and techniques in topological field
theory.  Our discussion focuses on finite symmetries, and we give indications
for a generalization to other symmetries.  We treat quotients and quotient
defects (often called ``gauging'' and ``condensation defects''), finite
electromagnetic duality, and duality defects, among other topics.  We include
an appendix on finite homotopy theories, which are often used to encode finite
symmetries and for which computations can be carried out using methods of
algebraic topology.  Throughout we emphasize exposition and examples over a
detailed technical treatment.

 \end{abstract}
\maketitle

The study of symmetry in quantum field theory is longstanding with many
points of view.  For a relativistic field theory in Minkowski spacetime, the
symmetry group of the theory is the domain of a homomorphism to the group of
isometries of spacetime; the kernel consists of \emph{internal} symmetries
that do not move the points of spacetime.  It is these internal
symmetries---in Wick-rotated form---that are the subject of this paper.
Higher groups, which have a more homotopical nature, appear in many recent
papers and they are included in our treatment.  The word `symmetry' usually
refers to invertible transformations that preserve structure, as in Felix
Klein's \emph{Erlangen program}, but one can also consider algebras of
symmetries---e.g., the universal enveloping algebra of a Lie algebra acting
on a representation of a Lie group---and in this sense symmetries can be
non-invertible.

Quantum field theory affords new formulations of symmetry beyond what one
usually encounters in geometry.  If a Lie group~$G$ acts as symmetries of an
$n$-dimensional field theory~$F$, then one expresses the symmetry as a larger
theory in which there is an additional background (nondynamical) field: a
connection on a principal $G$-bundle, i.e., a gauge field for the group~$G$.
This formulation resonates with geometry, where a $G$-symmetry is often
expressed as a fibering over a classifying space for the group~$G$.  But in
field theory one can go further and often express the symmetry on~$F$ in terms
of a boundary theory of an $(n+1)$-dimensional topological field theory~$\sigma
$.  This idea has been exploited in many contexts; a nonexhaustive list
includes~\cite{Wi1,BM,MS,KWZ,FT1,GK,ABEHS}.  In a related picture, following the
influential paper~\cite{GKSW}---for an early exploration in the context of
2-dimensional rational conformal field theory, see~\cite{FFRS}---symmetries in
field theory are usually expressed in terms of topological defects in the
theory.  These defects act as operators on state spaces, and defects can be
used in other ways too; their topological nature makes them flexible and
powerful.

  \begin{figure}[ht]
  \centering
  \includegraphics[scale=.4]{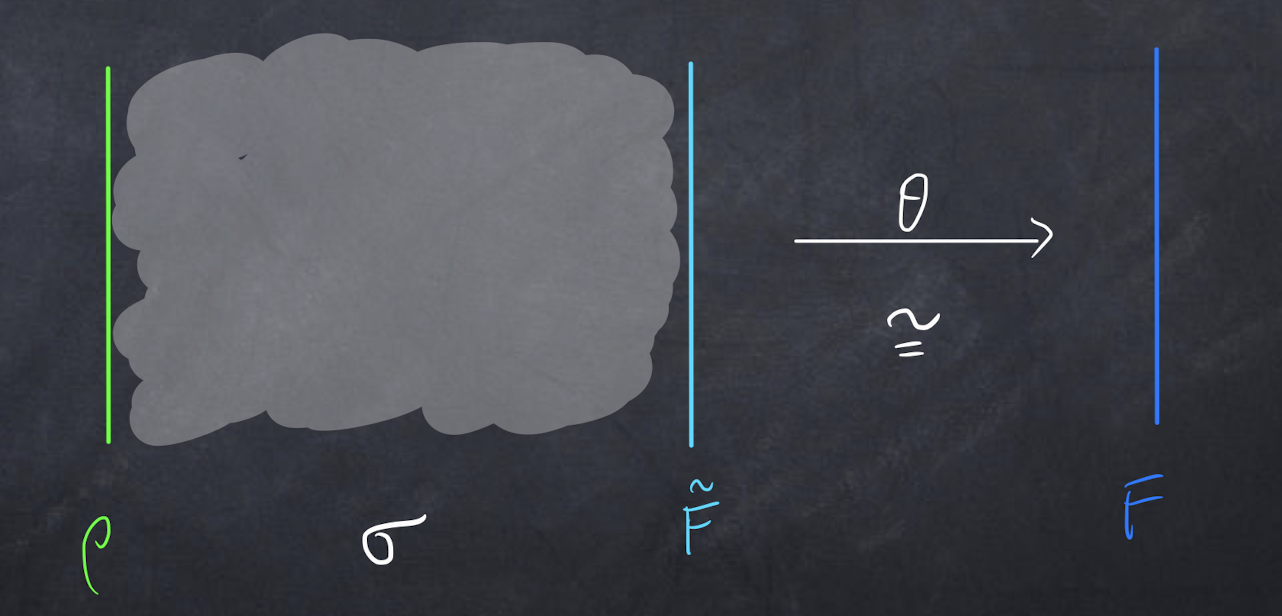}
  \vskip -.5pc
  \caption{The sandwich picture of $\sr$ acting on~$F$: the $\sr$-module
  structure on~$F$}\label{fig:33}
  \end{figure}

\begin{figure}[ht]
  \centering
  \includegraphics[scale=1.3]{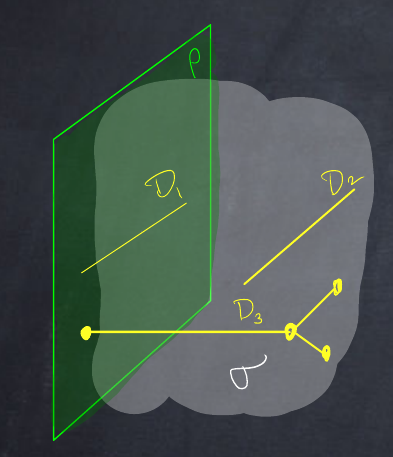}
  \vskip -.5pc
  \caption{The quiche~$\sr$, a $\rho $-defect~$D_1$ (on the boundary), and
  $\sr$-defects $D_2$ (in the bulk) and $D_3$ (emanating from the
  boundary)}\label{fig:37}  
  \end{figure}

Our starting point here is an old idea: the separation of an abstract symmetry
structure from a concrete realization as symmetries of some object.  The advent
of abstract groups~\cite{W} was a significant development in mathematics, as
was the advent of abstract algebras.  We offer an abstract symmetry structure
in the context of field theory as Definition~\ref{thm:9} and its concrete
realization on a field theory as Definition~\ref{thm:11}.  For broad conceptual
purposes one can analogize a field theory to a linear representation of a Lie
group or to a module over an algebra, and these analogies inspire some of our
nomenclature, for example the use of \emph{module} for a boundary theory.  The
essential content of our definition is that the action of a ``symmetry
algebra'' on an $n$-dimensional field theory~$F$ expresses~$F$ as a
\emph{sandwich} $\rho \otimes _\sigma \tF$ in which $\sigma $~is an
$(n+1)$-dimensional topological field theory; $\rho $~is a topological right
boundary theory of~$\sigma $, often assumed to be \emph{regular} or
\emph{Dirichlet}; and $\tF$~is a left boundary theory of~$\sigma $, which
typically is not topological.  The sandwich---the dimensional reduction
of~$\sigma $ on an interval with endpoints colored by~$\rho $
and~$\tF$---together with an isomorphism~$\theta $ to the original theory~$F$,
is depicted in Figure~\ref{fig:33}.  Defects supported away from
$\tF$-boundaries belong to the topological theory~$\sigma$ with its topological
right boundary theory~$\rho $, the pair~$\sr$ that comprises the abstract
symmetry structure.  We introduce the term \emph{$n$-dimensional
quiche}\footnote{\label{quiche} The term `quiche' stands in for the open-face
version of the sandwich in Figure~\ref{fig:33} with the boundary theory~$\tF$
removed; only $\sr$~remains as in Figure~\ref{fig:37}.  Defects can be embedded
in the filling, can stick to the crust, or can do both.  We use the phrase
`$n$-dimensional quiche' for both the case in which $\sigma $~is a full
$(n+1)$-dimensional topological field theory and the case in which $\sigma $~is
a once-categorified $n$-dimensional field theory; see footnote~\footref{OC}
below.} for the pair~$\sr$.  These topological defects act in the quantum field
theory~$F$ by transport via the isomorphism~$\theta $, but they can be
manipulated universally in the \emph{topological} field theory~$\sr$
independently of any particular \emph{$\sr$-module}.  In this sense
\emph{$\sr$-defects} are analogous to elements of an abstract algebra, a point
of view stressed in~\cite{F4}.  This also provides a connection to the work
of~\cite{GKSW}, in which the role of topological defects in implementing
symmetries is developed.  We go further and take the topological defects out of
a general quantum field theory and embed them into a topological field
theory~$\sr$, which opens up the application of powerful topological methods
and theorems.  In particular, we indicate how to use the cobordism hypothesis
to develop a complete calculus of topological defects.

Another aspect of our work is a clarification of the role of topological
defects and their relation to symmetries.  In \S\ref{sec:2} we give a
definition of local and global topological defects, together with a description
of background fields in the presence of these defects.  We also construct a
composition law on topological defects.  We stress that the composition law
preserves the codimension of defects: the composition of two codimension~$\ell
$ defects is a codimension~$\ell $ defect, notwithstanding claims one often
hears to the contrary.

The sandwich presentation of a theory appears in earlier talks and papers, such
as \cite{Te,GKSW,FT1,PSV,GK}.  Also, theories of defects in topological field
theory are not new, and the link between defects and symmetry is
well-established: \cite{FFRS,KaSa,FSV} is a small sample of older literature.
However, as we just stressed when introducing the notion of a quiche~$\sr$, one
of our main contributions is the development of a calculus of $\sr$-defects as
abstract symmetries that obey universal relations which apply to \emph{all}
$\sr$-module theories.  In previous works, equations among topological defects
are derived by computations in a particular theory, not in the abstract
symmetry structure~$\sr$.  Furthermore, we show how \emph{fully
local}\footnote{We use the more descriptive `fully local' for what is often
called `fully extended'.}  topological field theory provides powerful tools to
make these computations.

Topological field theory imposes strong finiteness constraints, called
\emph{dualizability}, but one can relax those constraints as follows.  In the
lingo, one takes~ $\sigma $ to be a \emph{once-categorified $n$-dimensional
topological field theory} and takes~$\rho $ and~$\tF$ to be a \emph{relative
field theory} to~$\sigma $.  We make comments in this direction
throughout,\footnote{\label{OC}See Remark~\ref{thm:4}(1),
Remark~\ref{thm:6}(3), Remark~\ref{thm:105}, Remark~\ref{thm:10}(8),
Remark~\ref{thm:12}(1,2), Example~\ref{thm:38}, and the introduction to
Appendix~\ref{sec:5}.} though in almost all of our examples $\sigma $~is a full
$(n+1)$-dimensional theory.  Under the basic analogy of field theory with Lie
group representations, $\sr$-modules with $\sigma $~a \emph{full}
$(n+1)$-dimensional theory correspond to representations of \emph{finite}
groups.  It is desirable to investigate in greater depth analogs of infinite
discrete group and compact Lie group symmetries.
 
We begin in Section~\ref{sec:1} with a quick exposition of groups and
algebras of symmetries.  The case of algebras~(\S\ref{subsec:1.2}) provides
the most direct motivation for our definitions.  We discuss quotients and
projective symmetries in these contexts; both have echos in field theory.  In
Section~\ref{sec:2} we review formal ideas in Wick-rotated field theory.
The basic framework sees a field theory as a linear representation of a
geometric bordism category, an idea most developed for \emph{topological}
theories.  We introduce domain walls, boundaries, and more general defects.
Our treatment here is quite heuristic, favoring exposition over precision; a
technically complete account is possible for topological theories.  As
already stated, our main definitions are in Section~\ref{sec:3}.  We
illustrate with a few examples in~\S\ref{subsec:3.3}, deferring more details
and more intricate examples to~\S\ref{sec:4}.  Section~\ref{sec:3} concludes
with a general discussion of quotients by symmetries and finite
electromagnetic duality, which realizes quotients for a special class of
symmetries.
 
Section~\ref{sec:4} illustrates our formulation of symmetry through a series of
examples.  The case of symmetries in quantum mechanics~(\S\ref{subsec:4.2}),
which we linger over, makes contact with the motivating scenario of modules
over an algebra and also provides valuable intuition for higher dimensional
theories.  From there we move on to examples in higher dimensions and examples
with higher symmetry.  We focus on the composition law for defects, which often
does not have a valid expression in classical terms; see~\S\ref{subsec:4.4}.
We conclude in Section~\ref{sec:6} with a discussion of quotient defects,
duality defects, and some applications thereof.
 
There is a class of topological field theories constructed by a finite version
of the Feynman path integral.  These \emph{finite homotopy theories} are the
subject of Appendix~\ref{sec:5}.  In the basic case one sums over maps into a
$\pi $-finite space.  (Significantly, one can drop $\pi $-finiteness and
construct a once-categorified theory from \emph{any} topological space.)  These
theories are a fertile laboratory for general concepts in field theory, and as
well they are often the basis of a symmetry structure which acts on quantum
field theories of interest.  By their nature they are amenable to computations
based on topological rather than analytic techniques.  We sketch how to
manipulate defects in such theories.
 
As already mentioned, our goal in this paper is to illustrate the sandwich
formulation of symmetries and the resulting topological calculus of defects
rather than to give a complete and rigorous development.  The recent
papers~\cite{BR,RV,SW} are both relevant and illustrate the substantial
technical work involved.  We also remark that unitarity is not brought in here,
but of course it is important for physical applications.  Our referencing is
hardly complete; we refer the reader to the recent Snowmass whitepaper on
generalized symmetries~\cite{CDIS} as well as the Snowmass whitepaper on
physical mathematics~\cite[\S2.5]{BFMNRS} for more perspective, examples, and
references.  The lecture notes~\cite{F3} cover much of the same ground, but
there are some different examples developed there as well.  See also the
conference proceedings~\cite{F4}, which contains additional motivation and an
application to line defects in 4-dimensional gauge theories.
 
We offer this work as a tribute to Vaughan Jones, whose untimely passing is a
great loss, both mathematically and personally.  We treasure the memories of
our interactions with Vaughan in the realms of mathematics, physics, and
well beyond. 
 
Ibou Bah, David Ben-Zvi, Mike Freedman, Dan Friedan, Mike Hopkins, Theo
Johnson-Freyd, Alexei Kitaev, Justin Kulp, Kiran Luecke, Ingo Runkel, Will
Stewart, and Jingxiang Wu offered valuable comments, for which we thank them
all.  This research is under the auspices of the Simons Collaboration on Global
Categorical Symmetry, and we thank our colleagues for their valuable feedback.
We are grateful to Andrew Moore for his expert rendering of \autoref{fig:38}.
We also thank the referee for a careful reading and many useful comments.

\bigskip\bigskip
{\small
\def\reftext{References}
\renewcommand{\tocsection}[3]{%
  \begingroup 
   \def\tmp{#3}%
   \ifx\tmp\reftext
  \indentlabel{\phantom{1}\;\;} #3%
  \else\indentlabel{\ignorespaces#1 #2.\;\;}#3%
  \fi\endgroup}
\tableofcontents
}

   \section{Groups and algebras of symmetries}\label{sec:1}

We review two settings for symmetry in mathematics: groups of symmetries~
(\S\ref{subsec:1.1}) and algebras of symmetries~(\S\ref{subsec:1.2}).
(Appendix~\ref{sec:5} generalizes the former in a topological setting.)  In
each instance we restrict our exposition for the most part to the simplest case
of \emph{finite} symmetries, though many considerations generalize beyond the
finite case.  In particular, the groups and algebras carry no topology.  For
each there is abstract ``symmetry data'' as well as concrete realizations of
that symmetry data.  The distinction between abstract symmetry and its concrete
realizations serves us well when we come to field theory in~\S\ref{sec:3}.
Here, in~\S\ref{subsec:1.3}, we discuss quotients in both the group and algebra
contexts.  We conclude with brief discussions of projective
symmetries~(\S\ref{subsec:1.4}) and higher algebras of
symmetries~(\S\ref{subsec:1.5}).

  \subsection{Fibering over~$BG$}\label{subsec:1.1}

Let $G$~be a finite group.\footnote{The discussion in this subsection
generalizes to a Lie group~$G$ acting on a smooth manifold~$X$, in which case
we incorporate connections, replacing~$BG$ with~$B\mstrut _{\nabla }G$, as
in~\cite{FH1}.}  A \emph{classifying space}~$BG$ is derived from a
contractible topological space~$EG$ equipped with a free $G$-action by taking
the quotient; the homotopy type of~$BG$ is independent of choices.  If $X$~is
a topological space equipped with a $G$-action, then the \emph{Borel
construction} is the total space of a fiber bundle
  \begin{equation}\label{eq:1}
     \begin{gathered} \xymatrix@C-28pt{X_G\ar[d]^{\pi }&=EG\times \mstrut _GX\\BG}
     \end{gathered} 
  \end{equation}
with fiber~$X$.  If $*\in BG$ is a chosen point, and we choose a basepoint in
the $G$-orbit in~$EG$ labeled by~$*$, then the fiber~$\pi \inv (*)$ is
canonically identified with~$X$.  We say the abstract (group) symmetry data
is the pair~$(BG,*)$, and a realization of the symmetry~$(BG,*)$ on~$X$ is a
fiber bundle~\eqref{eq:1} over~$BG$ together with an identification of the 
fiber over~$*\in BG$ with~$X$.

  \begin{remark}[]\label{thm:1}
 We use a pair~$(\sX,*)$ consisting of a $\pi $-finite topological space~$\sX$
and a basepoint~$*\in \sX$ as a generalization of~$(BG,*)$.  (See
\autoref{thm:53} for the definition of $\pi $-finite.)  In this context the
based loop space~$\Omega \sX$ is a higher, homotopical version of a finite
group: a \emph{grouplike $A_{\infty }$-space}~\cite{Sta}, which is the
generalization of the more classical \emph{$H$-group}~\cite[\S1.5]{Sp} that
takes into account higher coherence.  For simplicity we call these \emph{higher
finite groups}.
  \end{remark}

  \subsection{Algebras of symmetries}\label{subsec:1.2}
 
Let $A$~be an algebra, and for simplicity suppose that the ground field
is~$\CC$.  For our expository purposes it suffices to assume that $A$~and the
modules that follow are finite dimensional.  Let $R$~be the right
\emph{regular} $A$-module, i.e., the vector space~$A$ furnished with the
right action of~$A$ by multiplication.  The pair~$(A,R)$ is abstract
(algebra) symmetry data: the action of~$(A,R)$ on a vector space~$V$ is a
pair~$(L,\theta )$ consisting of a left $A$-module~$L$ together with an
isomorphism of vector spaces
  \begin{equation}\label{eq:2}
     \theta \:R\,\otimes  \mstrut _AL\xrightarrow{\;\;\cong
     \;\;}V.  
  \end{equation}

  \begin{example}[]\label{thm:2}
 Let $G$~be a finite group.  The \emph{group algebra} $A=\CC[G]$ is the free
vector space on the set~$G$, which is then a linear basis of~$A$; multiply
basis elements according to the group law in~$G$.  A left $A$-module~$L$ is
canonically identified as a linear representation of~$G$.  The tensor product
in~\eqref{eq:2} recovers the vector space which underlies the representation.
In the setup of~\S\ref{subsec:1.1}, take~$X=L$ to construct a vector bundle
$L_G\to BG$ whose fiber over~$*\in BG$ is~$L$.
  \end{example}

Observe that the right regular module satisfies the algebra isomorphism 
  \begin{equation}\label{eq:3}
     \End_A(R)\cong A, 
  \end{equation}
where the left hand side is the algebra of linear maps $R\to R$ that commute
with the right $A$-action.

  \subsection{Quotients}\label{subsec:1.3}

In the topological setting of~\S\ref{subsec:1.1}, the total space~$X_G$ of
the Borel construction plays the role of the quotient space~$X/G$.  Indeed,
if $G$~acts freely on~$X$, then there is a homotopy equivalence $X_G\simeq
X/G$; in general, $X_G$~is called the \emph{homotopy quotient}.   
 
For any map $f\:Y\to BG$ of topological spaces we form the \emph{homotopy
pullback}\footnote{In one model for the homotopy pullback, a point of~$Z$ is a
triple $(y,e,\gamma )$ in which $y\in Y$, $e\in X_G$, and $\gamma $~is a path
in~$BG$ from ~$f(y)$ to~$\pi (e)$.  If $\pi $~is a fiber bundle, then the usual
Cartesian pullback is a realization of the homotopy pullback.}
  \begin{equation}\label{eq:4}
     \begin{gathered}
     \xymatrix{&Z\ar@{-->}[dl]\ar@{-->}[dr]\\Y\ar[dr]_{f}&&X_G\ar[dl]^{\pi
     }\\&BG} 
     \end{gathered} 
  \end{equation}
If $Y$~is path connected and pointed, then there is a homotopy
equivalence\footnote{The based loop space~$\Omega Y$ is a grouplike $A_\infty
$-space, so it has a classifying space; see \autoref{thm:1} and the references
therein for more details.}  $Y\simeq B(\Omega Y)$.  If $BG$~also has a
basepoint, and if the map $f\:Y\to BG$ is basepoint-preserving, then $f$~ is
the classifying map\footnote{The classifying space construction is a functor
from groups and homomorphisms to topological spaces and continuous maps.} of a
homomorphism $\Omega Y\to G=\Omega BG$, at least in the $A_\infty $-homotopical
sense.  In this case $Z$~is the homotopy quotient of~$X$ by the action
of~$\Omega Y$.  As a special case, if $G'\subset G$~is a subgroup, and
$Y=BG'\to BG$ is the classifying map of the inclusion, then $Z$~is homotopy
equivalent to the total space of the Borel construction~$X_{G'}$.  Hence
\eqref{eq:4}~is a generalized quotient construction.  For~$G'=\{e\}$ we
have~$Y=*$ and we recover~$Z=X$, as in~\S\ref{subsec:1.1}.
 
There is an analogous story in the setting~\S\ref{subsec:1.2} of algebras.
An \emph{augmentation} of an algebra~$A$ is an algebra homomorphism $\epsilon
\:A\to \CC$.  Use~$\epsilon $ to endow the scalars~$\CC$ with a right
$A$-module structure: set $\lambda \cdot a=\lambda \epsilon (a)$ for $\lambda
\in \CC$,\,$a\in A$.  If $L$~is a left $A$-module, the vector space 
  \begin{equation}\label{eq:5}
     Q=\CC\otimes  \mstrut _AL=:\CC\otimes \mstrut _\epsilon
     L 
  \end{equation}
plays the role of the ``quotient'' of~$L$ by~$A$. 

  \begin{example}[]\label{thm:3}
 For the group algebra of a finite group~$G$, there is a natural augmentation 
  \begin{equation}\label{eq:6}
     \begin{aligned} \epsilon \:\CC[G]&\longrightarrow \CC \\
      \sum\limits_{g\in G}\lambda _gg&\longmapsto \sum\limits_{g\in G}\lambda
      _g\end{aligned} 
  \end{equation}
where $\lambda _g\in \CC$.  If $L$~is a representation of~$G$, extended to a
left $\CC[G]$-module, then the tensor product~\eqref{eq:5} is the vector space
of \emph{coinvariants} (i.e., the quotient of~$L$ by the subspace generated by
$g\cdot \ell -\ell $, where $g\in G$, $\ell \in L$); 
  \begin{equation}\label{eq:7}
     1\otimes \ell =1\otimes g\cdot \ell ,\qquad \ell
     \in L, \quad g\in G,  
  \end{equation}
in the tensor product with the augmentation.  More generally, an augmentation
of~$\GA$ is induced from a character of~$G$, i.e., a 1-dimensional linear
representation of~$G$.

As a particular case, let $S$~be a finite set equipped with a left $G$-action,
and let $L=\CC\langle S \rangle$ be the free vector space generated by~$S$.
Then for the natural augmentation~\eqref{eq:6}, the vector space $\CC\otimes
\mstrut _\epsilon L$ can be identified with~$\CC\langle S/G \rangle$, the free
vector space on the quotient set.  More generally, any character~$\chi \:G\to
\Cx$ induces a line bundle $L_\chi \to S\gpd G$ over the \emph{groupoid} or
\emph{stack} quotient, and for the associated augmentation~$\epsilon \:\GA\to
\CC$ the space of coinvariants $\CC\otimes _\epsilon L$ is~$H_0(S\gpd G;L_\chi
)$.  This is also isomorphic to the space of global sections of the line bundle
$L_\chi \to S\gpd G$.
  \end{example}

We can form the ``sandwich''~\eqref{eq:5} with any right $A$-module in place
of the augmentation.  For $A=\CC[G]$, if $G'\subset G$~is a subgroup, then
$\CC\langle G'\backslash G  \rangle$ is a right $G$-module; for~$G'=G$ it
reduces to the augmentation module~\eqref{eq:6}.  If $L$~is a
$G$-representation, then 
  \begin{equation}\label{eq:8}
     \CC\langle G'\backslash G \rangle\otimes \mstrut _{\CC[G]}L\cong \CC\otimes
     \mstrut _{\CC[G']}L 
  \end{equation}
is the vector space of coinvariants of the restricted $G'$-representation. 

  \begin{remark}[]\label{thm:77}
 There is a potential mismatch in our description of quotients in topology and
quotients in algebra.  To align our accounts, one should use \emph{derived}
quotients in algebra, and so replace the tensor product in~\eqref{eq:8} with
the (left) derived tensor product, i.e., with Tor.  Then one computes the
entire complex homology of the Borel quotient, not just the free vector space
generated by its components.  However, this mismatch does not occur for finite
groups in characteristic zero.
  \end{remark}

  \subsection{Projective symmetries}\label{subsec:1.4}

We begin with an example in the algebra framework~\S\ref{subsec:1.2}.  Let
$G$~be a finite group, and suppose 
  \begin{equation}\label{eq:9}
     1\longrightarrow \Cx\longrightarrow G^\tau \longrightarrow
     G\longrightarrow 1 
  \end{equation}
is a central extension.  Let $L^\tau \to G$ be the complex line bundle
associated to the principal $\Cx$-bundle~\eqref{eq:9}.  Define the
\emph{twisted group algebra} 
  \begin{equation}\label{eq:10}
     A^\tau =\bigoplus\limits_{g\in G}L^\tau _g. 
  \end{equation}
Then $A^\tau $~inherits an algebra structure from the group structure
of~$G^{\tau }$.  Furthermore, $G^\tau \subset A^\tau $ lies in the group of
units.  An $A^\tau $-module restricts to a linear representation of~$G^\tau $
on which the center~$\Cx$ acts by scalar multiplication, and vice versa.
Observe that there is no analog of the augmentation~\eqref{eq:6} unless the
central extension~\eqref{eq:9} splits; indeed, an augmentation induces a
splitting.  (Restrict $A^\tau \to \CC$ to $G^\tau \subset A^\tau $.)  More
generally, if $G'\subset G$ is a subgroup, then a splitting of the restriction
of~\eqref{eq:9} over~$G'$ induces an $A^\tau $-module structure on~$\CC\langle
G'\backslash G \rangle$, and we can use this to define the quotient by~$G'$, as
in~\eqref{eq:8}.  Absent the splitting, the projectivity obstructs the quotient
construction.  There is an analogous story in the context
of~\S\ref{subsec:1.1}; see Remark~\ref{thm:78}.

  \begin{remark}[]\label{thm:79}
 That central extensions obstruct augmentations has an echo in field theory:
't~Hooft anomalies obstruct the quotient operation (gauging) by a symmetry. 
  \end{remark}

  \subsection{Higher algebra}\label{subsec:1.5}

The higher versions of finite groups in Remark~\ref{thm:1} have an analog in
algebras as well.  For example, a \emph{fusion category}~$\sA$ is a ``once
higher'' version of a finite dimensional semisimple algebra, and there is a
well-developed theory of modules over a fusion category~\cite{EGNO}.  In
particular, $\sA$~is a right module over itself, the right regular module.  A
finite group~$G$ gives rise to the fusion category $\sA=\Vect[G]$ of finite
rank vector bundles over~$G$ with convolution product.  The analog of an
augmentation for a fusion category~$\sA$ is a \emph{fiber functor}---a tensor
functor $\sA\to \Vect$---and for $\sA=\Vect[G]$ the natural choice is
pushforward under the map $G\to *$ to a point.  Just as a character produces an
augmentation of~$\GA$, a central extension of~$G$ by~$\Cx$ produces a fiber
functor $\Vect[G]\to \Vect$.  If $\omega $~is a cocycle which represents an
element of~$H^3(G;\Cx)$, then there is a twisted variant $\Vect^\omega [G]$,
but there is no fiber functor if the cohomology class of~$\omega $ is nonzero;
see \autoref{thm:28}.
 
Higher categorical generalizations of fusion categories are a topic of much
current interest and development, and presumably have analogs of the
constructions presented above.

   \section{Formal structures in field theory}\label{sec:2}

We review basic notions in Wick-rotated field theory on compact manifolds.
Segal~\cite{S1} initiated this framework for 2-dimensional conformal field
theories.  Recently, Kontsevich-Segal~\cite{KS} discuss general quantum field
theories from this viewpoint.  The entire story is most developed for
topological field theories, beginning with Atiyah~\cite{A}, who made the
connection to Thom's theory of bordism; continuing with the introduction and
development of fully local field theory~\cite{F1,La,BD,L}; and then with the
connection of fully local field theory to defects~\cite{K}.  (We have only
skimmed the surface of relevant literature.)  Our exposition emphasizes the
metaphor:
  \begin{equation}\label{eq:95}
     \textnormal{field theory $\sim$  representation of a Lie
     group} 
  \end{equation}
We briefly touch on axioms~(\S\ref{subsec:2.1}), domain
walls~(\S\ref{subsec:2.2}), boundary theories and
anomalies~(\S\ref{subsec:2.3}), and general defects~(\S\ref{subsec:2.4}).
There is no pretense of rigor or completeness here.  For the topological case
there are rigorous definitions in the literature for most of what we write; a
few items are still under development.

  \subsection{Axioms}\label{subsec:2.1}

The discrete parameters that determine the ``type'' of field theory are a
nonnegative integer~$n$ and a collection~$\sF$ of $n$-dimensional
fields.\footnote{\label{field}\,`Field' has a precise meaning:
see~\cite{FT3,FH1} or~\cite{nLab}.  Let $\Man_n$ be the category whose objects
are smooth $n$-manifolds and whose morphisms are local diffeomorphisms.  There
is a notion of \emph{sheaves} on this category, with respect to the
\emph{Grothendieck topology} of open covers.  A \emph{field} in dimension~$n$
is a sheaf $\sF\:\Man_n\op\to \Set_\Delta $ with values in the category of
simplicial sets, i.e., a functor $\Man_n\op\to \Set_\Delta $ that satisfies the
sheaf condition.  (This could be a single field or a collection of fields; we
do not define irreducibility here.)  Heuristically, a field is a local object
one can attach to an $n$-manifold.  A field on an $n$-manifold~$X$ is a
0-simplex in~$\sF(X)$.} One thinks of~$n$ as the dimension of \emph{spacetime}
and $\sF$~as the collection of \emph{background} fields.  Some fields have a
topological flavor---orientations, spin structures, etc.---while others are
more geometric---Riemannian metrics, connections for a fixed gauge group,
scalar fields, spinor fields, etc.  More precisely, a field~$\sF$ is
topological if it is a locally constant sheaf, in which case it corresponds to
a \emph{tangential structure}~\cite{Las}.  (There are no fluctuating fields;
they have already been ``integrated out'' before the formulation in this axiom
system.  Nor is there spacetime; we work in the Wick-rotated setting in which
every nonzero tangent vector is spacelike.)  There is a bordism
category~$\Bord_n(\sF)$ of $n$-dimensional smooth manifolds~$M$ with corners
equipped with a choice of fields, i.e., an object in~$\sF(M)$.  We refer to the
literature for more details, say~\cite{L,CS} for the fully local topological
case and~\cite{KS} for the nonextended general case.  We assume that all
\emph{topological} theories are fully local (i.e., fully extended downward in
dimension), in which case $\Bord_n(\sF)$~is a symmetric monoidal $n$-category.
In the nonextended case, we interpret `$\Bord_n(\sF)$' as a 1-category
$\Bord_{\langle n-1,n \rangle}(\sF)$ whose objects are closed $(n-1)$-manifolds
and whose morphisms are bordisms between them.  Let $\sC$~be a symmetric
monoidal $n$-category.\footnote{\label{loop}One can replace `$n$-category' with
`$(\infty ,n)$-category' in our exposition.  Also, we implicitly assume that
$\Omega ^n\sC=\CC$ and $\Omega ^{n-1}\sC$ is equivalent to the category~$\Vect$
of vector spaces or to the category of $\zt$-graded vector spaces.  However,
these assumptions can be relaxed.  For the notation, recall that
looping~$\Omega \sC$ of the symmetric monoidal $n$-category~$\sC$ is the
symmetric monoidal $(n-1)$-category $\Hom(1,1)$ of endomorphisms of the tensor
unit.  We can iterate the looping construction.}  A topological field theory is
a symmetric monoidal functor
  \begin{equation}\label{eq:11}
     F\:\Bord_n(\sF)\longrightarrow \sC. 
  \end{equation}
Recall that the cobordism hypothesis~\cite{L} enables a calculus of such
functors in terms of duality data inside the codomain category~$\sC$.  Turning
to \emph{nontopological} theories, a similar calculus is not in place and is a
subject of wide interest.  In the meantime, we confine ourselves to
\emph{nonextended} nontopological theories, and so replace~$\sC$ by the
1-category~$t\!\Vect$ of suitable complex topological vector spaces under
tensor product~\cite{KS,Wed}.  Finally, a field theory may be evaluated in
smooth families parametrized by a smooth manifold~$S$, and it should behave
well under base change.  Therefore, \eqref{eq:11}~should be sheafified
over~$\Man$, the site of smooth manifolds and smooth diffeomorphisms~\cite{ST}.
This applies to both topological and nontopological field theories.

  \begin{remark}[]\label{thm:4}
 \ 
 \begin{enumerate}[label=\textnormal{(\arabic*)}]

 \item A topological field theory imposes strong finiteness.  In the
metaphor~\eqref{eq:95}, a \emph{topological} field theory is analogous to a
representation of a \emph{finite} group.  We also use the notion of a
\emph{once-categorified $n$-dimensional field theory}, which in the
topological case is a symmetric monoidal functor $\Bord_n(\sF)\to \sC$, where
$\sC$~is a symmetric monoidal $(n+1)$-category.  With typical choices of
codomain~$\sC$, in the top dimension such a theory assigns a vector space
rather than a complex number.  The finiteness conditions are more relaxed
than in a full topological field theory; for example, the vector spaces
attached to top dimensional closed manifolds need not be finite dimensional in
a once-categorified topological theory.

 \item The collection of field theories of a fixed dimension~$n$ on a fixed
collection~$\sF$ of background fields has an associative composition law:
juxtaposition of quantum systems with no interaction, sometimes called
`stacking'.  We denote this composition law as a tensor product.  For example,
on a closed $(n-1)$-manifold~$Y$ the state space~$(F_1\otimes F_2)(Y)$ is the
tensor product $F_1(Y)\otimes F_2(Y)$ of the state spaces of the constituent
systems.  There is a unit theory~$\bone$ for this operation.  For example, if
$F_1,F_2$~are theories, and $Y$~is a closed $(n-1)$-manifold with background
fields, then $(F_1\otimes F_2)(Y)=F_1(Y)\otimes F_2(Y)$.  The unit theory has
$\bone(Y)=\CC$; there is a single state on every space.  There is then a
subcategory of units for the composition law: \emph{invertible field theories}.

 \end{enumerate}
  \end{remark}

  \begin{figure}[ht]
  \centering
  \includegraphics[scale=.4]{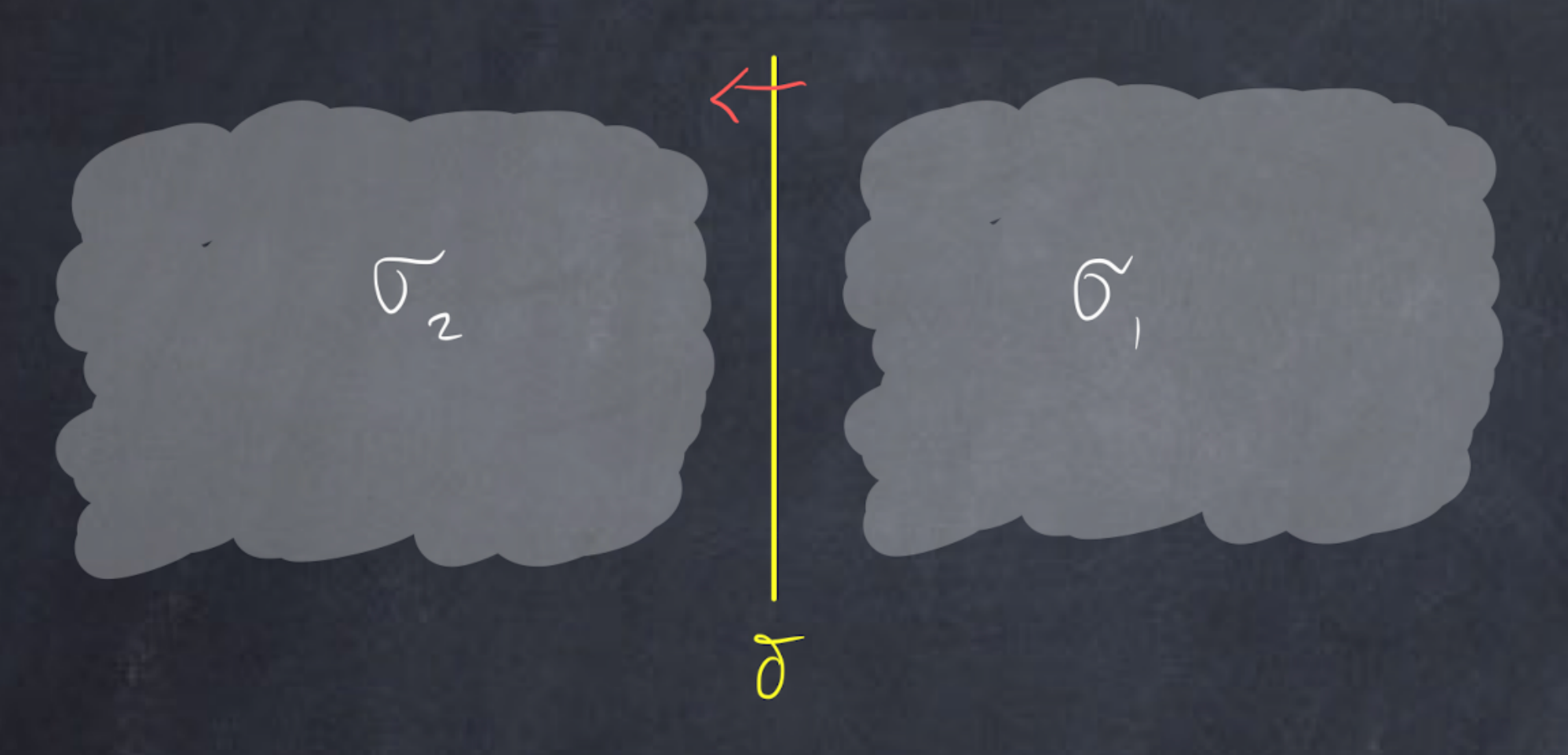}
  \vskip -.5pc
  \caption{A domain wall $\delta \:\sigma _1\to \sigma _2$}\label{fig:1}
  \end{figure}

  \subsection{Domain walls}\label{subsec:2.2}

\begin{figure}[ht]
  \centering
  \includegraphics[scale=1.7]{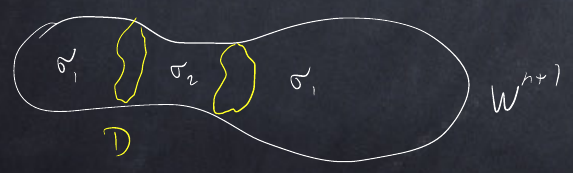}
  \vskip -.5pc
  \caption{Domain walls in the manifold~$W$}\label{fig:2}
  \end{figure}

Let $\sigma _1,\sigma _2$ be $(n+1)$-dimensional theories on background
fields~$\sF_1,\sF_2$ with common codomain~$\sC$.  (Recall the notation for
background fields in footnote~\footref{field}.  In the sequel `$\sigma $'
usually denotes a \emph{topological} field theory, but in this section the
theories~$\sigma_i$ need not be topological.) Morally, a {domain wall} $\delta
\:\sigma _1\to \sigma _2$ is the analog\footnote{However, $\sigma _1$~and
$\sigma _2$~need not be algebra objects in the symmetric monoidal category of
field theories.} of a bimodule, so we use the convenient terminology `$(\sigma
_2,\sigma _1)$-bimodule'; see Figure~\ref{fig:1} for a depiction.  In the
topological case this can be made precise: one can build a higher category in
which a domain wall is a sufficiently dualizable 1-morphism.  In the general
case, the triple $(\sigma _1,\sigma _2,\delta )$ is a functor with domain a
bordism category of smooth $(n+1)$-dimensional manifolds with corners, each
equipped with a partition into regions labeled~`1' and~`2' separated by a
codimension one submanifold (with corners) which is ``$\delta $-colored''; the
codomain of the functor is~$\sC$.  The bordism category is illustrated in
Figure~\ref{fig:2} in the top dimension.  See \cite[Example 4.3.23]{L} for the
topological case, though of course the notion transcends the purely
topological.  The background fields on the domain wall form a sheaf~$\sF$ over
a category whose objects are $n$-manifolds embedded in a germ of an
$(n+1)$-manifold.  Furthermore, there are maps to bulk fields
  \begin{equation}\label{eq:12}
     \begin{gathered} \xymatrix{&\sF\ar[dl]\ar[dr]\\\sF_1&&\sF_2}
     \end{gathered} 
  \end{equation}
when we restrict all sheaves to these germs with the $n$-dimensional domain
wall deleted.  (In the topological case, when $\sigma _1=\sigma _2$, this data
is spelled out further in~\S\ref{subsec:2.5}.)  The background fields and the
correspondence~\eqref{eq:12} control the nature of the domain wall.  Thus we
can have geometric domain walls which depend on a Riemannian metric between
topological theories, or in the purely topological case a spin domain wall
between oriented theories, etc.  As a special case, a domain wall from the
tensor unit theory~$\bone$ to itself is an $n$-dimensional (absolute,
standalone) theory, though with $(n+1)$-dimensional fields instead of
$n$-dimensional fields.  More generally, we can tensor any domain wall $\delta
\:\sigma _1\to \sigma _2$ with an $n$-dimensional theory to obtain a new domain
wall.

There is a composition law on \emph{topological} domain walls that are parallel
in the sense they bound a cylindrical region:
  \begin{equation}\label{eq:13}
     \begin{gathered} \xymatrix{\sigma _1\ar[r]^{\delta '}\ar@/_1pc/[rr]_{\delta
     ''\circ \,\delta '} & \sigma _2\ar[r]^{\delta ''}& \sigma _3}
     \end{gathered} 
  \end{equation}

  \subsection{Boundary theories, anomalies, and anomalous theories}\label{subsec:2.3}

\begin{figure}[ht]
  \centering
  \includegraphics[scale=.37]{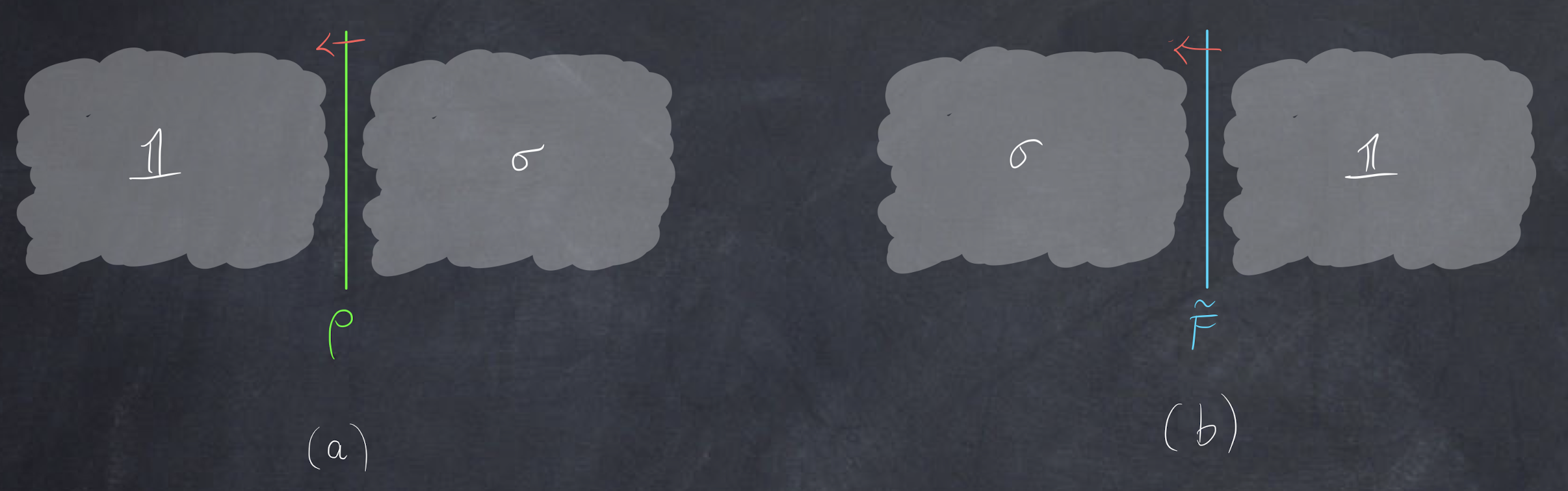}
  \vskip -.5pc
  \caption{(a) a right boundary theory (b) a left boundary
  theory}\label{fig:3} 
  \end{figure}

Following the metaphor of domain wall as bimodule, there are special cases of
right or left modules.  For field theory these are called \emph{right
boundary theories} or \emph{left boundary theories}, as depicted in
Figure~\ref{fig:3}.  (Normally, we omit the region labeled by the tensor unit
theory~`$\bone$' in the drawings.)  A right boundary theory of~$\sigma $ is a
domain wall $\sigma \to \bone$; a left boundary theory is a domain wall
$\bone\to \sigma $.  The nomenclature of right vs.~left may at first be
confusing; it does follow standard usage for modules over an algebra---the
direction (right or left) is that of the action of the algebra on the module.
In fact, following our general usage for domain walls, we sometimes use the
terms `right $\sigma $-module' and `left $\sigma $-module' for right and left
boundary theories.   
 
A special nomenclature is used for the special case in which the bulk theory
is invertible.  

  \begin{definition}[]\label{thm:5}
 Let $\alpha $~be an invertible $(n+1)$-dimensional field theory.  An
\emph{anomalous field theory}~$F$ with \emph{anomaly}~$\alpha $ is a left
$\alpha $-module.
  \end{definition}

\noindent
 The choice of left vs.~right is a convention we make.\footnote{Quite
generally in geometry, it is convenient to put structural actions on the
right (as, for example, the action of the structure group on a principal
bundle) and geometric actions on the left.}  We emphasize that the
background fields for~$\alpha $ and~$F$ may be different, as
in~\eqref{eq:12}.  For example, a free spinor field theory~$F$ in
3~dimensions is defined on spin Riemannian manifolds, whereas the associated
anomaly theory~$\alpha $ is topological: it is defined on a bordism category
of spin manifolds.  In other words, $\sF_F=\{\textnormal{metric,
spin structure}\}$ whereas $\sF_\alpha =\{ \textnormal{spin structure}\}$.

  \begin{remark}[]\label{thm:6}
 \ 
 \begin{enumerate}[label=\textnormal{(\arabic*)}]

 \item In the metaphor~\eqref{eq:95} of field theory as Lie group
representation, an anomalous field theory is a projective representation and
the anomaly is the cocycle that measures the induced central extension of
the Lie group.

 \item An $(n+1)$-dimensional topological field theory with a topological
boundary theory is defined as a functor out of a bordism category usually
denoted~$\Bord^\partial _{n+1}$; see~\cite[Example 4.3.22]{L}.

 \item A boundary theory of a once-categorified $n$-dimensional theory
(Remark~\ref{thm:4}(1)) is called a \emph{relative field theory}; it is
defined on a subcategory of~$\Bord_{n+1}$ which drastically constrains the
allowed $(n+1)$-manifolds~\cite{Ste}.  One can replace the
$(n+1)$-dimensional theories in~\S\ref{subsec:2.2} and~\S\ref{subsec:2.3}
with once-categorified $n$-dimensional theories.  Since finiteness conditions
for once-categorified topological field theories are relaxed, this leads to
wider applicability.
 \end{enumerate}
  \end{remark}

  \subsection{Defects}\label{subsec:2.4}
 
Domain walls and boundaries are special cases of the general notion of a
\emph{defect} in a field theory.  Our discussion here is specifically for
\emph{topological} theories; with modification, some aspects apply more
generally (see Remark~\ref{thm:104} below).  Defects are supported on
submanifolds, or more generally on stratified subsets.  Our goal in this
section is to outline a calculus of fully local topological defects based on
the cobordism hypothesis.  Just as fully local topological field theories are
generated by data associated to a point, so too can global defects be generated
by purely local data.  The nature of this local data depends on the codimension
of the defect, as we spell out below.
 
\begin{figure}[ht]
  \centering
  \includegraphics[scale=1.4]{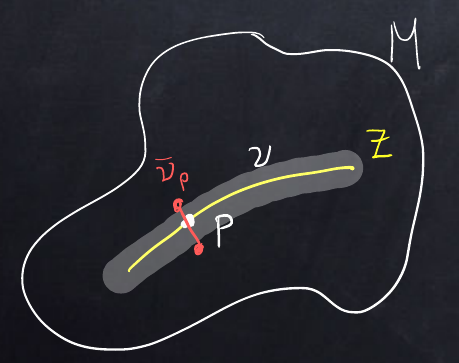}
  \vskip -.5pc
  \caption{The tubular neighborhood and link of a submanifold}\label{fig:4}
  \end{figure}

Suppose $m$~is a positive integer, $\sF$~is a collection of background
fields, and 
  \begin{equation}\label{eq:14}
     \sigma \:\Bord_m(\sF)\longrightarrow \sC 
  \end{equation}
is a topological field theory with values in a symmetric monoidal
$m$-category~$\sC$.  (In our application to quiche, $m=n+1$.)  We describe
defects of \emph{codimension}~$\ell $ in a $k$-dimensional manifold~$M$, where
$k\in \{1,\dots ,m\}$, $\ell \in \{1,\dots ,m\}$ and $\ell \le k$.  (There are
also defects of codimension~0, but they require a separate treatment which we
do not give here.)  Let $Z\subset M$ be a submanifold of codimension~$\ell $,
and let $\nu \subset M$ be an open tubular neighborhood of~$Z\subset M$; assume
the closure~$\bn$ is the total space of a fiber bundle $\bn\to Z$ with fiber
the closed $\ell $-dimensional disk.  The fiber over~$p\in Z$ is
denoted~$\bn_p$; its boundary~$\partial \bn_p$ is diffeomorphic to the $\ell
$-dimensional sphere~$S^{\ell -1}$.  It is the \emph{link} of~$Z\subset M$
at~$p$; see Figure~\ref{fig:4}.  Caution: the depicted point~$p\in Z$ is
\emph{not} an embedded point defect in~$Z$, but rather it is the support of the
local defect data.

  \begin{remark}[]\label{thm:95}
 Since $\sigma $~is a \emph{topological} field theory, we may assume that the
sheaf~$\sF$ is locally constant.  In the presence of a defect supported on~$Z$,
the sheaf~$\sF\res M$ is refined to a \emph{constructible} sheaf relative to
the stratification~$Z\subset M$.  We elaborate in~\S\ref{subsec:2.5}.  

  \end{remark}

  \begin{definition}[]\label{thm:7}
 Assume that $M$~is a closed manifold and $Z\subset M$ is a closed
submanifold. 

 \begin{enumerate}[label=\textnormal{(\arabic*)}]

 \item A \emph{local defect} at~$p\in Z$ is a morphism
  \begin{equation}\label{eq:15}
     \delta _p\in \Hom\bigl(1_{\Omega ^{\ell -1}\sC},\sigma (\partial \bn_p) \bigr). 
  \end{equation}
Observe that $\sigma (\partial \bn_p)$~is an object in~$\Omega ^{\ell -1}\sC$
(see footnote~\footref{loop} for the definition of the loop category),
$1_{\Omega ^{\ell -1}\sC}$~is the tensor unit object in~$\Omega ^{\ell -1}\sC$,
and $\delta _p$~is a 1-morphism in~$\Omega ^{\ell -1}\sC$.

 \item The \emph{transparent \textnormal{(}local\textnormal{)} defect} is
$\delta _p=\sigma (\bn_p)$, where we regard~$\bn_p$ as a bordism $\emptyset
^{\ell -1}\to \partial \bn_p$.

 \item A \emph{global defect} on~$Z$ is a morphism
  \begin{equation}\label{eq:16}
     \delta _Z\in \Hom\bigl(1_{\Omega ^{m -1}\sC},\sigma (\partial \bn) \bigr);
  \end{equation}
if $\Omega ^{m -1}\sC$~is a category of vector spaces, then $\delta _Z$~ is a
vector in a vector space.

 \item The \emph{transparent \textnormal{(}global\textnormal{)} defect} is
$\delta _Z=\sigma (\bn)$. 

 \end{enumerate} 
  \end{definition}

\noindent
 The transparent defects can be erased safely.

 We make several comments about this definition.

  \begin{remark}[]\label{thm:8}
 $W=M\setminus \nu $~is a compact manifold with boundary~$\partial
\bn$.  Define the bordism $W\:\partial \bn\to \emptyset $ by letting the
boundary be incoming.  If $\delta _Z$~is a global defect, evaluate the
theory on~$(M,Z,\delta _Z)$ as $\sigma (W)(\delta _Z)$.  This is of the same
type as the value~$\sigma (M)$ on the closed manifold~$M$: a complex number
if $\dim M=m$, a complex vector space if $\dim M=m-1$, etc.
  \end{remark} 

  \begin{remark}[]\label{thm:96}
 As written, \autoref{thm:7} does not take into account background fields, a
defect that we ameliorate in~\S\ref{subsec:2.5}.  The main idea is to replace
the single datum in~\eqref{eq:15}, \eqref{eq:16} with a family of data
parametrized by a space of background fields.  For the local
defect~\eqref{eq:15} it is the space $\sF\Bigl(\germ\bigl(\{p\}\subset \nu
_p\bigr) \Bigr)$, the value of the constructible sheaf~$\sF$ on the restriction
of $Z\subset M$ to a neighborhood of $\{p\}\subset \nu _p$.  For the global
defect~\eqref{eq:16} it is the space $\sF\bigl(\germ(\partial \bn) \bigr)$,
the value of the sheaf~$\sF$ on a germ of a neighborhood of $\partial \bn$
in~$M$.
  \end{remark}

  \begin{remark}[]\label{thm:106}
 \autoref{thm:7}(1) defines a local defect at the point~$p$ in the particular
submanifold~$Z\subset M$.  There is also the notion of a local defect
\emph{theory}; it is defined in~\S\ref{subsec:2.5} below.
  \end{remark}

  \begin{remark}[]\label{thm:97}
 Local defects can be ``integrated'' to global defects.  The general case is
discussed in~\S\ref{subsec:2.5}.  As an illustration in a special case, suppose
$Z\subset W$ is equipped with a normal framing.  This identifies each
link~$\partial \bn_p$, $p\in Z$, with the standard sphere~$S^{\ell -1}$.  In
this situation it makes sense to assign a single local defect
  \begin{equation}\label{eq:18}
     \delta \in \Hom\bigl(1_{\Omega ^{\ell -1}\sC},\sigma (S^{\ell -1}) \bigr) 
  \end{equation}
to~$Z$.  Now $\sigma (S^{\ell -1})\in \Omega ^{\ell -1}\sC$ defines an $(m-\ell
+1)$-dimensional field theory~$\sigma ^{(\ell -1)}$---the dimensional reduction
of~$\sigma $ along~$S^{\ell -1}$---and a local defect~ $\delta $~determines a
left boundary theory~$\delta ^{(\ell -1)}$ for~$\sigma ^{(\ell -1)}$, if we
assume sufficient finiteness (dualizability).  Note that the cobordism
hypothesis with singularities~\cite[\S4.3]{L} is used to define the boundary
theory~$\delta ^{(\ell -1)}$.  In turn, that boundary theory is used to
integrate the local defect~\eqref{eq:18} to the global defect
  \begin{equation}\label{eq:17}
     \delta _Z = \bigl(\sigma ^{(\ell -1)},\delta  ^{(\ell
     -1)}\bigr)\bigl(\zo\times Z\bigr)\; \in 
     \Hom\bigl(1_{\Omega ^{m -1}\sC},\sigma (Z\times S^{\ell -1}) \bigr),
  \end{equation}
where $\{0\}\times Z$ is colored with the boundary theory~$\delta ^{(\ell -1)}$
and $\{1\}\times Z$ is outgoing.  
 
So far we have not specified background fields.  To begin that discussion,
observe that the theory~$\sigma ^{(\ell -1)}$ takes values in~$\Omega ^{\ell
-1}\sC$.  If $S^{\ell -1}$~is not equipped with any background fields, then the
background fields of the reduced theory are encoded in the sheaf (see
footnote~\footref{field})
  \begin{equation}\label{eq:105}
     \Man_{m-\ell +1}\op\xrightarrow{\;\;-\times S^{\ell
     -1}\;\;}\Man_m\op\xrightarrow{\;\;\;\sF\;\;\;}\Set_\Delta \;.
  \end{equation}
Depending on~$\sF$ we might be able to endow~$S^{\ell -1}$ with some fields to
simplify~\eqref{eq:105}.  For example, if $\sF$~is an $m$-dimensional
orientation, and we orient~$S^{\ell -1}$, then we can take the background field
of the dimensionally reduced theory~$\sigma ^{(\ell -1)}$ to be an $(m-\ell
+1)$-dimensional orientation.  In the case at hand let $\sF$~be an
$m$-framing---a trivialization of the $m$-dimensional tangent bundle---and
supply $S^{\ell -1}$ with an $\ell $-framing.  Then we can truncate the
dimensionally reduced theory~$\sigma ^{(\ell -1)}$ to a once-categorified
$(m-\ell )$-dimensional theory whose background field is an $(m-\ell
)$-framing.  The local defect~\eqref{eq:18} then also uses an $(m-\ell
)$-framing, and the last $(m-\ell )$~vectors of the bulk $m$-framing are
required to restrict to the $(m-\ell )$-framing of the defect.
  \end{remark} 

  \begin{remark}[]\label{thm:107}
 If the bulk theory~$\sigma $ is the trivial---tensor unit---theory, then a
local defect~\eqref{eq:15} is an object in~$\Omega ^\ell \sC$ and a global
defect~\eqref{eq:16} is a number.  By the cobordism hypothesis, a local defect
determines an $(m-\ell )$-dimensional topological field theory, once background
fields are appropriately accounted for as in~\S\ref{subsec:2.5}.  In this case,
local-to-global integration computes the partition function of this field
theory on~$Z$.
  \end{remark}

  \begin{remark}[]\label{thm:98}
 A defect on~$Z$ may be tensored with a standalone field theory on~$Z$
to obtain a new defect.  This corresponds to composing with an element
of~$\Hom(1,1)$ in~\eqref{eq:15} or~\eqref{eq:16}.
  \end{remark}

  \begin{remark}[]\label{thm:99}
 As in~\eqref{eq:12}, the sheaf of background fields on a defect need
not agree with the sheaf of background fields in the bulk; the former need
only map to the latter.  Thus we can have a spin defect in an oriented
topological field theory.
  \end{remark}

  \begin{remark}[]\label{thm:100}
 Defects can also be defined for manifolds~$M$ with boundaries and corners, and
in the standard situation $Z\subset M$ is a submanifold.  An example is
depicted in Figure~\ref{fig:13}, in which the interval is a submanifold of the
closed strip.
  \end{remark}

  \begin{remark}[]\label{thm:101}
 If $M$~is a manifold with boundary, and $Z\subset \partial M$ is a
submanifold of the boundary, then this too is allowed since we could
extend~$Z$ away from the boundary via a transparent defect, thereby bringing
us the situation envisioned in \autoref{thm:100}.
  \end{remark}

  \begin{remark}[]\label{thm:102}
 We also use a generalization in which boundaries, corners, and singularities
are allowed in~$Z$.  Then different strata of~$Z$ have different links, and we
compute them and assign (local) defects working from the lowest codimension to
the highest.  We give several illustrations, for example in~\S\ref{subsec:4.2}
and at the end of~\S\ref{subsubsec:4.4.1}.
  \end{remark}

  \begin{remark}[]\label{thm:103}
 If $M$~is a closed manifold of dimension~$m-1$, then $V=\sigma (M)$~is
a vector space and $\sigma \bigl([0,1]\times M \bigr)$ is the identity
map~$\id_V$.  A defect supported in the interior of $[0,1]\times M$ evaluates
under~$\sigma $ to a linear operator on~$V$.  In this situation the terms
`operator' and `observable' are often used in place of `defect'.
  \end{remark}

  \begin{remark}[]\label{thm:104}
 There are also (nontopological) defects in nontopological theories.  For
positive dimensional \emph{local} defects we need an extension beyond a
two-tier theory, but it is not needed for \emph{global} defects.  In
nontopological theories local defects take values in a limit as the radius of
the linking sphere shrinks to zero.  For $\dim M=m$ and $\dim Z=0$ the
resulting \emph{point defects} are often called `local operators';
see~\eqref{eq:32} for the case of quantum mechanics.  For $\dim Z=1$ they are
\emph{line defects}.
  \end{remark}

  \begin{remark}[]\label{thm:105}
 In a once-categorified $(m-1)$-dimensional theory
  \begin{equation}\label{eq:106}
     \sigma \:\Bord_{m-1}(\sF)\longrightarrow \sC,
  \end{equation}
if the manifold~$M$ in Definition~\ref{thm:7} is $m$-dimensional, then the
prescriptions~\eqref{eq:15} and~\eqref{eq:16} for labeling defects require
evaluation on manifolds not in the domain of~$\sigma $.  We indicate the
necessary modification to the prescriptions in case~$m=2$.  If $\sigma $~were a
full 2-dimensional theory, say a \emph{2-framed} theory, then a point defect
would be labeled by an element of $\Hom\bigl(1,\sigma (S^1) \bigr)$, where the
circle~$S^1$ has the constant 2-framing.  Suppose $x=\sigma (\pt)\in \sC$ is
the value of~$\sigma $ on the standard 2-framed point.  Let $c\:1\to
x\dual\otimes x$ be the coevaluation 1-morphism in the duality data for~$x$.
Then $\sigma (\cir)=c^R\circ c$, where $c^R$~is the right adjoint of~$c$.  (See
\cite[Figs.~24--25]{FT2} for a similar computation in the framed bordism
category with evaluation in place of coevaluation, and also \cite[\S2.2]{FT2}
for a general discussion of the algebra $\End^R(c):=c^R\circ c$.)  The right
adjoint~$c^R$ exists if $\sigma $~is a full 2-dimensional theory, and in that
case we have the adjunction isomorphism
  \begin{equation}\label{eq:107}
     \Hom(1,c^R\circ c)\cong \Hom(c,c).
  \end{equation}
The right hand side, $\End(c)$, is defined even if $\sigma $~is only a
once-categorified 1-dimensional theory.  Therefore, we replace
$\Hom\bigl(1,\sigma (\cir) \bigr)$---the space of point defects in a full
2-dimensional theory---with $\End(c)$ in a once-categorified 1-dimensional
theory.  Since $\End(c)$~is an algebra, point defects in a once-categorified
1-dimensional theory have a composition law, just as they do in a full
2-dimensional theory (see below).  One can imagine extending the domain
of~\eqref{eq:106} and thus consider~ $\End(c)$ as the value of~$\sigma $ on the
non-Hausdorff manifold obtained by identifying two intervals on the complement
of a point, as in Figure~\ref{fig:36}.  This 1-framed 1-manifold acts as a
substitute for the 2-framed~$S^1$ in a full 2-dimensional 2-framed theory.  The
reader can draw the corresponding non-Hausdorff substitute for~$S^2$ to see
that the appellations \emph{raviolo} or \emph{UFO} are apposite for these
non-Hausdorff manifolds.  The 2-dimensional version of ravioli/UFOs appear in
algebraic geometry in relation to Hecke correspondences on the moduli stack of
vector bundles on an algebraic curve.  A recent application in the context of
topological field theory is contained in~\cite{BFN}; see also~\cite{BDGHK}.

  \begin{figure}[ht]
  \centering
  \includegraphics[scale=.7]{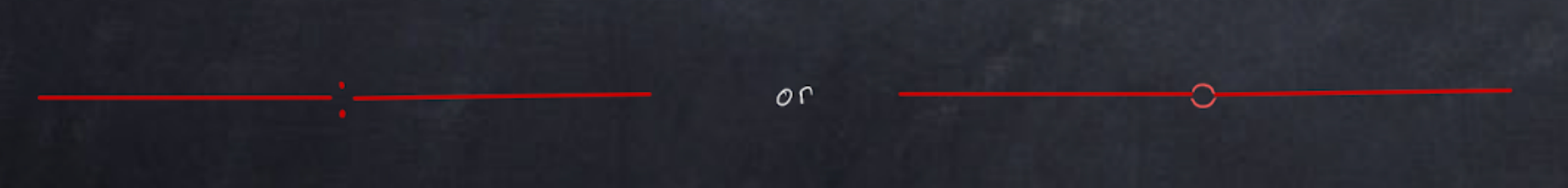}
  \vskip -.5pc
  \caption{Two possible depictions of the 1-dimensional
  \emph{raviolo}}\label{fig:36}
  \end{figure}
  \end{remark} 

  \begin{remark}[]\label{thm:108}
 The value of a topological field theory~$\sigma $ on~$S^{\ell -1}$ is an
$E_{\ell }$-algebra.  This leads to a composition law on defects, either for
local defects~ \eqref{eq:15} or global defects~\eqref{eq:16}.  If $Z$~is
normally framed, one can consider two parallel copies~$Z',Z''$, and then a
normal slice of the complement of open tubular neighborhoods of~$Z',Z''$ inside
a closed tubular neighborhood of~$Z$ is the ``pair of pants'' which defines the
composition law of the $E_\ell $-structure.  The composition law on topological
point defects is a topological version of the usual operator product expansion.
The composition law gives rise to the dichotomy between \emph{invertible
defects} and \emph{noninvertible defects}.
  \end{remark}

  \subsection{Tangential structures and the passage from local to global defects}\label{subsec:2.5}

Our goals in this section are: (1)~to explain how to include background fields
into the discussion of defects in~\S\ref{subsec:2.4}, and (2)~to set up the
application of the cobordism hypothesis to integrate a local defect to a global
defect.  Detailed arguments are not provided; the adventurous reader is invited
to work them out.
 
To begin we briefly discuss background fields in {topological} field theory.
Recall from footnote~\footref{field} that background fields are encoded in a
simplicial sheaf
  \begin{equation}\label{eq:112}
     \sF\:\Man_m\op\longrightarrow  \Set_\Delta 
  \end{equation}
on the category of smooth $m$-manifolds and local diffeomorphisms.  Since an
$m$-manifold is locally diffeomorphic to an $m$-ball, a sheaf is determined by
its values on balls in affine $m$-space, together with its values on local
diffeomorphisms of $m$-balls.  The colimit as the radius of the ball shrinks to
zero is the \emph{stalk}, and the sheaf is determined by the {stalk} and the
action of a group of germs of diffeomorphisms acting on it.  A field theory is
\emph{topological} if it factors through a sheaf of fields that is
\emph{locally constant}, which means that the inclusion map from a small ball
to a larger ball maps under the sheaf to a \emph{weak equivalence} of
simplicial sets.  At this point we shift from the world of simplicial sets to
the more intuitive world of topological spaces~\cite{BK}.  Furthermore, for a
locally constant sheaf the action of diffeomorphisms is equivalent to the
action of the general linear group~$\GL_m \!\RR$, and up to homotopy this is an
action of the orthogonal group~$\O_m$.  Replacing a simplicial set by a space,
a topological background field is a topological space equipped with an
$\O_m$-action, and hence there is an associated fibration
  \begin{equation}\label{eq:113}
     B\longrightarrow \BO_m.
  \end{equation}
The data~\eqref{eq:113} is often called an $m$-dimensional \emph{tangential
structure}: if $M$~is a smooth $m$-manifold, and if $M\to \BO_m$ is a
classifying map of its tangent bundle, then a lift 
  \begin{equation}\label{eq:114}
     \begin{gathered} \xymatrix{&B\ar[d]\\M\ar@{-->}[ur]\ar[r] & \BO_m}
     \end{gathered} 
  \end{equation}
is a $B$-structure on the manifold~$M$.  Example: $B=B\!\SO_m$ and the lift is
an orientation.  Tangential structures in this form were introduced into
bordism theory by Lashof~\cite{Las}.  A fibration~\eqref{eq:113} gives rise to
a simplicial\footnote{We use the equivalence already cited between topological
spaces and simplicial sets.} sheaf~\eqref{eq:112}: to an $m$-manifold~$M$ we
assign the space of lifts in~\eqref{eq:114}.

  \begin{figure}
  \centering
	\begin{tikzpicture}[
		cross/.style={->,preaction={draw=white, -,line width=6pt}},
		snake/.style={decorate, decoration={snake, amplitude = 0.4mm, segment length = 2mm, post length = 2mm}},
		shift left/.style ={commutative diagrams/shift left={#1}},
  		shift right/.style={commutative diagrams/shift right={#1}}
		]
		
		\definecolor{byzantine}{rgb}{0.74, 0.2, 0.64}
		\definecolor{orange}{rgb}{1, 0.55, 0}
		\definecolor{blue}{rgb}{0.25, 0.4, 0.96}
		\definecolor{red}{rgb}{.8, 0.0, 0.0}
		\definecolor{yellow}{rgb}{0.82, 0.63, 0.04}
		\definecolor{green}{rgb}{0.13, 0.55, 0.13}
		\definecolor{grey}{rgb}{0.6, 0.6, 0.6}



		\node (bo) at (1,0) {$\BO_m$};
		\node (bobo1) at (6,0) {$\BO_{m-\ell}\times \BO_{\ell}$};
		\node (bobo2) at (11,0) {$\BO_{m-\ell}\times \BO_{\ell-1}$};
		\node (b) at (1,3) {$B$};
		\node (bprime) at (6,3) {$B'$};
		\node (bhat) at (11,3) {$\widehat{B}$};

		\node[yellow] (e1) at (3, 3.9) {$\widetilde{\sE}$};
		\node[yellow] (e2) at (3, 0.6) {$\sE$};
		\node[yellow] (e3) at (12.5, 3.9) {$\widehat{\sE}$};

		\node (mz) at (-1.5, -1.5) {$M-Z$};
		\node (z) at (3.5, -1.5) {$Z$};
		\node (sv) at (8, -1.5) {$S(\nu )$};
		\node (d) at (3.5, 1.5) {$D'$};
		\node (dhat) at (8, 1.5) {$\widehat{D}$};
		\node (m) at (-1.5, 0.5) {$M$};
	
		\begin{scope}[font = \scriptsize]


		\draw[->] (d) -- (bobo1);

		\node[scale=.9][black, align=center, rectangle, draw] (bulk) at
(1.8, 1.7) {Bulk\\Tangential\\Structure};
		\node[scale=.65][black, align=center, rectangle, draw] (defect) at (4.68, .8) {Defect\\Tangential\\Structure};
		\node[red, align=center, rectangle, draw] (dtb) at (8.3, 2.45) {Defect-to-Bulk};
		\node[green, align=center, rectangle, draw] (thy) at (4.05, 3.35) {Bulk Theory};


		\draw[->, yellow, shift right] (e1) edge node[above, pos=0.5] {$\sCt$} (b);
		\draw[->, yellow] (e2) -- node[above, pos=0.5] {$\sCt$}(bo);
		\draw[->, yellow, shift right] (e3) edge (bhat);

		\draw[->, green, shift right] (b) edge node[right, pos=0.5] {$\,\sigma($pt$)$} (e1);
		\draw[->, green, shift right] (bhat) edge node[right, pos=0.5] {$\,\sigma($pt$)$} (e3);

		\draw[->] (b) -- (bo);
		\draw[->, cross] (bprime) -- (bobo1);
		\draw[->] (bhat) -- node[right, midway] {$\beta$} (bobo2);
		\draw[->] (bhat) -- node[above, midway] {$S^{\ell-1}$} (bprime);
		\draw[->] (bprime) -- (b);
		\draw[->, cross] (bobo2) -- node[above, pos=0.4] {$S^{\ell-1}$} node[below, pos=0.45] {$\pi$} (bobo1);
		\draw[->, cross] (bobo1) -- (bo);

		\draw[->, cross] (dhat) -- node[above right, pos=0.3] {} (bobo2);
		\draw[->, blue] (mz) -- (bo);
		\draw[->, byzantine] (mz) -- (b);
		\draw[->, blue] (z) -- (bobo1);
		\draw[->, byzantine, cross] (z) -- (d);
		\draw[->, cross] (dhat) -- node[above, pos=0.3] {$S^{\ell-1}$} (d);
		\draw[->] (sv) -- node[above, midway] {$S^{\ell-1}$} (z);
		\draw[->, byzantine,cross] (sv) -- (dhat);
		\draw[->, byzantine, cross] (sv) -- (bhat);
		\draw[->, blue] (sv) -- (bobo2);
		\draw[->, red] (dhat) -- (bhat);
		\draw[right hook->] (mz) -- (m);
		\draw[right hook->] (sv) to[bend left=15] (mz);
		

		\draw (5.4,2.8) -- (5.7, 2.8) -- (5.7, 2.5);
		\draw (10.2,2.8) -- (10.5, 2.8) -- (10.5, 2.5);
		\draw (7.36,1.3) -- (7.65, 1.3) -- (7.85, 1.15);

		\end{scope}
		

		\draw[very thick, ->, snake, grey] (8, -2.3) --  (9, -3.2);



		\node (D) at (9.5, -5.5) {$D'$};
		\node (Z) at (9.5, -8.5) {$Z$};
		\node (gamma) at (12.5, -4.5) {$\Gamma_\pi(\beta)$};
		\node[yellow] (hom) at (7.5, -4.1) {$\sH$};
		\node[yellow] (Hom) at (7.5, -7.1) {$\sH$};
		\node (bobo3) at (12.5, -7.1) {$\BO_{m-\ell}\times \BO_\ell$};
		\node[yellow] (int) at (10.5, -3) {$\int_{S^{\ell-1}} \widehat{\sE}$};
		\draw[->, byzantine] (Z) -- (D);
		\draw[->, blue] (Z) -- (bobo3);

		\begin{scope}[font = \scriptsize]
		
		\node[cyan, align=center, rectangle, draw] (defthy) at (7.18, -5.1) {Defect Theory};
		
		\draw[->, red] (D) -- (gamma);
		\draw[->] (D) -- (bobo3);
		\draw[->] (gamma) -- (bobo3);
		\draw[->,cyan, shift left] (D) edge node[midway, below] {$\delta$} (hom);
		\draw[->,cyan, shift left] (Z) edge node[midway, below] {} (Hom);
		\draw[->,yellow, shift left] (hom) edge
		node[midway,above,scale=.9,pos=.7] 
		{$\qquad\qquad \quad \qquad
		\text{Hom}\bigl(1,\sigma(S^{\ell-1})\bigr)^{\text{fd}}$} (D); 
		\draw[->,yellow, shift left] (Hom) edge (Z); 
		\draw[->,green, shift right] (gamma) edge node[midway, right] {$\;\sigma(S^{\ell-1})$} (int);
		\draw[->,yellow, shift right] (int) edge node[midway, left] {$\Omega^{\ell-1} \sCt\;$} (gamma);
		\end{scope}
	\end{tikzpicture}
  \caption{Local defect data, including tangential structures}\label{fig:38}
  \end{figure}

We turn now to the diagram in \autoref{fig:38}, which will occupy us for the
rest of this section.  The diagram is built from the following
data---systematic explanations follow---which define a fully local bulk
topological field theory~$\sigma $:

 \begin{enumerate}[label=\textnormal{(B\arabic*)}]

 \item the fibration $B\to \BO_m$

 \item the (yellow) fibration $\sE\to \BO_m$

 \item the (green) section~$\sigma (\pt)\:B\to \widetilde{E}$ 
  \end{enumerate}

\noindent
 The additional data that define the defect are:

 \begin{enumerate}[label=\textnormal{(D\arabic*)}]

 \item the fibration $D'\to \BO_{m-\ell }\times \BO_\ell $

 \item the (red) map $\hD\to \hB$

 \item the (cyan) section~$\delta \:D'\to \sH$
 \end{enumerate}

First, the data of the bulk theory.  The fibration~(B1) is the tangential
structure of the bulk theory~$\sigma $, as just explained.  The theory~$\sigma
$ takes values in an $(\infty ,m)$-category~$\sC$.  Form the space~$\sCt$ by
removing from~$\sC$ all non-fully-dualizable objects and all noninvertible
morphisms.  The cobordism hypothesis~\cite{L} implies that $\sCt$~carries an
$\O_m$-action; this action defines the fibration~(B2) by the Borel
construction~\eqref{eq:1}.  The cobordism hypothesis asserts that the fully
local topological field theory~$\sigma $ is determined by and can be defined by
a section~(B3) of the pullback of the fibration~(B2) to the total space
of~(B1).

Heuristically, the space~$\BO_m$ parametrizes a universal family of
$m$-dimensional vector spaces; $\BO_m$~ can be modeled as a Grassmann manifold.
In terms of the bordism category, $\BO_m$~parametrizes a universal family of
points embedded in a germ of an $m$-manifold.  Up to homotopy, a point embedded
in a germ of an $m$-manifold is equivalent to an $m$-dimensional vector space,
namely the tangent space.  The total space~$B$ of the fibration~(B1)
parametrizes the universal family of points equipped with a $B$-structure.
(This can be taken to be the definition of a $B$-structure.)  It is for each
point in the universal family that we must specify the value of~$\sigma $; that
is the section~(B3).
 
We map into this universal data a particular smooth $m$-manifold~$M$ with a
codimension~$\ell $ submanifold~$Z$.  The classifying map of the tangent bundle
to the complement~$M\setminus Z$ of~$Z$ is depicted in the diagram, as is a
lift of that classifying map to~$B$.  This encodes a $B$-structure on the
complement~$M\setminus Z$.  Depending on the nature of the defect, the
$B$-structure may or may not extend over its support~$Z$.  For example, if
$B=B\!\Spin_m$ encodes a spin structure, then there may be codimension~2
defects on a spin manifold over which the spin structure does not extend.
 
Now we turn to the defect data (D1)--(D3).  The fibration~(D1) defines the
tangential structure along the support of the defect.  The map~(D2) is gluing
data from the defect tangential structure to the bulk tangential structure.
The section~(D3) is the data that determines the defect.
 
In more detail, the space $\BO_{m-\ell }\times \BO_\ell $ parametrizes
$m$-dimensional real vector spaces equipped with an $(m-\ell )$-dimensional
subspace and an $\ell $-dimensional complement.  This is the structure of the
tangent space to an $m$-manifold along a codimension~$\ell $ submanifold.  The
tangential structure~(D1) along the defect may use both the tangent and normal
spaces to the support of the defect.  The gluing to the bulk tangential
structure takes place on a deleted neighborhood of the support of the defect: a
tubular neighborhood minus the zero section.  This deformation retracts to the
sphere bundle of the normal bundle.  The arrow 
  \begin{equation}\label{eq:117}
     \pi \:\BO_{m-\ell }\times \BO_{\ell -1}\longrightarrow  \BO_{m-\ell }\times
     \BO_\ell 
  \end{equation}
is the universal normal sphere bundle.  (The typical fiber is the \emph{link}
of the submanifold.)  In the diagram both the bulk tangential structure~(B1)
and the defect tangential structure~(D1) have been pulled back to the total
space of the universal normal sphere bundle, thereby defining the spaces~$\hB$
and~$\hD$.  (The black brackets in the diagram indicate pullback squares.)  The
map~(D2) is the defect-to-bulk arrow that compares the two tangential
structures.  In the diagram the arrow $Z\to D'$ encodes the defect tangential
structure on the particular defect~$Z\subset M$.  The bulk and defect
tangential structures have been lifted to maps out of the total space~$S(\nu )$
of the sphere bundle of the normal bundle $\nu \to Z$.  The compatibility of
the bulk and defect tangential structures is the homotopy commutation data of
the triangle
  \begin{equation}\label{eq:116}
     \begin{gathered} \xymatrix{\widehat{D}\ar[r]&\widehat{B}\\ S(\nu
     )\ar[u]\ar[ur]} \end{gathered} 
  \end{equation}
 
The local defect data~\eqref{eq:15} uses the bulk theory~$\sigma $ evaluated on
the linking spheres of the submanifold.  The lower diagram in \autoref{fig:38}
includes (green arrow) the value of~$\sigma $ on the universal linking spheres,
which are parametrized by $\BO_{m-\ell }\times \BO_\ell $.  The fiber bundle
$\Gamma _\pi (\beta )\to \BO_{m-\ell }\times \BO_\ell $ is defined by
specifying its fiber: the space of $B$-structures on the corresponding $(\ell
-1)$-sphere.  The value of~$\sigma $ on the linking sphere takes values
in~$\Omega ^{\ell -1}\sCt$, and as we move over the parameter space
$\BO_{m-\ell }\times \BO_\ell $ of spheres the categories $\Omega ^{\ell
-1}\sCt$ form a local system over~$\Gamma _\pi (\beta )$ (pulled back from a
local system over the base $\BO_{m-\ell }\times \BO_\ell $).  The fully
dualizable subcategories of the hom categories $\Hom\bigl(1,\sigma (S^{\ell
-1}) \bigr)$ form a local system $\int_{S^{\ell-1}} \widehat{\sE}\to \Gamma
_\pi (\beta )$.  The map $D'\to \Gamma _\pi (\beta )$ is constructed using the
Defect-to-Bulk map, and the local system~$\sH\to D'$ is constructed by pullback
along this map.  Finally, a (universal) local defect theory~(D3) is a
section~$\delta $ of $\sH\to D'$.
  (A particular defect on~$Z$, as in~\eqref{eq:15}, is a section of the
pullback of $\sH\to D'$ over~$Z$ via $Z\to D'$.)
 
Finally, at the bottom of \autoref{fig:38} the general local defect
theory~$\delta $ has been pulled back to the particular defect with
support~$Z$.  The cobordism hypothesis with singularities integrates this local
defect to a global defect on~$Z$.

  \begin{remark}[]\label{thm:117}
 In contrast to defects of higher co-dimension, domain walls have the unique
feature of allowing two different theories $\sigma_1, \sigma_2$ to live on
opposite sides of the wall. However, this variation may be incorporated in the
same diagram \autoref{fig:38}: the bulk tangential structure $B$ will have two
different components $B_1, B_2$ representing the tangential structures of
$\sigma_{1,2}$.  The following extended exercise, which the authors found
illuminating, uses this variation.
  \end{remark}

  \begin{exercise}[]\label{thm:118}
 This extended exercise illustrates how \autoref{fig:38} works for some simple
choices of tangential structures.  In particular, note that whereas boundaries
of morphisms in a bordism category must be labeled as incoming or
outgoing---i.e., they carry a time direction---domain walls do not
intrinsically carry this structure.  On the other hand, we first introduced
domain walls in~\S\ref{subsec:2.2} with a coorientation, i.e., a time
direction.  In this exercise we show how that coorientation is part of the
choice of tangential structure, and that other choices are possible.

For a domain wall we have~$\ell =1$, and the reader can profitably set~$m=1$ as
well to simplify.

This exercise considers two choices of Bulk Tangential Structure and two
choices of Defect Tangential Structure, hence four combinations: 
 
 \begin{enumerate}[label=\textnormal{(\arabic*)}]

 \item bulk theories oriented; domain wall cooriented

 \item bulk theories unoriented; domain wall cooriented

 \item bulk theories oriented; domain wall not cooriented

 \item bulk theories unoriented; domain wall not cooriented 
 \end{enumerate}

\noindent
 The reader is invited to construct Defect-to-Bulk maps in each case and,
for~$m=1$, to compute what data defines a domain wall theory.  First, we spell
out in detail and for general~$m$ how the various tangential structure are
encoded in \autoref{fig:38}.

The two choices of Bulk Tangential Structure that we consider lead to different
fibrations $B\to \BO_m$.  View~$\BO_m$ as the (moduli) space of $m$-dimensional
real inner product spaces~$\Pi $; the fiber of~$B\to \BO_m$ over~$\Pi $
parametrizes additional geometric structures on~$\Pi $.  If the ambient
theories~$\sigma _1,\sigma _2$ are unoriented, then we set
  \begin{equation}\label{eq:131}
     B=\BO_m\,\sqcup \,\BO_m\longrightarrow \BO_m .
  \end{equation}
Here the tangential structure merely labels the two theories~$\sigma _1,\sigma
_2$.  View $\BO_{m-1}\times \BO_1$ in the upper part of \autoref{fig:38} as the
(moduli) space of $m$-dimensional real inner product spaces~$\Pi $ equipped
with a codimension one subspace $\Pi '\subset \Pi $, and view $\BO_{m-1}\times
\BO_0$ as additionally encoding a choice of component of~$\Pi \setminus \Pi '$.
Then compute the fibration $\beta \:\hB\to \BO_{m-1}\times \BO_0\simeq
\BO_{m-1}$ to be
  \begin{equation}\label{eq:135}
     \beta \:\hB=\BO_{m-1}\,\sqcup \,\BO_{m-1}\longrightarrow \BO_{m-1}. 
  \end{equation}
If the ambient theories~$\sigma _1,\sigma _2$ are oriented, then the fibration
$B\to \BO_m$ is
  \begin{equation}\label{eq:132}
     B=\BSO_m\,\sqcup \,\BSO_m\longrightarrow \BO_m,
  \end{equation}
and compute the fibration $\beta$  to be
  \begin{equation}\label{eq:136}
     \beta \:\hB=\BSO_{m-1}\,\sqcup \,\BSO_{m-1}\longrightarrow \BO_{m-1}.
  \end{equation}

We also consider two choices of Defect Tangential Structure $D'\to \BO_{m-1
}\times \BO_{1}$.  Note that this is a choice of a tangential structure along
the defect as well as a structure on the normal bundle.  We leave the domain
wall unoriented, and the first choice is to coorient it, in which case the
fibration $D'\to \BO_{m-1 }\times \BO_{1 }$ is the nontrivial double cover
  \begin{equation}\label{eq:133}
     D'=\BO_{m-1}\times \BSO_1\longrightarrow \BO_{m-1}\times \BO_1 .
  \end{equation}
(Note $\BO_1\simeq \RP^{\infty}$ and $\BSO_1\simeq *$, and a model for the
double cover $\BSO_1\to \BO_1$ is the infinite sphere covering the infinite
real projective space.)  In that case compute the fibration $\hD\to
\BO_{m-1}\times \BO_0$ to be
  \begin{equation}\label{eq:137}
     \hD=\BO_{m-1}\,\sqcup \,\BO_{m-1}\longrightarrow \BO_{m-1}. 
  \end{equation}
Here is an interpretation of~\eqref{eq:137}. Fix an $m$-dimensional real inner
product space~$\Pi $ and a codimension one subspace~$\Pi '\subset \Pi $.  A
choice of component of $\Pi \setminus \Pi '$ orients the normal line~$\Pi /\Pi
'$.  So too does the choice of coorientation of $\Pi '\subset \Pi $.  Then the
fiber of~\eqref{eq:137} are the two possibilities: the orientations of~$\Pi
/\Pi '$ agree or they disagree.  Our second choice of Defect Tangential
Structure is not to coorient the domain wall, in which case \eqref{eq:133}~is
replaced by the identity map
  \begin{equation}\label{eq:134}
     D'=\BO_{m-1}\times \BO_1\longrightarrow \BO_{m-1}\times \BO_1 ,
  \end{equation}
and so too \eqref{eq:137}~is replaced by the identity map
  \begin{equation}\label{eq:138}
     \hD=\BO_{m-1}\longrightarrow \BO_{m-1}. 
  \end{equation}

The final piece of tangential structure data is the Defect-to-Bulk map $\hD\to
\hB$ in \autoref{fig:38}.  Observe that if the domain wall is not cooriented,
then according to~\eqref{eq:138} the space $\hD\simeq \BO_{m-1}$ is connected,
hence so too is its image in~$\hB$; by~\eqref{eq:135} and~\eqref{eq:136} the
latter space~$\hB$ is not connected.  This means that such domain walls cannot
transition between two different theories; on both sides of the wall sit the
same theory~$\sigma _1$ or~$\sigma _2$.  In this regard, contemplate a domain
wall supported on $\RP^{m-1}\subset \RP^{m}$. 
 
Now set~$m=1$ and take the codomain of the bulk theories to be the category of
vector spaces.  Then each bulk theory~$\sigma _i$, $i=1,2$, evaluates on a
(positively oriented) point to be a complex vector space~$V_i$.  If the bulk
theory is unoriented, as in~\eqref{eq:131}, then there is additional data which
we leave as an exercise.  (Hint: see~\cite[Example 2.4.28]{L}.)  The main
exercise is, for each of the four cases listed above, contemplate different
Defect-to-Bulk maps and for each compute what data specifies the Defect
Theory~$\delta $ in \autoref{fig:38}.  For example, in~(1) there is a
Defect-to-Bulk map so that the data is an element of~$\Hom(V_1,V_2)$.  What are
other possibilities?
  \end{exercise}

   \section{Symmetry in field theory}\label{sec:3}

We begin with the definitions, first of abstract topological symmetry
data---a \emph{quiche}---in field theory~(\S\ref{subsec:3.1}) and then of a
realization in quantum field theory~(\S\ref{subsec:3.2}).  We give some
variations, most notably a relaxation of finiteness conditions (see
Remark~\ref{thm:10}(8)), and also to symmetries of anomalous field theories
(see Remark~\ref{thm:12}(3)).  Section~\ref{subsec:3.3} illustrates with a
few examples; more are developed in~\S\ref{sec:4}.  In~\S\ref{subsec:3.4} we
discuss the quotient of a field theory by a symmetry: the gauging operation.
It is expressed in terms of an augmentation of the field theory that encodes
the symmetry.  In~\S\ref{subsec:3.5} we describe a dual symmetry which is
induced on a quotient, at least in the situation of finite electromagnetic
duality.

  \subsection{Abstract topological symmetry data in field theory}\label{subsec:3.1}

This discussion is inspired by the considerations in~\S\ref{subsec:1.1}
and~\S\ref{subsec:1.2}.  The crucial notion of a `regular boundary theory' is
given immediately after the following.  See footnote~\footref{quiche} for an
explanation of the choice of terminology.

  \begin{definition}[]\label{thm:9}
 Fix~$n\in \ZZ^{\ge0}$.  An \emph{$n$-dimensional quiche} is a pair~$(\sigma
,\rho )$, where $\sigma \:\Bord_{n+1}(\sF)\to \sC$ is an $(n+1)$-dimensional
topological field theory and $\rho $~is a right topological $\sigma $-module.
  \end{definition}

\noindent
 The dimension~$n$ pertains to the theories on which~$(\sigma ,\rho )$ acts,
not to the dimension\footnote{The dimension \emph{does} pertain to~$\sigma $ if
$\sigma $~is a once-categorified $n$-dimensional theory.} of the field
theory~$\sigma $.  One might want to assume that $\rho $~is nonzero if the
codomain~$\sC$ is a \emph{linear} $n$-category; this is true for the particular
boundaries in Definition~\ref{thm:13} below.  Note too that we can relax the
condition that $\sigma $~be a full $(n+1)$-dimensional field theory; see
Remark~\ref{thm:10}(8).
 
This definition is extremely general.  The following singles out a class of
boundary theories which more closely models the discussions
in~\S\ref{subsec:1.1} and~\S\ref{subsec:1.2}.  Recall that if $\sC'$~is a
symmetric monoidal $n$-category, then there is a symmetric monoidal Morita
$(n+1)$-category $\Alg(\sC')$ whose objects are algebra objects in~$\sC'$ and
whose 1-morphisms $A_0\to A_1$ are $(A_1,A_0)$-bimodules~$B$; we write
`${}\mstrut _{A_1}\hneg B\mstrut _{A_0}$' to emphasize the bimodule structure.
If $A_1=1$, then we write the resulting right module as $B\mstrut _{\hneg A_0}$. 
See~\cite{L,JS,Hau,GS,BJS} for a development of Morita theory in higher
categories as well as for discussions of dualizability.

  \begin{definition}[]\label{thm:13}
 Suppose $\sC'$~is a symmetric monoidal $n$-category and $\sigma $~is an
$(n+1)$-dimensional topological field theory with codomain $\sC=\Alg(\sC')$.
Let $A=\sigma (\pt)$.  Then $A$~is an algebra in~$\sC'$ which, as an object
in~$\sC$, is $(n+1)$-dualizable.  (We can relax to $n$-dualizable; see
Remark~\ref{thm:10}(8).)  Assume that the right regular module~$\Ar$ is
$n$-dualizable as a 1-morphism in~$\sC$.  Then the boundary theory~$\rho $
determined by~$\Ar$ is the \emph{right regular boundary theory} of~$\sigma $,
or the \emph{right regular $\sigma $-module}.
  \end{definition}

\begin{figure}[ht]
  \centering
  \includegraphics[scale=1.6]{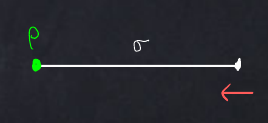}
  \vskip -.5pc
  \caption{The bordism that computes~$A_A$}\label{fig:5}
  \end{figure}

\noindent
 We use an extension of the cobordism hypothesis \cite[Example~4.3.22]{L} to
generate the boundary theory~$\rho $ from the right regular module~$\Ar$.
Observe that $\Ar$~is the value of the pair~$(\sigma ,\rho )$ on the bordism
depicted in Figure~\ref{fig:5}; the white point is incoming, so the
depicted bordism maps  $\pt\to \emptyset $.

  \begin{remark}[]\label{thm:10}
 \ 

 \begin{enumerate}[label=\textnormal{(\arabic*)}]

 \item The right regular $\sigma $-module~$\rho $ satisfies $\End_\sigma (\rho
)\cong \sigma $, as follows from the cobordism hypothesis by the corresponding
statement for algebras.  See \autoref{fig:35}.

 \item The regular boundary theory is often called a \emph{Dirichlet boundary
theory}.

 \item For arbitrary $\sr$ acting on a field theory~$F$ as in the next
section, we can replace~$\sr$ with an algebra and its right regular module at
the price of losing some dualizability; see Remark~\ref{thm:12}(2).

 \item Not every topological field theory~$\sigma $ can appear in
Definition~\ref{thm:9}.  For example, consider a 3-dimensional
Reshetikhin-Turaev theory, which we assume has been extended to a fully local
theory, i.e., a $(0,1,2,3)$-theory.  (Usually one takes `Reshetikhin-Turaev
theory' to mean a $(1,2,3)$-theory, but in fact it can be made fully
local~\cite{FST}.)  The main theorem in~\cite{FT2} asserts that ``most'' such
theories do not admit any nonzero topological boundary theory, hence they
cannot act as symmetries of a 2-dimensional field theory. (If we only assume
that the Reshetikhin-Turaev theory is a $(1,2,3)$-theory, then there are
possible boundary theories, at least if we include
$\zt$-gradings.\footnote{For example, take $H^{\bullet }(\cir\times
\cir;\CC)$ as a $\zt$-graded Frobenius algebra and tensor with the algebra
object~1 in the modular tensor category to produce a $(1,2)$~oriented
boundary theory.}) On the other hand, the Turaev-Viro theory~$\sP $ formed
from a (spherical) fusion category~$\Phi $ takes values in the
3-category~$\Alg(\Cat)$ for a suitable 2-category~$\Cat$ of linear
categories.  Thus $\sP$~admits the right regular $\sigma $-module defined by
the right regular $\Phi $-module~$\Phi \mstrut _{\hneg\Phi }$; the necessary
dualizability of~$\Phi \mstrut _{\hneg\Phi }$ is proved in~\cite{DSS}. 

 \item Let $\sX$~be a $\pi $-finite space, as in Definition~\ref{thm:53}, and
let $\sXn$ be the associated $(n+1)$-dimensional topological field theory.  A
basepoint $*\to \sX$ determines a right regular $\sigma $-module;
see~Definition~\ref{thm:73}(1).  This holds even if $\sX$~is equipped with a
reduced cocycle, i.e., a cocycle on the pair~$(\sX,*)$.  Finite group
symmetries are of this type, as are finite higher group symmetries and finite
2-group symmetries.

 \item Let $G$~be a finite group.  Then $G$-symmetry in an $n$-dimensional
quantum field theory is realized via $(n+1)$-dimensional finite gauge theory.
The partition function counts principal $G$-bundles, weighted by the reciprocal
of the order of the automorphism group.  The regular boundary theory has an
additional fluctuating field: a section of the principal $G$-bundle.  Finite
$G$-gauge theory can be realized as in~(5) with $\sX=BG$ a classifying space
of~$G$.

 \item A variation on~(6) is a Dijkgraaf-Witten theory~\cite{DW} in which the
counting of bundles is also weighted by a characteristic number defined by a
cohomology class in $H^{n+1}(BG;\Cx)$.  For~$n=1$ this class is represented by
a central extension~\eqref{eq:9} of~$G$, and the fully local field theory with
values in the Morita 2-category $\Alg(\Vect)$ is generated by the twisted group
algebra~\eqref{eq:10}.  Observe that the passage from linear to projective
symmetries (see~\S\ref{subsec:1.4}) is not a structural change in the
framework, but rather is a different choice of~$(\sigma ,\rho )$.

 \item As in Remark~\ref{thm:6}(2), the topological field theory~$\sigma $
need only be a once-categorified $n$-dimensional theory, not a full
$(n+1)$-dimensional theory; see Example~\ref{thm:38}.  However, a full theory
may allow the possibility of more defects; see Example~\ref{thm:40}.

 \end{enumerate}
  \end{remark}

  \subsection{Concrete realization of topological symmetry in field theory}\label{subsec:3.2}

Let $\sigma$~be an $(n+1)$-dimensional topological field theory and let $\rho
$~be a right topological $\sigma $-module.  We now define a realization of
the quiche~$\sr$ as symmetries of a quantum field theory.

  \begin{figure}[ht]
  \centering
  \includegraphics[scale=1.6]{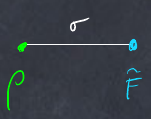}
  \vskip -.5pc
  \caption{The sandwich}\label{fig:6}
  \end{figure}

  \begin{definition}[]\label{thm:11}
 Let $(\sigma ,\rho )$ be an $n$-dimensional quiche.  Let $F$~be an
$n$-dimensional field theory.  A \emph{$\sr$-module structure} on~$F$ is a
pair~$\tFt$ in which $\tF$~is a left $\sigma $-module and $\theta $~is an
isomorphism
  \begin{equation}\label{eq:21}
     \theta \:\rho \otimes \mstrut _{\sigma }\tF\xrightarrow{\;\;\cong \;\;}F 
  \end{equation}
of absolute $n$-dimensional theories. 
  \end{definition}

\noindent
 Here `$\rho \otimes \mstrut _{\sigma }\tF$' notates the dimensional
reduction of~$\sigma $ along the closed interval with boundaries colored
with~$\rho $ and~$\tF$; see Figure~\ref{fig:6}.  The bulk theory~$\sigma $ with
its right and left boundary theories~$\rho $ and~$\tF$ is sometimes called a
\emph{sandwich}.   

  \begin{remark}[]\label{thm:12}
  \  
 \begin{enumerate}[label=\textnormal{(\arabic*)}]

 \item As in Remark~\ref{thm:10}(8), $\sigma $~need only be a once-categorified
$n$-dimensional theory.  In that case $\rho $~and $\tF$ are relative field
theories~\cite{Ste}.

  \begin{figure}[ht]
  \centering
  \includegraphics[scale=.5]{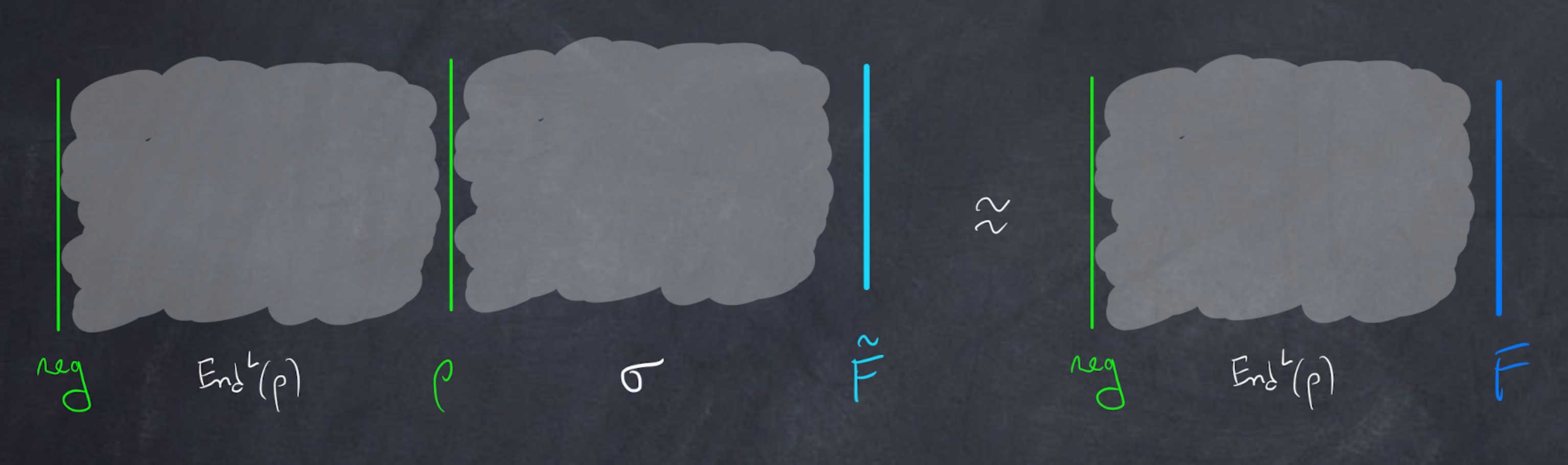}
  \vskip -.5pc
  \caption{The algebra~$\End^L(\rho )$ acting on~$F$}\label{fig:35}
  \end{figure}

 \item As alluded to in Remark~\ref{thm:10}(3), we can replace an
arbitrary~$\sigma $ with a theory whose value on a point is an algebra, as in
\autoref{thm:13}, as follows.  Define the 1-morphism $\rho _1\:\sigma (\pt)\to
1$ as the value of~$\sr$ on the bordism in Figure~\ref{fig:5}.  The composition
$\End^L(\rho _1)=\rho _1\circ \rho _1^L$ of $\rho _1$ with its left adjoint is
an algebra object in~$\Omega \sC$, and $\rho _1$~is a left $\End^L(\rho
_1)$-module; see \cite[\S2.2]{FT2}.  Assuming $\End^L(\rho _1)$~is
$n$-dualizable, it determines a once-categorified $n$-dimensional topological
field theory $\End^L(\rho )\:\Bord_{n}(\sF)\to \Alg(\Omega \sC)$.  If we
furthermore assume that the right regular module of $\End^L(\rho _1)$ is
$n$-dualizable, then it determines a right relative field theory `reg'
over~$\End^L(\rho )$.  Then as depicted in Figure~\ref{fig:35}, if $F$~has a
$\sr$-module structure, it also acquires a $\bigl(\End^L(\rho
),\textnormal{regular} \bigr)$-module structure.  These dualizability
assumptions hold in many examples.

  \begin{figure}[ht]
  \centering
  \includegraphics[scale=1.6]{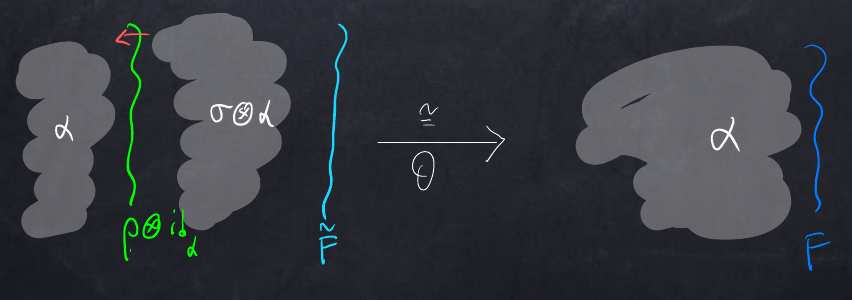}
  \vskip -.5pc
  \caption{The symmetry~$\sr$ acting on an anomalous theory}\label{fig:7}
  \end{figure}

 \item Definition~\ref{thm:11} extends to anomalous theories~$F$, or more
generally to left boundary theories from some $(n+1)$-dimensional theory, as
illustrated in Figure~\ref{fig:7}.  In this case $\tF$~is a left $(\sigma
\otimes \alpha )$-module, and the right $(\rho \otimes \id_{\alpha })$-module
completes the sandwich.   

 \item The theory~$F$ and so the boundary theory~$\tF$ may be topological or
nontopological, and we allow it to be not fully local (in which case we
use truncations of~$\sigma $ and~$\rho $).  We caution that there could be
more topological symmetries if we do not insist on full locality of~$\sr $,
and this can even happen if $F$~is a topological theory; see
Remark~\ref{thm:10}(4) for an example. 

 \item The sandwich picture Remark~\ref{thm:6} separates out the topological
part~$\sr$ of the theory from the potentially nontopological part~$\tF$ of
the theory.  This is advantageous, for example in the study of
defects~(\S\ref{sec:4}).  It allows general computations in the
$n$-dimensional quiche which apply to every realization as a symmetry of a
field theory.

 \item Typically, in the physics literature, symmetry persists under
renormalization group flow, hence a low energy approximation to~$F$ should also
be a $\sr$-module.  If $F$~is gapped, then at low energies we expect a
topological theory (up to an invertible theory), so we can bring to bear
powerful methods and theorems in topological field theory to investigate
\emph{topological} left $\sigma $-modules.  This leads to dynamical
predictions; see~\S\ref{sec:6}.

 \end{enumerate}
  \end{remark}

  \subsection{Examples}\label{subsec:3.3}

  \begin{example}[quantum mechanics $n=1$]\label{thm:14}
 Consider a quantum mechanical system defined by a Hilbert space~$\sH$ and a
time-independent Hamiltonian~$H$.  The Wick-rotated theory~$F$ is regarded as
a map with domain $\Bord_{\langle 0,1 \rangle}(\sF)$ for
  \begin{equation}\label{eq:23}
      \sF=\{\textnormal{orientation}, \textnormal{Riemannian metric}\}. 
  \end{equation}
Roughly speaking, $F(\pt_+)=\sH$ and $F(X)= e^{-\tau H/\hbar}$ for~$\tau \in
\RR^{>0}$ and $X=[0,\tau ]$ with the standard orientation and Riemannian
metric.  We refer to~ \cite[\S3]{KS}, \cite{Wed}, and~\cite{S2} for more
precise statements.
 
Now suppose $G$~is a finite group equipped with a unitary representation
$S\:G\to U(\sH)$, and assume that the $G$-action commutes with the
Hamiltonian~$H$.  To express this symmetry in terms of Definition~\ref{thm:9}
and Definition~\ref{thm:11}, let $\sigma $~be the 2-dimensional finite gauge
theory with gauge group~$G$.  If we were only concerned with~$\sigma $ we
might set the codomain of~$\sigma $ to be $\sC=\Alg(\sC')$ for $\sC'$~the
category of finite dimensional complex vector spaces and linear maps.  But to
accommodate the boundary theory~$\tF$ for quantum mechanics, we let~$\sC'$ be
a suitable category of topological vector spaces, as in~\cite[\S3]{KS}.  The
quiche~$\sr$ is defined on $\Bord_2=\Bord_{\langle 0,1,2 \rangle}$ with no
background fields.  Then $\sigma (\pt)=\GA$ is the complex group algebra
of~$G$, and $\rho (\pt)$~is its right regular module.
 
  \begin{figure}[ht]
  \centering
  \includegraphics[scale=1.5]{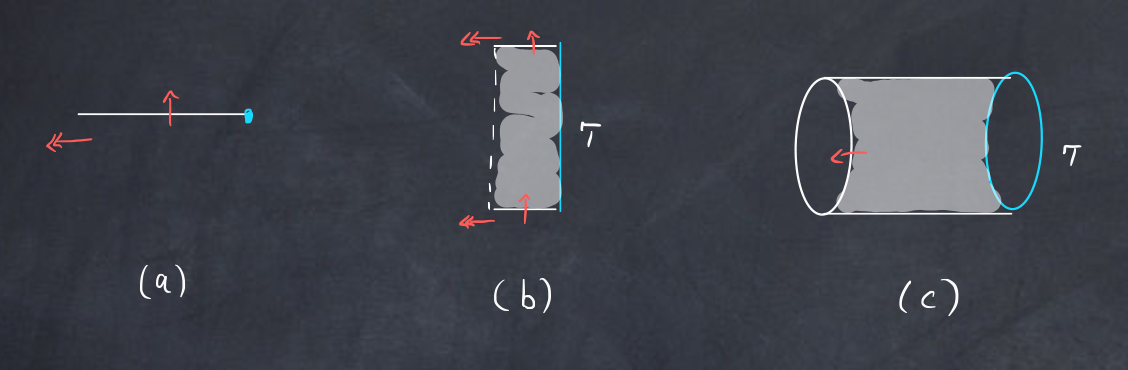}
  \vskip -.5pc
  \caption{Three bordisms evaluated in~\eqref{eq:22} in the theory~$(\sigma ,\tF)$}\label{fig:8}
  \end{figure}

Now we describe the left boundary theory~$\tF$, which has as background
fields~\eqref{eq:23}, as does the (absolute) quantum mechanical theory~$F$.
Observe that by cutting out a collar neighborhood it suffices to define~$\tF$
on cylinders (products with~$[0,1]$) over $\tF$-colored boundaries.  The
bordisms in Figure~\ref{fig:8} do not have a well-defined width since there
is a Riemannian metric only on the colored boundary.  That boundary has a
well-defined length~$\tau $ in~(b) and~(c).  The ``arrows of time''
distinguish incoming from outgoing boundaries in codimension one; we defer
to~\cite[\S2.1.1]{FT2} for the conventions in higher codimension and for the
constancy condition encoded in the dotted line in~(b).  Evaluation of these
bordisms under~$(\sigma ,\tF)$ gives:
  \begin{equation}\label{eq:22}
     \begin{aligned} &\textnormal{(a)~the left module }\sHl \\
      &\textnormal{(b) }e^{-\tau H/\hbar}\:\sHl\longrightarrow \sHl \\
      &\textnormal{(c)~the central function
      $g\longmapsto\Tr\mstrut _{\sH}\bigl(S(g)e^{-\tau H/\hbar}\bigr)$ on~$G$}
      \end{aligned} 
  \end{equation}
Assertions~(a) and~(b) are part of the definition of~$\tF$; it is the essential
data needed to construct the nontopological $\sigma $-module~$\tF$.  For~(c),
diagonalize~$H$ and decompose the group action to reduce to the tensor product
of (i)~ the bulk theory with an irreducible finite dimensional module defining a
topological boundary theory, and (ii)~an invertible nontopological standalone
1-dimensional theory.  That (i)~computes the character can be found
in~\cite{MS}, for example.
  \end{example}

  \begin{remark}[]\label{thm:15}
 As already mentioned in Remark~\ref{thm:10}(6), the finite gauge
theory~$\sigma $ can be constructed via a finite path integral from the $\pi
$-finite space~$BG$.  Similarly, the boundary theory~$\rho $ can be
constructed from a basepoint $*\to BG$: the principal $G$-bundles are
equipped with a trivialization on $\rho $-colored boundaries.  A traditional
picture of the $G$-symmetry of the theory~$F$ uses this \emph{classical}
picture: the background fields~$\sF$ are augmented to
$\tsF=\{\textnormal{orientation}, \textnormal{Riemannian metric},
\textnormal{$G$-bundle}\}$, which fibers over the sheaf
$\{\textnormal{$G$-bundle}\}$, so in that sense fibers over~$BG$ as
in~\S\ref{subsec:1.1}.  There is an absolute field theory on~$\tsF$ which is
the ``coupling of~$F$ to a background gauge field'' for the symmetry
group~$G$.  The framework we are advocating here of $F$~as a $\sr$-module
uses the \emph{quantum} finite gauge theory~$\sigma $.
  \end{remark}

  \begin{remark}[]\label{thm:16}
 The finite path integral construction of the regular (Dirichlet) boundary
theory makes the isomorphism~$\theta $ in~\eqref{eq:21} apparent.  Namely, to
evaluate~$(\sigma ,\rho )$ we sum over $G$-bundles equipped with a
trivialization on $\rho $-colored boundaries.  Since the trivialization
propagates across an interval, the sandwich theory (Figure~\ref{fig:6}) is
the original theory~$F$ without the explicit $G$-symmetry.
  \end{remark}

  \begin{example}[a once-categorified symmetry theory]\label{thm:38}
 Let $G$~be an infinite discrete group and let $\GA$~be its group algebra,
which we treat as untopologized.  As an object in the Morita
2-category~$\Alg(\Vect)$ of algebras in vector spaces, $\GA$~is 1-dualizable
but not 2-dualizable.  By the cobordism hypothesis, it determines a
once-categorified 1-dimensional topological field theory~$\sigma $ with
$\sigma (\pt)=\GA$.  Furthermore, the right regular module~$\GA_{\GA}$,
regarded as a 1-morphism $\GA\to 1$ in~$\Alg(\Vect)$, has a right adjoint but
not a left adjoint.  Hence it determines a right relative field theory~$\rho
$.  The pair~$\sr$ is a valid 1-dimensional quiche (see
Remark~\ref{thm:10}(8)).
 
A similar story works for $G$~a compact Lie group.  By the Peter-Weyl theorem
the space of $L^2$~functions on~$G$ is a completion of a direct sum~$A_G$ of
matrix algebras; the sum is indexed by the set of equivalence classes of
irreducible representations of~$G$.  (If $\dim G>0$ then the direct sum~$A_G$
is not unital; adjoin a unit to obtain a unital algebra.  Its ``regular''
module is taken to be the direct sum without the unit.)  There is a
once-categorified 1-dimensional topological field theory with values
in~$\Alg(\Vect)$ whose value on a point is~$A_G$.  In this theory the circle
maps to an infinite dimensional vector space which is a sum of lines, one line
for each irreducible representation of~$G$.
 
If $G$~is an infinite discrete group or a compact Lie group, and if $\sH$~is a
Hilbert space equipped with a linear $G$-action, and $H$~is a $G$-invariant
Hamiltonian, then as in Example~\ref{thm:14} we can construct a left
$\sr$-module structure on the 1-dimensional quantum mechanical theory~$F$ built
from~$\sH,H$.  This illustrates Remark~\ref{thm:12}(1).
  \end{example}

  \begin{example}[full WZW]\label{thm:17}
 As mentioned earlier (Remark~\ref{thm:10}(4)), many 3-dimensional
Chern-Simons theories~$\sigma '$ do not admit nonzero fully local topological
boundary theories, hence cannot act as symmetries on 2-dimensional field
theories.  But the doubled theory $\sigma =|\sigma ' |^2$ is a Turaev-Viro
theory, and it can be realized with codomain $\Alg(\sC')$ for $\sC'$~a
suitable 2-category of linear categories; see \cite[\S1.3]{FT2} for example.
In particular, $\sigma $~admits a right regular boundary theory~$\rho $.
Then the full \emph{nonchiral} Wess-Zumino-Witten model~$F$ (with the same
group and level as the Chern-Simons theory~$\sigma '$) carries a $\sr$-module
structure.\footnote{Analogously to Example~\ref{thm:14}, we must
augment~$\sC'$ to include linear categories enriched over suitable
topological vector spaces.}
  \end{example} 

  \begin{figure}[ht]
  \centering
  \includegraphics[scale=.35]{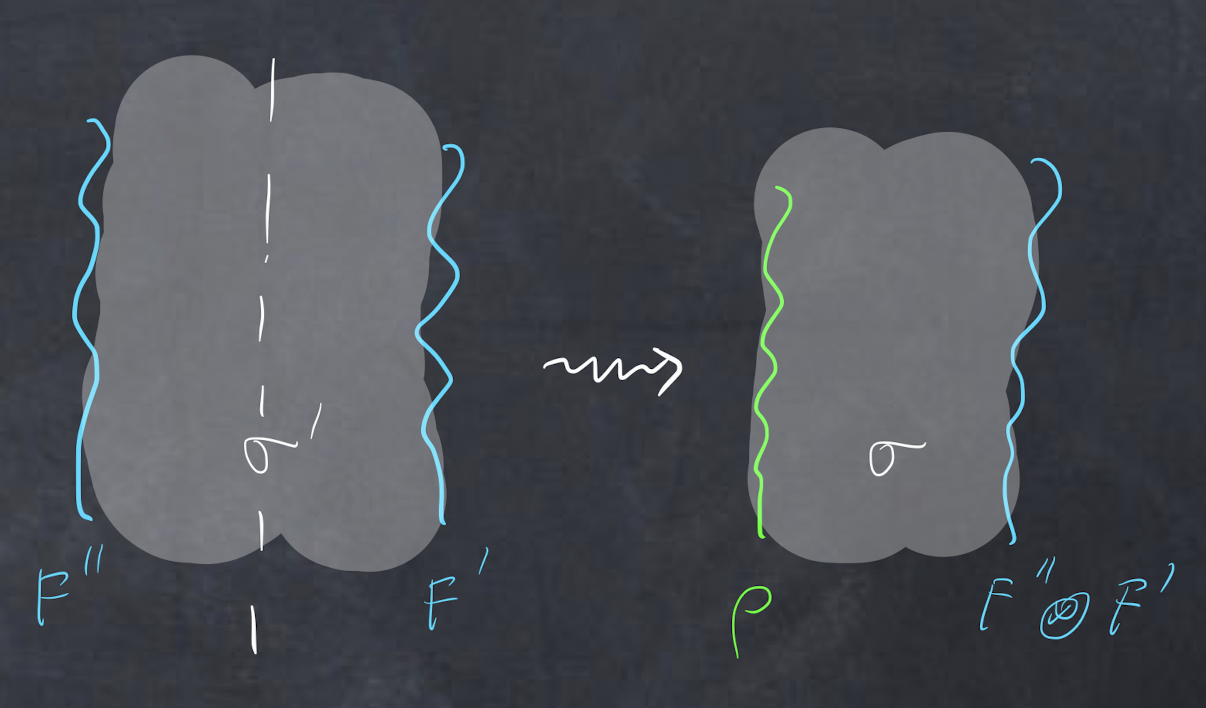}
  \vskip -.5pc
  \caption{Folding the chiral-antichiral WZW picture to obtain symmetry of
full WZW}\label{fig:26}
  \end{figure}

  \begin{remark}[]\label{thm:39}
 Frequently a \emph{chiral} 2-dimensional rational conformal field theory~$F'$,
such as a chiral WZW model, is viewed as a left boundary theory of a
3-dimensional topological field theory~$\sigma '$;
see~\cite{Wi2,MSei2,EMSS,ADW,MSei3}.  There is a conjugate anti-chiral
theory~$F''$ which is a right boundary theory of~$\sigma '$.  There is a
canonical nonchiral theory~$F$ formed as the sandwich $F''\otimes _{\sigma
'}F'$.  This is called the \emph{diagonal} combination of the chiral and
antichiral theories.  The setup in Example~\ref{thm:17} is a folding of this
diagonal combination in the middle, which doubles~$\sigma '$ to~$\sigma $ with
left boundary theory~$F''\otimes F'$; see Figure~\ref{fig:26} in which the
right regular boundary theory~$\rho $ is also depicted.  (Note that whereas a
modular tensor category is used in the construction of~$\sigma '$, only the
underlying fusion category is retained under doubling to form~$\sigma $: the
braiding is lost.)  The right regular boundary theory~$\rho $ produces the
diagonal combination; \emph{any} topological right boundary theory can be
substituted in place of~$\rho $ to form a sandwich which is a 2-dimensional
conformal field theory.  In some cases topological right $\sigma $-modules can
be classified, and this leads to a classification of full conformal field
theories obtained by combining a fixed chiral rational conformal field theory
with its conjugate anti-chiral theory.  The traditional approach does not use
\emph{full locality}, but rather uses single-valuedness of correlation
functions in genera~0 and~1 (which follows automatically in our setup for
oriented boundary theories); see~\cite{CIZ,MSei1,DV,KaSa,FRS} and also more
recent papers~\cite{D,E-M}.  We hope to elaborate on our approach elsewhere.
  \end{remark}

  \begin{example}[a homotopical symmetry]\label{thm:18}
 Let $G$~be a connected compact Lie group, and suppose $A\subset G$ is a finite
subgroup of the center of~$G$.  Let $\bG=G/A$.  Then a $G$-gauge theory in
$n$~dimensions---for example, pure Yang-Mills theory---often has a
$BA$~symmetry.  In this case we take $\sigma =\sXd{n+1}{\BtA}$ to be the
$A$-gerbe theory based on the $\pi $-finite space~$\BtA$, and we take~ $\rho $
to be the regular boundary theory constructed from a basepoint $* \to \BtA$.
(See Appendix~\ref{sec:5} for finite homotopy theories.)  The left $\sigma
$-module~$\tF$ is a $\bG$-gauge theory: given an $A$-gerbe in the bulk, on the
$\tF$-boundary we sum over pairs consisting of a $\bG$-connection and an
isomorphism of the restricted $A$-gerbe with the obstruction to lifting the
principal $\bG$-bundle to a principal $G$-bundle.  (The isomorphism is a
fluctuating field in~$\tF$.)  Aspects of this example are discussed in more
detail in~\cite[\S4]{FT3}, and it is taken up again in~\cite{F4}.

This remark clarifies an issue that comes up in many physics papers, beginning
with~\cite{tH}.  In a pure Yang-Mills theory with gauge group $G$ (or in a
gauge theory, such as Donaldson-Witten theory, with all fields in the adjoint
representation) the partition function on a manifold is constructed from the
sum over principal $G$-bundles with fixed \emph{'t~Hooft flux}. Indeed, as
pointed out by 't~Hooft, the partition function makes sense for such bundles,
which are principal $\bG$-bundles that do not lift to principal $G$-bundles.
In our picture this corresponds to the insertion of a codimension two defect on
the $\rho $-boundary.  If we consider the same field representations, but take
the gauge group to be $\overline G$, then when defining the partition function
we sum over 't Hooft fluxes.  In the current picture, the 't Hooft flux of the
$G$-theory is fixed because of the boundary theory on the topological side of
the quiche, coupled to the $\overline G$ theory on the right.

  \end{example}

These examples only scratch the surface; we offer additional illustrations
in~\S\ref{sec:4} below.  Many more examples appear in the literature.

  \subsection{Quotienting by a symmetry in field theory}\label{subsec:3.4}

Section~\ref{subsec:1.3} is motivation for the following. 

  \begin{definition}[]\label{thm:19}
 Let $\sC'$ be a symmetric monoidal $n$-category, and set $\sC=\Alg(\sC')$.  An
\emph{augmentation} of $A\in \sC$ is an algebra homomorphism $\eA\:A\to 1$
from~$A$ to the tensor unit~$1\in \sC$.
  \end{definition}

\noindent
 Thus $\eA$~is a 1-morphism in~$\sC'$ equipped with data that exhibits the
structure of an algebra homomorphism.  Augmentations may not exist, as
in~\S\ref{subsec:1.4}.  

  \begin{remark}[]\label{thm:20}
 A general 1-morphism $A\to 1$ in~$\sC$ is an object of~$\sC'$ equipped with
a right $A$-module structure.  An augmentation is a right $A$-module
structure on the tensor unit~$1\in \sC'$. 
  \end{remark}

  \begin{definition}[]\label{thm:21}
 Let $\sC'$ be a symmetric monoidal $n$-category, and set $\sC=\Alg(\sC')$.
Let $\sF$~be a collection of $(n+1)$-dimensional fields, and suppose $\sigma
\:\Bord_{n+1}(\sF)\to \sC$ is a topological field theory.  A right boundary
theory~$\epsilon $ for~$\sigma $ is an \emph{augmentation} of~$\sigma $ if
$\epsilon (\pt)$~is an augmentation of~$\sigma (\pt)$ in the sense of
Definition~\ref{thm:19}.  
  \end{definition}

\noindent
 An augmentation in this sense is often called a \emph{Neumann boundary
theory}. 

  \begin{remark}[]\label{thm:22}
 In this context, if $\rho $~is the right regular boundary theory of~$\sigma $
and $\epsilon $~is an augmentation of~$\sigma $, then we can use the
homomorphism $\epsilon (\pt)\:A\to 1$ to make~$1$ into a left $A$-module, where
$A=\sigma (\pt)$, and so construct a dual \emph{left} boundary theory~$\epsilon
^L$.  Then the sandwich $\rho \otimes _\sigma \epsilon ^L$ is the trivial
theory, as follows from the cobordism hypothesis since its value on a point is
$A\otimes _A1\cong 1$.
  \end{remark}

  \begin{example}[finite path integrals]\label{thm:23}
 Let $\sX$~be a $\pi $-finite space, and let $\sigma =\sXn$ be the associated
topological field theory.  There is a canonical Neumann boundary theory; it
is the quantization of $\id_{\sX}\:\sX\to \sX$.  See~\autoref{thm:73}.
  \end{example}

  \begin{example}[twisted version]\label{thm:24}
 Continuing, suppose $\lambda \in Z^{n+1}(\sX;\Cx)$ is a cocycle for ordinary
cohomology with $\Cx$~coefficients.\footnote{We can use ``cocycles'' in
generalized cohomology theories as well.}  Recall (Definition~\ref{thm:72})
that a right boundary theory may be constructed from a pair~$(p,\mu )$ of a
map $p\:\sY\to \sX$ of $\pi $-finite spaces and a cochain $\mu \in
C^n(\sY;\Cx)$ such that $\delta \mu =-p^*\lambda $.  For~$\sY=\sX$ and
$f=\id_{\sX}$ the cochain~$\mu $ exists iff the cohomology class $[\lambda
]\in H^{n+1}(\sX;\Cx)$ vanishes.  For example, a Dijkgraaf-Witten theory with
nontrivial twisting does not admit an augmentation.  If $[\lambda ]=0$, and
even if~$\lambda = 0$, then different choices of~$\mu $ (up to coboundaries)
yield different Neumann boundary theories.  The general definition of an
augmentation in this context is Definition~\ref{thm:73}(2).
  \end{example}

We use notations in Definition~\ref{thm:9} and Definition~\ref{thm:11} in the
following.  

  \begin{definition}[]\label{thm:25}
 Suppose given an $n$-dimensional quiche~$\sr$ and a $\sr$-module structure
$\tFt$ on a quantum field theory~$F$.  Suppose $\epsilon $~is an augmentation
of~$\sigma $.  Then the \emph{quotient} of~$F$ by the symmetry~$\sigma $ with
augmentation~$\epsilon $ is  
  \begin{equation}\label{eq:24}
     F\bs :=\epsilon  \otimes _\sigma \tF. 
  \end{equation}
  \end{definition}

  \begin{figure}[ht]
  \centering
  \includegraphics[scale=1.3]{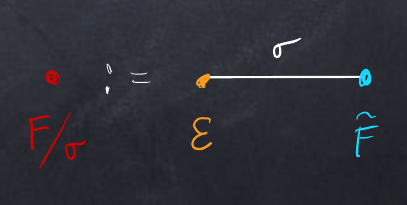}
  \vskip -.5pc
  \caption{The quotient $F\bs$}\label{fig:9}
  \end{figure}

\noindent
 The right hand side of~\eqref{eq:24} is the sandwich in Figure~\ref{fig:9}.

  \begin{example}[]\label{thm:26}
 Let $G$~be a finite group, and let $\sigma =\sXd{n+1}{BG}$ be the associated
finite gauge theory.  Use the canonical Neumann boundary theory of
Example~\ref{thm:23}.  In the semiclassical picture this corresponds to
summing over all principal $G$-bundles with no additional fields on the
$\epsilon $-colored boundaries.  This is the usual quotienting operation, oft
called `gauging'.
  \end{example}

  \begin{remark}[]\label{thm:27}
 A nontrivial $n$-cocycle~$\mu $, as in Example~\ref{thm:24}, induces a sum
over $G$-bundles with weights, so a twisted version of the usual quotient.
This twist goes by various names: `discrete torsion', `$\theta $-angles',
etc., depending on the context.  The following example is an illustration. 
  \end{remark} 

  \begin{example}[]\label{thm:93}
 Picking up on \autoref{thm:38}, consider the quantum mechanical system of a
particle on the Euclidean line~$\RR_x$ with Hamiltonian the Laplace operator
$-d^2/dx^2$.  The system is invariant under the action of the infinite discrete
group~$\ZZ$ by translations.  We realize it as a theory relative to
1-dimensional once-categorified $\ZZ$-gauge theory.  The group
algebra~$\CC[\ZZ]$ has a natural augmentation~\eqref{eq:6}, and the
quotient~\eqref{eq:24} relative to this augmentation is\footnote{One cannot
simply use the Hilbert space of states, since there are no $\ZZ$-invariant
$L^2$~functions on~$\RR$, but rather one uses a \emph{rigging} that locates the
Hilbert space between two nuclear spaces; see~\cite[\S3]{KS} and~\cite{Wed}.}
the particle on the circle~$\RR/\ZZ$.  Now for~$\theta \in \RR/2\pi \ZZ$
consider the character $n\mapsto e^{in\theta }$ of~$\ZZ$.  The quotient by the
augmentation that corresponds to this character is the particle on the circle
in the presence of a constant magnetic field.
  \end{example} 

  \begin{example}[]\label{thm:113}
 Picking up on \autoref{thm:18}, if we replace the regular boundary theory by
the augmentation~$\epsilon $, then the sandwich $\epsilon \otimes _\sigma \tF$
is the $\bG$-gauge theory. 
  \end{example}

  \begin{example}[]\label{thm:28}
 For~$n=2$ if the codomain of~$\sigma $ is a 3-category $\sC=\Alg(\Cat)$ of
tensor categories, then an augmentation $A\to\Vect$ of the tensor
category~$A=\sigma (\pt)$ is called a `fiber functor'.  Fiber functors need not
exist.  For example, suppose $\lambda $~is a cocycle which represents the
nonzero cohomology class in~$H^3(B\zt;\Cx)$.  The fusion category
$\Vect^\lambda [\zt]$ is a twisted categorified group ring of~$\zt$ with
coefficients in~$\Vect$: see \cite[Example~2.3.8]{EGNO}.  (An alternative
description of~$\Vect^\lambda [\zt]$ is in~\cite[\S4]{FHLT}.)  This tensor
category does not admit a fiber functor \cite[Example~5.1.3]{EGNO}.  The
associated 3-dimensional topological field theory---a Dijkgraaf-Witten
theory---with its right regular module encodes \emph{anomalous} group actions
on quantum field theories.  This is an example of an \emph{'t Hooft anomaly}.
  \end{example} 

Now suppose that the codomain of~$\sigma $ has the form $\sC=\Alg(\sC')$, as in
Definition~\ref{thm:21}, let $\rho $~be the right regular boundary theory, and
suppose $\epsilon $~is a right boundary theory which is an augmentation
of~$\sigma $.  Then there is a preferred domain wall~$\delta $ from~$\rho $
to~$\epsilon $ as well as a preferred domain wall~$\delta ^*$ from~$\epsilon $
to~$\rho $.  Namely, $\Hom_{\sigma (\pt)}\bigl(\epsilon (\pt),\rho (\pt)
\bigr)$ has a distinguished element which corresponds to the tensor unit in
$\rho \otimes _\sigma \epsilon ^L\cong 1$; see Remark~\ref{thm:22}.  

  \begin{definition}[]\label{thm:86}
 $\delta $ ~is the \emph{Dirichlet-to-Neumann domain wall}, and $\delta
^*$~is the \emph{Neumann-to-Dirichlet domain wall}. 
  \end{definition}

\begin{figure}[ht]
  \centering
  \includegraphics[scale=.6]{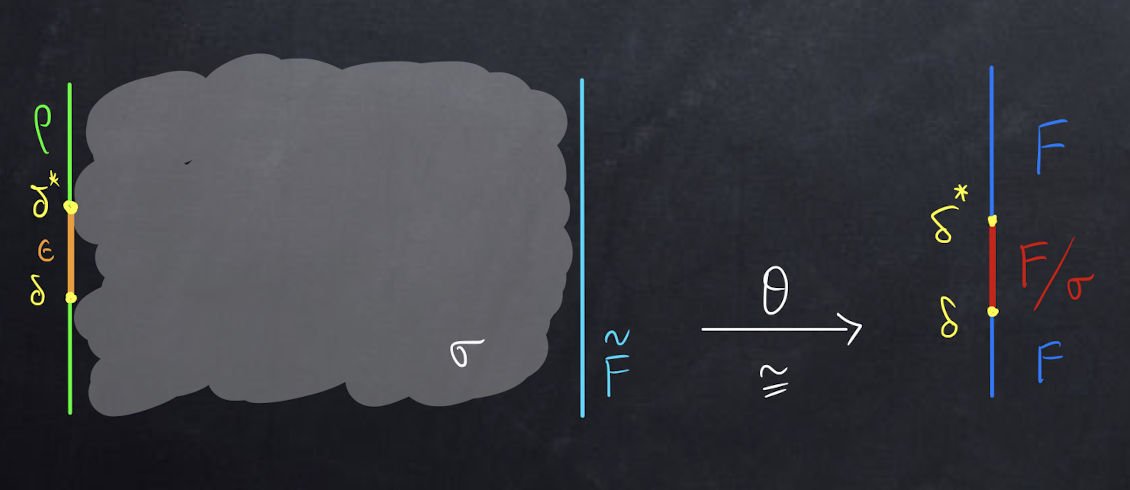}
  \vskip -.5pc
  \caption{The canonical domain walls $\delta \:F\to F\bs$ and $\delta
  ^*\:F\bs\to F$}\label{fig:34} 
  \end{figure}

Let $F$~be an $n$-dimensional field theory equipped with a $\sr$-module
structure.  Then $\delta $~and $\delta ^*$~determine canonical domain walls
$\delta \:F\to F\bs$ and $\delta ^*\:F\bs\to F$, as depicted in
Figure~\ref{fig:34}.

  \subsection{Dual symmetry on a quotient; finite electromagnetic duality}\label{subsec:3.5}

In special situations a quotient theory~$F\bs $ inherits a module structure for
a dual quiche to the original quiche~ $\sr$.  One situation in which this
occurs is when $\sigma $~is the field theory of a $\pi $-finite infinite loop
space, or equivalently a connective $\pi $-finite \emph{spectrum}; see
Definition~\ref{thm:53}.  Examples include symmetries by (higher) finite
\emph{abelian} groups as well as by 2-groups whose $k$-invariant is a
\emph{stable} cohomology class.  This dual symmetry is well-known in the
physics literature: in low dimensions there is a precise analog for nonabelian
groups~\cite{AG}, \cite[\S4.1.2]{DGNO}, and higher dimensional generalizations
have appeared recently~\cite{BSW,BBFP}.  Also, see~\cite{PSV} for
electromagnetic duality in the sandwich picture.

  \begin{remark}[]\label{thm:92}
 Although our exposition is confined to $\pi $-finite spectra, this duality
holds more generally: for example, electromagnetic duality for general finite
groups.  The expectation is that the dual symmetry to a general quiche~$\sr$
with augmentation~$\epsilon $ is the quiche~$\sdrd$ which comprises
$\sd=\End_\sigma (\epsilon )$ with its regular module~$\rho ^\vee$; it has an
augmentation $\epsilon ^\vee=\Hom_{\sigma }(\rho ,\epsilon )$.
  \end{remark}

Recall that if $A$~is a finite abelian group, then its Pontrjagin dual is the
finite abelian group $A\dual=\Hom(A,\TT)$.  There is a similar
\emph{character dual}\footnote{We could use $\QQ/\ZZ$ in place of~$\TT$, in
which case we obtain the \emph{Brown-Comenetz dual}~\cite{BC}.} for $\pi
$-finite spectra.  First, define the spectrum~$I\TT$ by the universal
property
  \begin{equation}\label{eq:96}
     [\bX,I\TT]\cong (\pi _0\bX)\dual 
  \end{equation}
for all spectra~$\bX$.  (Here $[\bX,\bX']$~denotes the abelian group of
homotopy classes of spectrum maps $\bX\to \bX$.)  The \emph{spectrum} of maps
$\sT\to I\TT$ is the character dual spectrum~$\sTd$ of the $\pi $-finite
spectrum~$\sT$; the spectrum~$\sTd$ is also $\pi $-finite.

Fix~$n\in \ZZ^{>0}$ and suppose $\sT$~is a $\pi $-finite spectrum with
0-space the \emph{pointed} topological space~$\XT$.  Let $\sigma
=\sXd{n+1}{\sT}$ be the corresponding $(n+1)$-dimensional topological field
theory.  The basepoint $*\to \XT$ determines a Dirichlet boundary
theory~$\rho $.  The homotopy class of the duality pairing
  \begin{equation}\label{eq:25}
     \STd\times \sT\longrightarrow \Sigma ^nI\TT 
  \end{equation}
is an $I\TT$-cohomology class on $\XTd\times \XT$; let $\mu $~be a cocycle
representative.\footnote{In many cases of interest the pairing~\eqref{eq:25}
factors through a simpler cohomology theory.  For example, if $\sT=\Sigma
^sHA$ is a shifted Eilenberg-MacLane spectrum of a finite abelian group, then
\eqref{eq:25}~factors through~$\Sigma ^nH\TT$ and we can represent~$\mu $ as
a singular cocycle with coefficients in~$\sT$.  Recall that we use the word
`cocycle' for any geometric representative of a generalized cohomology
class.}

  \begin{definition}[]\label{thm:29}
 \ 

 \begin{enumerate}[label=\textnormal{(\arabic*)}]

 \item The \emph{dual quiche}~$\sdrd$ to~$\sr$ is the finite homotopy
theory $\sd=\sXd{n+1}{\XTd}$ with Dirichlet boundary theory~$\rd$ defined by
the basepoint $*\to \XTd$.

 \item The \emph{canonical domain wall}~$\zeta $ between~$\sd$ and~$\sigma
$---i.e., $(\sd,\sigma )$-bimodule---is the finite homotopy theory
constructed from the correspondence of $\pi $-finite spaces
  \begin{equation}\label{eq:26}
     \begin{gathered} \xymatrix{&(\XTd\times \XT,\mu )\ar[dl]\ar[dr] \\
     \XTd&&\XT} \end{gathered} 
  \end{equation}
in which the maps are projections onto the factors in the Cartesian product.
There is a similar canonical domain wall $\zeta \dual\:\sigma \dual\to \sigma
$.

 \item The \emph{canonical Neumann boundary theories}~ $\epsilon ,\ed$ are
the finite homotopy theories induced from the identity maps on~$\XT,\XTd$,
respectively. 

 \end{enumerate} 
  \end{definition} 

 \noindent
 Our formulation emphasizes the role of~$\sigma $ as a symmetry for another
quantum field theory.  But $\sigma $~is a perfectly good $(n+1)$-dimensional
field theory in its own right.  From that perspective $\sd$~is the
$(n+1)$-dimensional \emph{electromagnetic dual} theory.  See~\cite{Liu} for
more about electromagnetic duality in this context. 

  \begin{remark}[]\label{thm:30}
 As usual, we have not made explicit the background fields for~$\sigma $,
$\sd$, and~$\zeta $, $\zeta\dual$.  In fact, the theories $\sigma $~and
$\sd$~are defined on bordisms unadorned by background fields: they are
``unoriented theories''.  For~$\zeta $ we need a set of (topological)
background fields which orient manifolds sufficiently to integrate~$\mu $.
For example, if $\mu $~is a singular cocycle with coefficients in~$\TT$, then
we need a usual orientation.  For~$I\TT$ we would need framings.
  \end{remark}

  \begin{proposition}[]\label{thm:31}
 There is an isomorphism of right $\sigma $-modules 
  \begin{equation}\label{eq:27}
     \psi \:\rd\otimes _{\sd}\zeta \xrightarrow{\;\;\cong \;\;}\epsilon 
  \end{equation}
  \end{proposition}

  \begin{figure}[ht]
  \centering
  \includegraphics[scale=1.3]{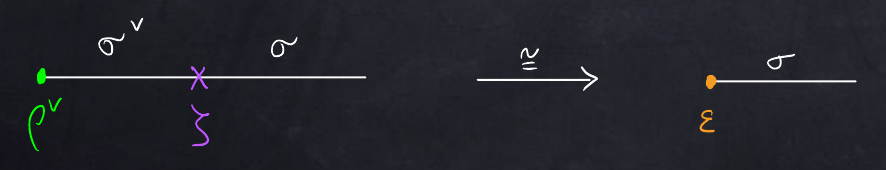}
  \vskip -.5pc
  \caption{An isomorphism of right $\sigma $-modules}\label{fig:10}
  \end{figure}

\noindent
 This isomorphism is depicted in Figure~\ref{fig:10}.  In words,
(generalized) electromagnetic duality swaps Dirichlet and Neumann boundary
theories. 

  \begin{proof}
 We use the calculus of $\pi $-finite spectra, as described
in~\S\ref{subsubsec:5.3.1}---see especially the composition
law~\eqref{eq:85}.  The theory~$\sigma $ is induced from~$\XT$, the
theory~$\sd$ from~$\XTd$, the boundary theory~$\rd$ from $*\to \XTd$, and the
domain wall~$\zeta $ from the correspondence diagram~\eqref{eq:26}.  Hence
$\rd\otimes _{\sd}\zeta $ is induced from the homotopy fiber product:
  \begin{equation}\label{eq:28}
     \begin{gathered}
     \xymatrix{\ast\ar[dr]&(\XT,0)\ar@{-->}[l]\ar@{-->}[r]&(\XTd\times \XT,\mu
     )\ar[dl]\ar[dr] \\ &\XTd&&\XT} \end{gathered} 
  \end{equation}
Here we use that the restriction of~$\mu $ to~$*\times \XT$ is zero.   So the
sandwich is the right $\sigma $-module induced from the composition 
  \begin{equation}\label{eq:29}
     \begin{gathered} \qquad \qquad \xymatrix{(\XT,0)\ar@{-->}[r] &(\XTd\times
     \XT,\mu )\ar[dr] \\&&\XT} \end{gathered} 
  \end{equation}
which is~$\id_{\XT}$.  That theory is the Neumann boundary theory~$\epsilon
$.  
  \end{proof}

  \begin{corollary}[]\label{thm:32}
 Let $F$~be a quantum field theory equipped with a $\sr$-module structure.
Then the quotient~$F\bs$ carries a canonical $\sdrd$-module structure. 
  \end{corollary}

  \begin{figure}[ht]
  \centering
  \includegraphics[scale=1.25]{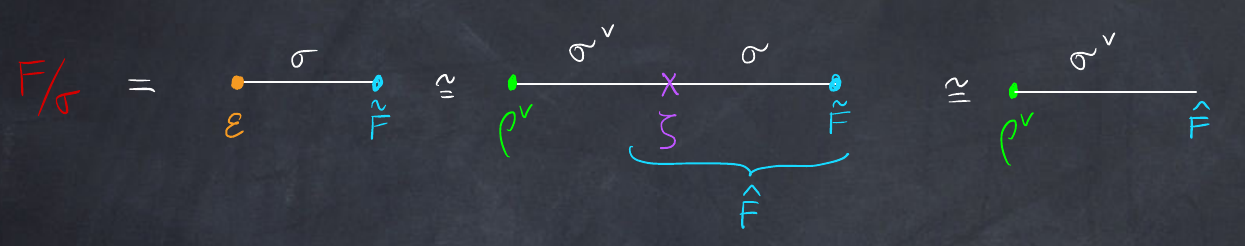}
  \vskip -.5pc
  \caption{The dual symmetry on the quotient~$F\bs$}\label{fig:11}
  \end{figure}

  \begin{proof}
 The proof is contained in Figure~\ref{fig:11}.  In words: Let $\tFt$ be the
$\sr$-module data, as in Definition~\ref{thm:11}.  Define the left
$\sd$-module 
  \begin{equation}\label{eq:30}
     \hF=\zeta \otimes _\sigma \tF 
  \end{equation}
and the isomorphism 
  \begin{equation}\label{eq:31}
     \hth\:\rd\otimes _{\sd}\hF = \rd\otimes _{\sd}\zeta \otimes _{\sigma
     }\tF\xrightarrow{\;\;\psi \,\otimes\, \theta \;\;} \epsilon \otimes _\sigma
     \tF = F\bs
  \end{equation}
Then $(\hF,\hth)$ is the desired $\sdrd$-module structure. 
  \end{proof}

  \begin{remark}[]\label{thm:80}
 The domain wall~$\zeta \:\sigma \to \sd$ maps left $\sigma $-modules to left
$\sd$-modules; this is the effect of electromagnetic duality (on left
modules).  It follows from the previous that the transform of~$F$ under
electromagnetic duality is the quotient~$F\bs$.  This duality is involutive
up to a multiplicative constant: the Euler theory; see~\cite{Liu} for
details. 
  \end{remark}

  \begin{example}[]\label{thm:89}
 Let $n=2$ and let $A$~be a finite abelian group.  As explained in~\cite{FT1}
and the references therein, given an appropriately admissible real-valued
function on~$A$ there is a corresponding \emph{Ising model}.  It can be
viewed as a 2-dimensional field theory on manifolds equipped with a
\emph{lattice} (appropriately defined).  The group~$A$ acts as a symmetry on
this theory: the Ising model has a $\sr$-module structure for $\sigma
=\sXd3{BA}$~the 3-dimensional $A$-gauge theory.  Finite electromagnetic
duality maps~$\sXd3{BA}$ to 3-dimensional $A\dual$-gauge
theory~$\sXd3{BA\dual}$.  The effect on the boundary Ising model is called
\emph{Kramers-Wannier duality}.  For $A=\bmut$ the admissible function is
parametrized by an inverse temperature~$\beta \in \RR^{>0}$ and, under the
canonical identification $A=A\dual$, Kramers-Wannier duality amounts to an
involution $\beta \leftrightarrow\beta \dual$ of $\RR^{>0}$.  The unique
fixed point~$\beta ^c$ is the critical temperature; it is the unique
temperature at which the Ising model is not gapped.  As another example, if
$A=\bmu5$ then there is a distinguished line in the space of admissible
functions (modulo uniform scaling), which is the line of five-state Potts
models.  (One can replace~5 with any integer~$\ge5$ in this discussion.)
There is again an involution on this line with a unique fixed point, but now
the model is gapped everywhere on the line; there is a first-order phase
transition at the fixed point, such as in the recent paper~\cite{D-CGHMT}.
  \end{example}

   \section{Symmetries, defects, and composition laws}\label{sec:4}

Elements of an abstract algebra~$A$ act as operators on any (left) module~$L$,
and any equation in~$A$ holds for the corresponding operators on~$L$.  The
analogs for the quiche~$\sr$ in field theory are \emph{defects} in~$\sr$ and
the relations among them.  Hence we begin in~\S\ref{subsec:4.1} with an
exposition of these defects and how they transport to topological defects in a
$\sr$-module theory.  We illustrate this concretely for finite\footnote{The
discussion extends to infinite discrete and compact Lie groups; see
Example~\ref{thm:38}.} groups of symmetries acting in quantum mechanics.  We
found this simple example to be quite instructive for the general story, which
explains the length of our treatment in~\S\ref{subsec:4.2}.
In~\S\ref{subsec:4.3} we move one dimension higher, where with extra room there
are new phenomena: the difference between local and global defects
(Remark~\ref{thm:41}), defects supported on singular sets
(Figure~\ref{fig:27}), etc.  These examples focus on ordinary finite groups of
symmetries.  Our formalism easily incorporates higher groups of symmetries, as
we take up in~\S\ref{subsec:4.4}.  The twistings in a higher group make the
composition laws for defects more complicated than might be suspected, as we
illustrate in \S\ref{subsubsec:4.4.1}.  (There are many theories with a 2-group
of symmetries as described in~\S\ref{subsubsec:4.4.1}; see~\cite{KT,Ta,CDI,BCH}
for example.)  More exotic phenomena can be exhibited with a simple 2-stage
spectrum, as we touch upon in~\S\ref{subsubsec:4.4.2}.

  \subsection{Generalities}\label{subsec:4.1}

\begin{figure}[ht]
  \centering
  \includegraphics[scale=.9]{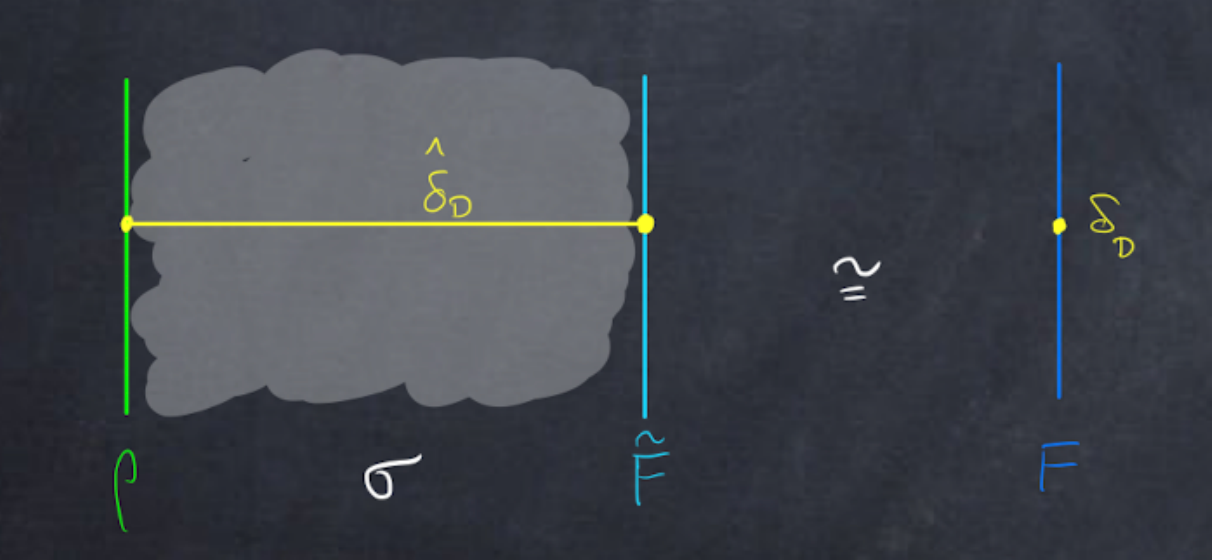}
  \vskip -.5pc
  \caption{Transporting a defect under the isomorphism~$\theta $ in~\eqref{eq:21}}\label{fig:12}
  \end{figure}

Fix a positive integer~$n$.  Suppose $\sr$~is an $n$-dimensional quiche and
$F$~is an $n$-dimensional quantum field theory equipped with a left
$\sr$-module structure~$\tFt$.  Assume, as in~\S\ref{subsec:2.4}, that $M$~is
a $k$-dimensional manifold or bordism, $k\in \{0,1,\dots ,n\}$, and $D\subset
M$~is a submanifold or a stratified subspace that is the support of a
defect~$\delta _D$.  Use the isomorphism~\eqref{eq:21} to transport the
defect~$\delta _D$ to a defect~$\hdD$ supported on $[0,1]\times D\subset
[0,1]\times M$ for the theory~$\srtF$, where $\{0\}\times M$ is $\rho
$-colored and $\{1\}\times M$ is $\tF$-colored; see Figure~\ref{fig:12}.

Conversely, defects in the theory~$\srtF$ transport to defects in~$F$, but
the possibilities are richer as we illustrate below.  We first single out a
collection of defects associated to the $\sr$-symmetry.

  \begin{definition}[]\label{thm:33}
 A \emph{$\sr$-defect} is a defect in the topological field theory~$\sr$.  We
call it a \emph{$\rho $-defect} if its support lies entirely in a $\rho
$-colored boundary.
  \end{definition}

\noindent
 These are defects in the abstract symmetry theory.  If $F$~is a quantum
field theory equipped with an $\sr$-module structure $\tFt$, then a
$\sr$-defect induces a defect in the theory~$\srtF$ and hence a defect in the
theory~$F$.  Since the defect in the sandwich picture is supported away from
$\tF$-colored boundaries, it is a \emph{topological} defect in the
theory~$F$.

  \begin{remark}[]\label{thm:34}
 Computations with $\sr$-defects, such as compositions, are carried out in
the topological field theory~$\sr$.  They apply to the induced defects in any
$\sr$-module.  
  \end{remark}

  \begin{remark}[]\label{thm:35}
 Pictures such as Figure~\ref{fig:12} are interpreted as a schematic for a
tubular neighborhood of the support~$D\subset M$ of the defect (and its
Cartesian product with~$[0,1]$).  Also, unless otherwise stated, for ease of
exposition we often implicitly assume a normal framing to ~$D$ so that its
link may be identified with a standard sphere.
  \end{remark} 

The image in~$F$ of a defect in the $\srtF$-theory might not be apparent;
this is a significant advantage of the sandwich picture of~$F$.

  \begin{example}[]\label{thm:40}
 Let $n=3$ and consider a 3-dimensional quantum field theory~$F$ on~$S^3$,
and assume $F$~has an $\sr$-module structure.  In the corresponding
$\srtF$-theory we can contemplate a defect supported on a 2-disk~$D$ in
$[0,1)\times S^3$ whose boundary $K=\partial D\subset \{0\}\times S^3$ is a
knot in the Dirichlet boundary.  (Such a knot is termed `slice'.)  It is
possible that $K$~does not bound a disk in~$S^3$---its Seifert genus may be
positive.  In this case the projection of the slice disk~$D$ to a defect in
the theory~$F$ on~$S^3$ is at best an immersed disk with boundary~$K$, and it
appears that such a topological defect is difficult to describe directly in
the theory~$F$.  More generally, it is an open question whether all operators
generated by topological defects in a full $(n+1)$-dimensional topological
field theory~$\sigma $ can be replicated in a once-categorified
$n$-dimensional theory, even allowing for ``raviolization''
(Remark~\ref{thm:105}). 
  \end{example}

  \subsection{Finite symmetry in quantum mechanics}\label{subsec:4.2}

We resume consideration of the quantum mechanical theory~$F$ in
Example~\ref{thm:14}: $n=1$, the state space is a Hilbert space~$\sH$
equipped with a Hamiltonian~$H$, and $H$~is invariant under the linear action
of a finite group~$G$ on~$\sH$.  Then $\sigma $~is the 2-dimensional finite
gauge theory which counts principal $G$-bundles, $\rho $~is the Dirichlet
boundary theory which sums over sections of the $G$-bundle on $\rho $-colored
boundaries, and $\tF$~is constructed from the left module~$\!{}\mstrut
_{A}\sH$ over the group algebra~$A=\GA$.  We take the codomain of~$\sigma $
to be the Morita 2-category of algebras in vector spaces.
 
Recall from \eqref{eq:22}(b) that $F\bigl([0,\tau ] \bigr)=e^{-\tau H/\hbar}$
for the standard Riemannian metric on~$[0,\tau ]$.  Now let $D=\{t\}\subset
(0,\tau )$ and suppose $\delta _t$~is a point defect.  The link of $D\subset
[0,\tau ]$ is a 0-sphere $S^0_\epsilon =\{t-\epsilon ,t+\epsilon \}$, and
according to Remark~\ref{thm:104} the point defect~$\delta _t$ lies in the
nuclear Fr\'echet space computed as the inverse limit
  \begin{equation}\label{eq:32}
     \lim\limits_{\epsilon \to 0}F(S^0_\epsilon ); 
  \end{equation}
see~\cite[\S3]{KS}.  The topological vector space~\eqref{eq:32} can be realized
as a space of operators which may be highly singular.  This is what is exactly
what is expected for observables in quantum theory.  For our more formal
purposes, we can simply treat~$\delta _t$ as a bounded operator on~$\sH$.

  \begin{figure}[ht]
  \centering
  \includegraphics[scale=.5]{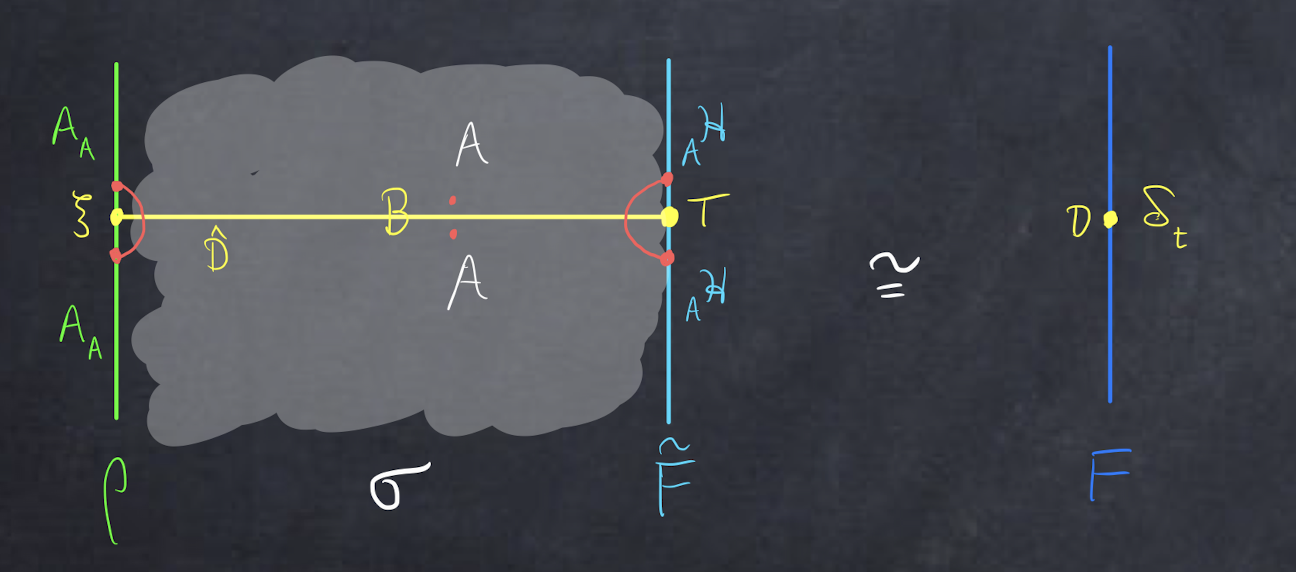}
  \vskip -.5pc
  \caption{The point defect~$\delta _t$ transported under the isomorphism~\eqref{eq:21}}\label{fig:13}
  \end{figure}

\begin{figure}[ht]
  \centering
  \includegraphics[scale=.5]{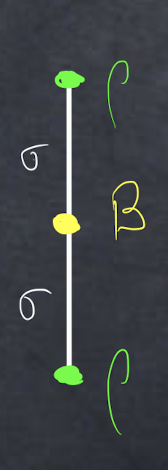}
  \vskip -.5pc
  \caption{The link of the $\rho $-boundary point of~$\widehat D$ in
Figure~\ref{fig:13}}\label{fig:14}
  \end{figure}

\begin{figure}[ht]
  \centering
  \includegraphics[scale=.5]{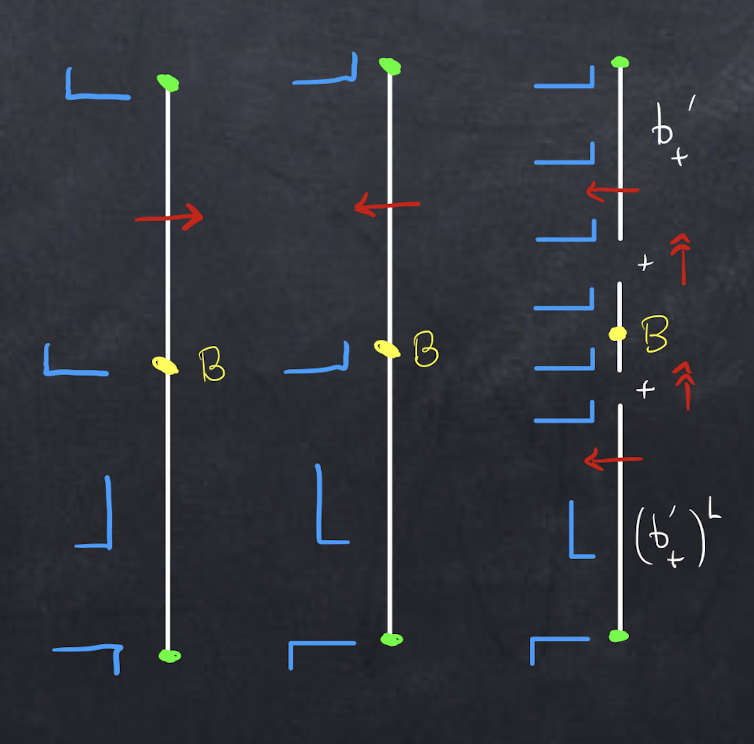}
  \vskip -.5pc
  \caption{Evaluation of the link}\label{fig:18}
  \end{figure}

Transport~$\delta _t$ to a defect in the sandwich theory, as in
Figure~\ref{fig:13}.  We obtain a defect supported on the
manifold-with-boundary $\widehat{D}=[0,1]\times \{t\}$, i.e., a domain wall.
Treat~$\widehat{D}$ as a stratified manifold and work in order of increasing
codimension.  First, the link of a point in the interior of~$\widehat{D}$
is~$S^0$, and $\Hom\bigl(1,\sigma (S^0) \bigr)$ is the category of left
$A\otimes A\op$-modules, or equivalently of $(A,A)$-bimodules.  So the label of
the defect along the interior is an $(A,A)$-bimodule~$\!{}\mstrut _{A}B\mstrut
_{\hneg A}$\,.  Next, the link of the endpoint on the $\rho $-colored boundary
is a closed interval with interior colored with~$\sigma $, boundary colored
with~$\rho $, and an interior point defect colored with~$B$; see
Figure~\ref{fig:14}.  To evaluate this under~$\sr$, we use the rules and
conventions laid out in~\cite{FT2}; they are used here in Figure~\ref{fig:18}.

\noindent
 (For comparison, see Figures~26 and~27 in \cite{FT2}.)  In the first picture
in Figure~\ref{fig:18}, then, the arrow of time points into the complement of
the endpoint of the defect; see Remark~\ref{thm:8}.  The 2-framing~$f_1,f_2$ is
depicted as long, short.  That framing is the transport of the constant framing
in Figure~\ref{fig:13}; the $\pi $-rotation of the framing is a consequence of
the unbending of the link.  Note that our conventions require that the last
framing vector point out at the $\rho $-boundary.  The second picture in
Figure~\ref{fig:18} is obtained from the first by reflection in a vertical
axis, so it is equivalent.  The third picture expresses this bordism as a
composition of bordisms glued at the $+$~point.  The bordism $(b'_+)^L$ on the
bottom is the left adjoint of the bordism~$b'_+$ on the top; see \cite[\S
A.2.5]{FT2}.  Under~$\sr$, the bordism $b'_+\:+\to \emptyset ^0$ evaluates to
the right regular module $A_A\:A\to 1$ in the Morita 2-category.  Its left
adjoint is the left $A$-module $\Hom_A(A,A)=A$.  Therefore, under~$\sr$ the
link depicted in \autoref{fig:14} evaluates to the vector space
  \begin{equation}\label{eq:33}
     A\otimes _AB\otimes _AA\cong B. 
  \end{equation}
Hence the label at the left endpoint of the defect in Figure~\ref{fig:13} is
a vector~$\xi \in B$.  At the right endpoint on the $\tF$-colored boundary we
must take a limit as the link shrinks, as in~\eqref{eq:32}.  A similar
analysis to Figure~\ref{fig:18} computes the value of this link under~$\sr$
as
  \begin{equation}\label{eq:37}
     \sH^*\otimes _AB\otimes _A\sH\cong \Hom_{(A,A)}\bigl(B,\End(\sH)\bigr), 
  \end{equation}
the space of $(A,A)$-bimodule maps $B\to \End(\sH)$; as remarked previously we
must interpret~`$\End(\sH)$' as a space of unbounded linear operators.  Let
$T\:B\to \End(\sH)$ be a choice of label at that endpoint.  Then the image of
the interval defect in the sandwich picture with labels $(\xi ,B,T)$ is the
point defect in the theory~$F$ labeled by the operator~$T(\xi )$.  (It is
instructive to consider the special case~$B=A$---the transparent defect in the
interior---in which case the defect illustrated in \autoref{fig:13} reduces to
two point defects.)

  \begin{figure}[ht]
  \centering
  \includegraphics[scale=.4]{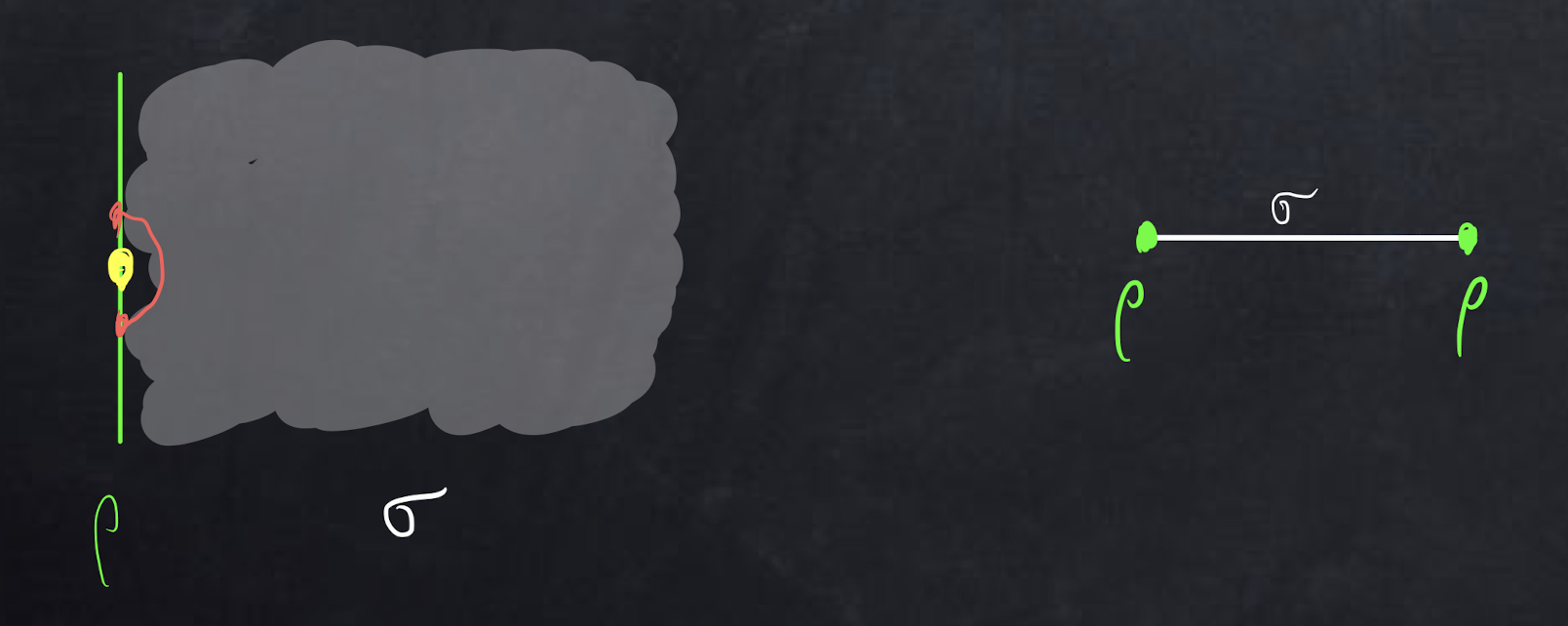}
  \vskip -.5pc
  \caption{A point $\rho$-defect and its link}\label{fig:15}
  \end{figure}

\begin{figure}[ht]
  \centering
  \includegraphics[scale=.4]{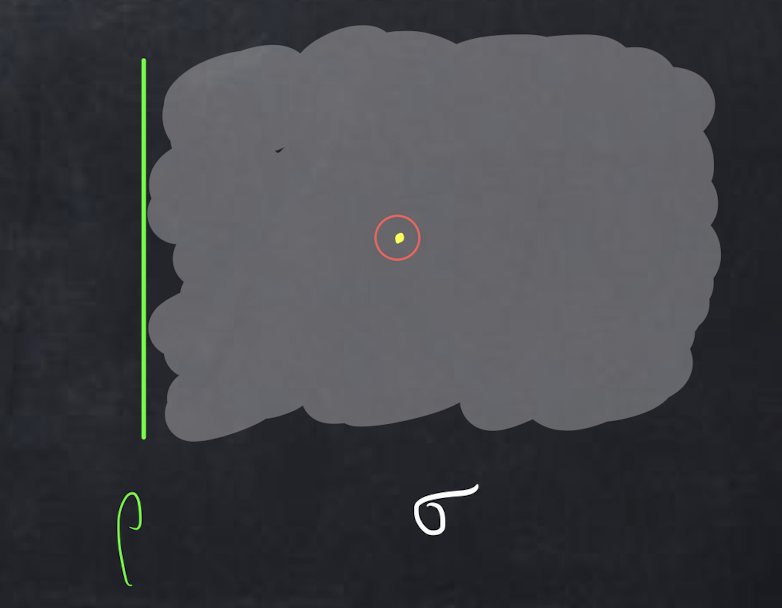}
  \vskip -.5pc
  \caption{An interior point defect and its linking circle}\label{fig:16}
  \end{figure}

The $\widehat{D}$-defects which specialize to $\sr$-defects have support
disjoint from the $\tF$-colored boundary, and so for these ${}\mstrut
_{A}B\mstrut _{\hneg A}=\!{}\mstrut _{A}A\mstrut _{\hneg A}$ is the identity
$(A,A)$-bimodule and $T=\id_{\sH}$ is the identity operator.  Then $\xi \in
A=\GA$.  In particular, we have a defect~$g\in G$ for each group element; see
Figure~\ref{fig:15}.  (The notion of a \emph{classical label} for a defect in a
finite homotopy theory is defined in \autoref{thm:115}.  The label~$g\in G$ is
an example.)  Note that the link of a point defect supported on the $\rho
$-colored boundary---a point $\rho $-defect---evaluates to the vector space
$A\otimes _AA=A$.  Next, consider a point $\sr$-defect with support in the
interior, as in Figure~\ref{fig:16}.  The link is a circle~$\cir$, and $\sigma
(\cir)$~is the center of the group algebra~$\GA$.  Here the classical labels
are conjugacy classes in~$G$: the label is the sum of the group elements in a
conjugacy class.  In particular, central elements of~$G$ can label interior
point defects.

  \begin{remark}[]\label{thm:36}
 \ 

 \begin{enumerate}[label=\textnormal{(\arabic*)}]

 \item The point defects depicted in Figures~\ref{fig:15} and ~\ref{fig:16}
have a clear geometric interpretation in the semiclassical construction of
finite gauge theory.  For the point $\rho $-defect labeled by a group
element~$g$, the principal $G$-bundle has a trivialization on the complement
of the point defect in the $\rho $-colored boundary, and the trivialization
on the $\rho $-colored boundary jumps by the group element~$g$ (relative to a
coorientation of the point defect in the $\rho $-colored boundary).  For the
interior point defect labeled by a conjugacy class, the principal $G$-bundle
is defined on the complement of the point and has holonomy in that conjugacy
class (again relative to a coorientation of the point defect).

 \item The $\sr$-defects that are usually associated with $G$-symmetry are
those supported on the $\rho $-colored boundary.  This observation applies
quite generally.  Observe that these $\rho $-defects commute with defects
whose support is disjoint from the $\rho $-colored boundary, since they are
topological and can be homotoped on that boundary without crossing the other
defects.  Similarly, the defects supported in the interior commute with $\rho
$-defects; this exhibits their central nature.  In this example the center is
smaller, so there are in a sense fewer interior $\sr$-defects than there are
$\rho $-defects.  This is not true in higher dimensions;
see~\S\ref{subsec:4.3}.

 \end{enumerate}
  \end{remark}

  \begin{figure}[ht]
  \centering
  \includegraphics[scale=.375]{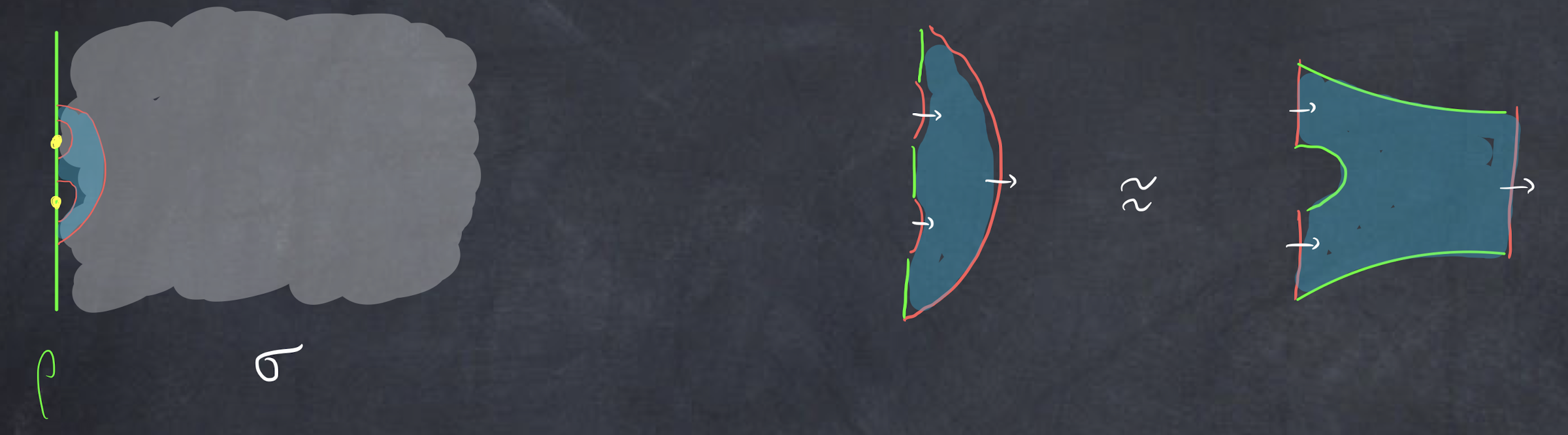}
  \vskip -.5pc
  \caption{The composition law by evaluation on a pair of chaps}\label{fig:17}
  \end{figure}

The composition law on point $\rho$-defects is computed by evaluating
the\footnote{This particular bordism is also known as
\emph{Gumby}:\raisebox{-2pc}{\includegraphics[scale=.15,trim=0 0 0
2pc]{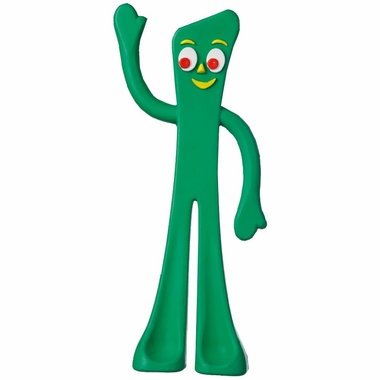}}} ``pair of chaps'' in
Figure~\ref{fig:17}.  This works out to be the multiplication map $A\otimes
A\to A$ of the group algebra; see~\eqref{eq:36} below.  In particular, on
classical labels in~$G$ it restricts to the group product $G\times G\to G$.

  \begin{remark}[]\label{thm:37}
 We make several observations that we invite the reader to apply to
subsequent examples as well. 

 \begin{enumerate}[label=\textnormal{(\arabic*)}]

 \item We can evaluate bordisms in the theory~$\sr$ by regarding $\sigma
=\sXd2{BG}$ as the finite homotopy theory built from~$BG$ with its basepoint
$*\to BG$.  So, for example, the mapping space of the link in
Figure~\ref{fig:15} is 
  \begin{equation}\label{eq:34}
     \Map\bigl(([0,1],\{0,1\})\,,\,(BG,*) \bigr)\simeq \Omega BG\simeq G, 
  \end{equation}
which quantizes to the vector space of functions on~$G$.  This is canonically
isomorphic to the vector space underlying the group algebra~$\GA$.
Similarly, the mapping space of the link in Figure~\ref{fig:16} is the free
loop space 
  \begin{equation}\label{eq:35}
     \Map(\cir,BG)\simeq \bigsqcup\limits_{[g]} BZ_g ,
  \end{equation}
where the disjoint union runs over conjugacy classes in~$G$ and $Z_g$~is the
centralizer of a chosen element~$g$ in the conjugacy class.  The mapping
space of the pair of chaps~$C$ in Figure~\ref{fig:17} fits into the
correspondence diagram
  \begin{equation}\label{eq:36}
     \begin{gathered} \xymatrix{&\Map\bigl((C,\partial C_\rho ),(BG,*)
     \bigr)\ar[dl]\ar[dr]\\ \Omega BG\times \Omega BG&&\Omega BG}
     \end{gathered} 
  \end{equation}
that encodes restriction to the incoming and outgoing boundaries.  Here
$\partial C_\rho $~is the $\rho $-colored portion of~$\partial C$.  The left
arrow in~\eqref{eq:36} is a homotopy equivalence and the right arrow is
composition of loops.  Hence the quantization of~\eqref{eq:36} is the
convolution product of functions on~$G$.

 \item The computation in~\eqref{eq:36} generalizes to any pointed $\pi
$-finite space $(\sX,*)$ in place of $(BG,*)$.  (We encounter this when
composing codimension one $\rho $-defects in any dimension.)  Then the
correspondence is multiplication on the group~$\OX$, and the quantization is
pushforward under multiplication, i.e., a convolution product.  If the codomain
of~$\sigma $ has the form~$\Alg(\sC')$, then compute~$\sigma (\pt)$ as follows:
(1)~quantize~$\OX$ to an object in~$\sC'$, and (2)~induce the algebra structure
from pushforward under multiplication $\OX\times \OX\to \OX$.

 \item In Figure~\ref{fig:13}, if ${}\mstrut _{A}B\mstrut _{\hneg
A}=\!{}\mstrut _{A}A\mstrut _{\hneg A}$ is the identity $(A,A)$-bimodule and
$\xi \in A$ is the unit, then $T\subset \Hom_A(\sH,\sH)$ maps to a point defect
in the theory~$F$ which commutes with the $G$-symmetry.  (After erasing
transparent defects, in the sandwich picture we have a point defect supported
on the $\tF$-boundary.)  It therefore commutes with all $\sr$-defects, which
can be seen in the sandwich picture by moving defects up and down without
collision.  Similarly, an interior point defect in Figure~\ref{fig:16} commutes
with a point $\rho $-defect in Figure~\ref{fig:15}: the interior point and
boundary point move freely up and down without intersection.  This makes the
topological nature and symmetry properties of $\sr$-defects manifest.

 \item Even if we begin with a group symmetry, as in this example, there are
noninvertible topological $\sr$-defects.  Here elements of the group
algebra~$\GA$ label point defects on the $\rho $-colored boundary, and the
algebra~$\GA$ contains noninvertible elements.  Also, central defects are
generally noninvertible.  This fits general quantum theory, which produces
algebras rather than groups.

 \item $\sr$-defects give rise to structure in any $\sr$-module: linear
operators on vector spaces of point defects and on state spaces, endofunctors
on categories of line defects and categories of superselection sectors, etc.
These can be used to explore dynamics.

 \end{enumerate}
  \end{remark}

  \subsection{2-dimensional theories with finite symmetry}\label{subsec:4.3}

Let $G$~be a finite group and let $\sigma $~be finite pure 3-dimensional
$G$-gauge theory.  As a fully local field theory, $\sigma $~can take values
in~$\Alg(\Cat)$, a suitable\footnote{See~\cite[\S1.2]{FT2} for one possible
choice.} 3-category of tensor categories, in which case $\sigma (\pt)$~is the
fusion category $\sA=\Vect[G]$ introduced in~\S\ref{subsec:1.5}.  We can
construct~$\sigma $ as the finite homotopy theory~$\sXd3{BG}$ based on the
$\pi $-finite space~$BG$.  This is convenient for computations.  The right
regular boundary theory~$\rho $ is constructed using the right regular
module~$\sA_{\sA}$.  There are no background fields for~$\sigma $ or~$\rho $:
the quiche~$\sr$ is an unoriented theory.

  \begin{figure}[ht]
  \centering
  \includegraphics[scale=.25]{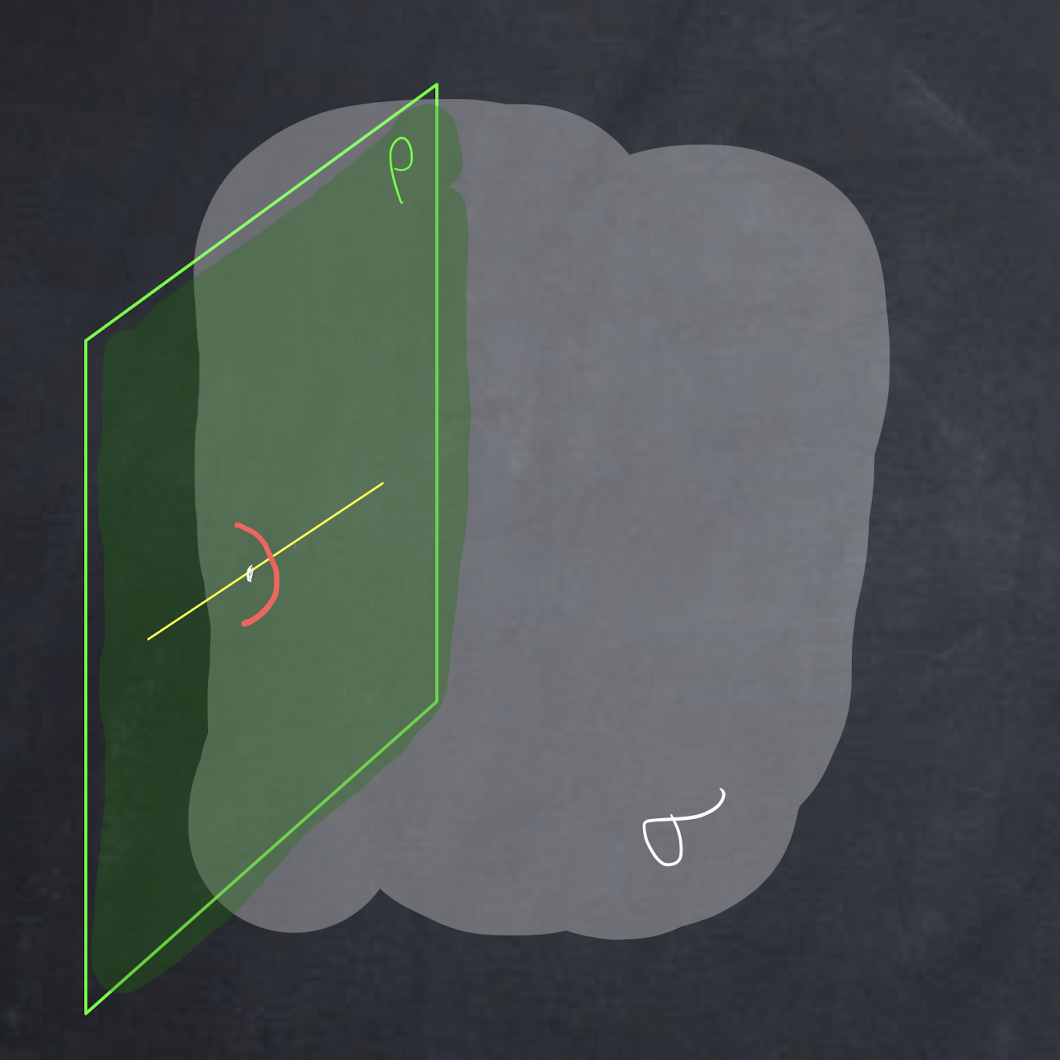}
  \vskip -.5pc
  \caption{A line defect supported on the $\rho $-colored boundary}\label{fig:19}
  \end{figure}

\begin{figure}[ht]
  \centering
  \includegraphics[scale=.25]{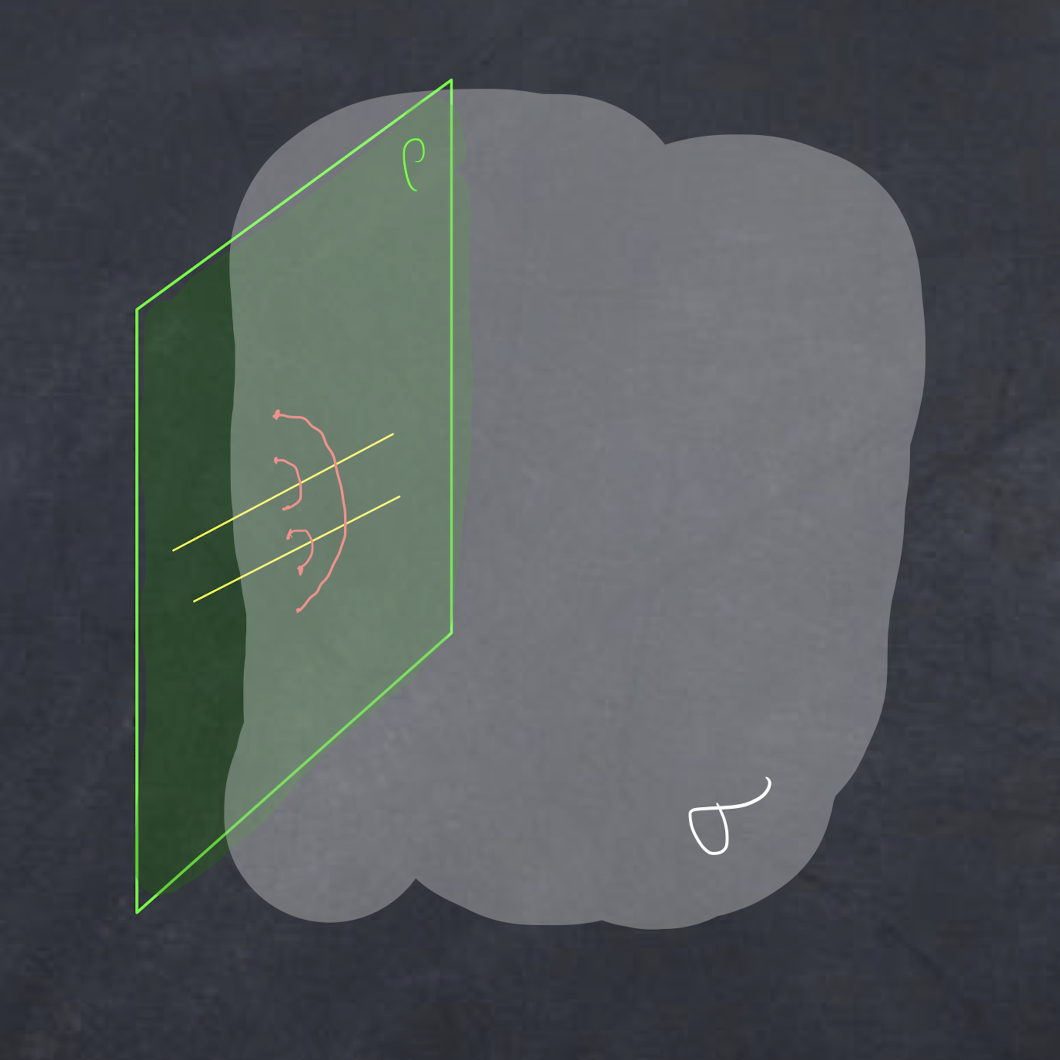}
  \vskip -.5pc
  \caption{Fusion of line defects}\label{fig:20}
  \end{figure}

The most familiar $\sr$-defects are the codimension~1 defects supported on
the $\rho $-colored boundary, as depicted in Figure~\ref{fig:19}.  The link
maps under~$\sr$ to the quantization of the mapping space~\eqref{eq:34}.  (It
is the same mapping space for the link of a codimension~1 defect in finite
gauge theory of \emph{any} dimension.)  That quantization is a linear
category, the category~$\Vect(G)$ of vector bundles over~$G$; it is the
linear category which underlies the fusion category~$\sA$.  The fusion
product---computed from the link in Figure~\ref{fig:20}, which is the same as
the link in Figure~\ref{fig:17}---is derived from the
correspondence~\eqref{eq:36} and is the fusion product of~$\sA$.  Each~$g\in
G$ gives rise to an \emph{invertible} defect, labeled by the vector bundle
over~$G$ whose fiber is~$\CC$ at~$g$ and is the zero vector space away
from~$g$. 

  \begin{remark}[$\RP^1\subset \RP^2$]\label{thm:41}
 As an illustration of how global defects may differ from local defects and
from classical labels, consider the theory~$\sr$ on $[0,1)\times \RP^2$ with a
defect supported on $\{0\}\times \RP^1$.  The category attached to the local
link is~$\Vect(G)$, as above, but globally the link twists---the normal bundle
to $\RP^1\subset \RP^2$ is the non-product real M\"obius line bundle---so the
local links quantize to a local system of categories over~$\RP^1$ with each
isomorphic to~$\Vect(G)$.  The twist is by the involution induced from
reflection on the linking interval.  Classically this reflection inverts the
parallel transport of a $G$-bundle trivialized at the endpoints of the link.
Denote inversion as $\iota \:G\to G$.  The induced involution $\iota
^*\:\Vect(G)\to \Vect(G)$ is pullback of vector bundles.  Defects supported
on~$\RP^1$ have a global label which is a section of this local system, or
equivalently a (homotopy) fixed point of the involution~$\iota ^*$, i.e., an
$\iota $-equivariant vector bundle over~$G$.  The invertible defects are trivial
lines supported at elements of order dividing two.
  \end{remark}

  \begin{figure}[ht]
  \centering
  \includegraphics[scale=.25]{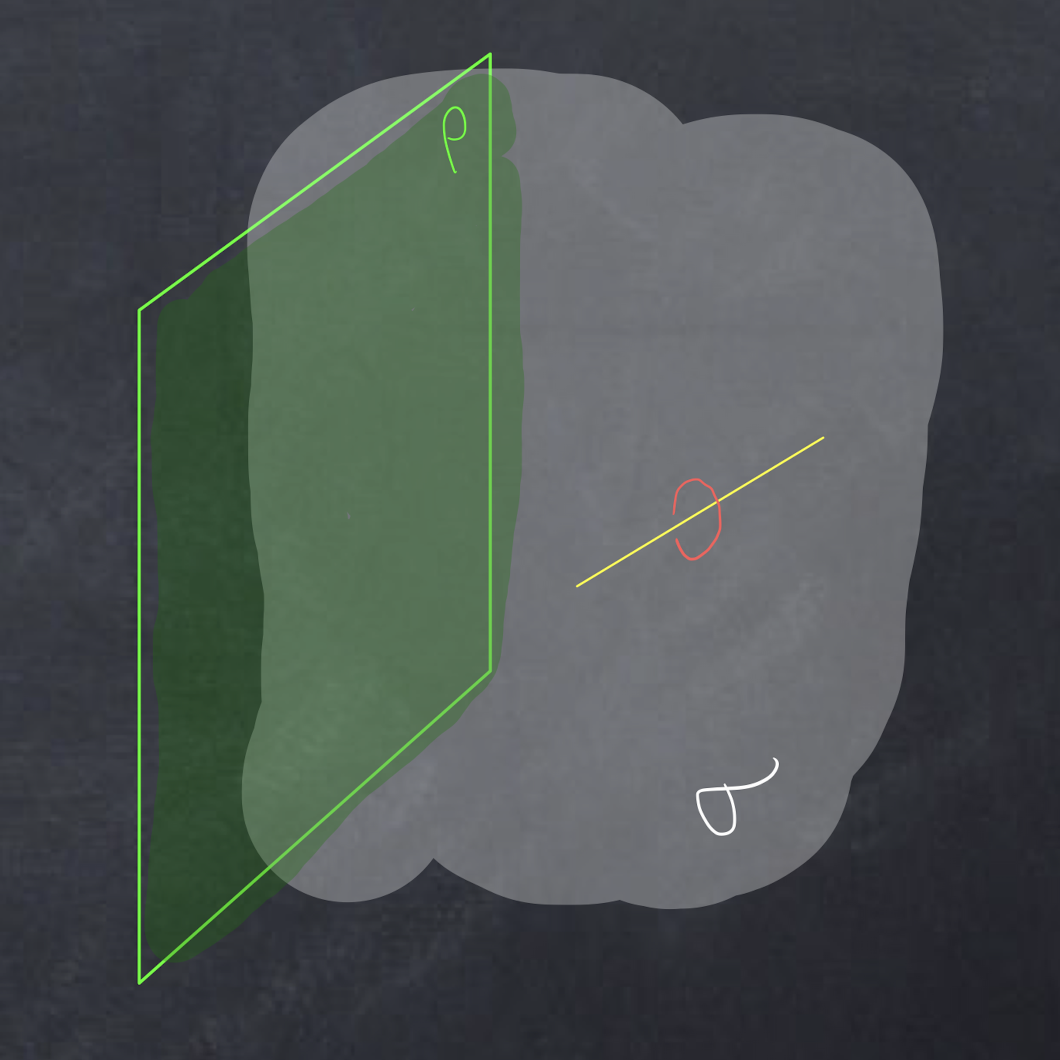}
  \vskip -.5pc
  \caption{A line defect supported in the bulk}\label{fig:21}
  \end{figure}

Now consider a line defect supported in the bulk, as in Figure~\ref{fig:21}.
The link is a circle, and so a local defect is an object in the category
$\sigma (\cir)=\Vect_G(G)$ of $G$-equivariant vector bundles over~$G$.  (Here
$G$~acts on itself via conjugation.)  This is the (Drinfeld) center of~$\sA$.
Note that unlike the case~$n=2$---see Remark~\ref{thm:36}(2)---the center
here is ``larger'' than the algebra~$\sA$.  The simple objects of the center
are labeled by a pair consisting of a conjugacy class and an irreducible
representation of the centralizer of an element in the conjugacy class.  The
corresponding defect is invertible iff the representation is 1-dimensional.
Among these defects are the Wilson and 't~Hooft lines of the 3-dimensional
$G$-gauge theory.  There is a rich set of topological defects that goes
beyond those labeled by group elements.

  \begin{remark}[]\label{thm:42}
 The reader can check that the only non-transparent point defects are scalar
multiples of the identity. 
  \end{remark}

  \begin{remark}[]\label{thm:43}
 A variation includes a twist of the pure $G$-gauge theory via a cocycle
representing a class in $H^3(BG;\Cx)$.  This is also a finite homotopy
theory, first studied by Dijkgraaf-Witten~\cite{DW}.  There is a regular
boundary theory~$\rho $, but there is not a fiber functor, as already mentioned
in~\S\ref{subsec:1.5}.   
  \end{remark}

  \begin{figure}[ht]
  \centering
  \includegraphics[scale=.35]{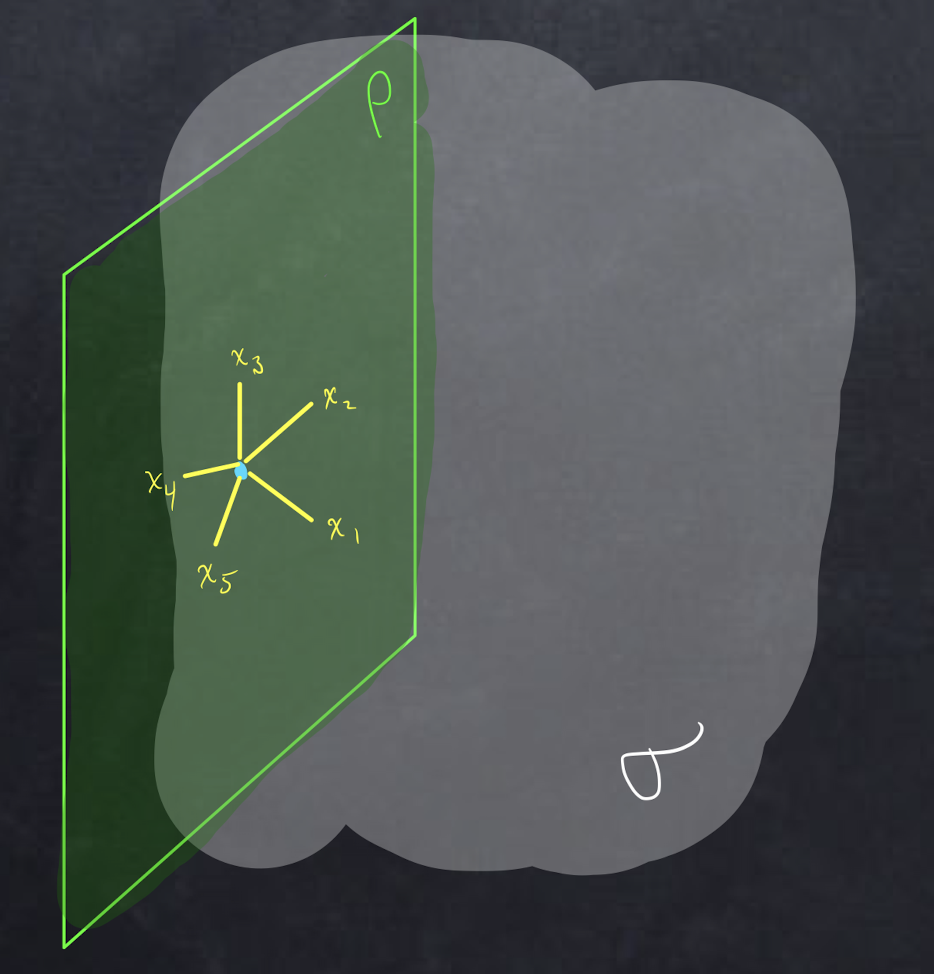}
  \vskip -.5pc
  \caption{A stratified $\rho $-defect}\label{fig:27}
  \end{figure}

The example of finite $G$-gauge theory generalizes to arbitrary Turaev-Viro
theories.  Let $\Phi $~be a spherical fusion category, let $\sigma $ be the
induced 3-dimensional topological field theory (of oriented bordisms) with
$\sigma (\pt)=\Phi $, and define the regular boundary theory~$\rho $ via the
right regular module~$\Phi _\Phi $.  The category of line $\rho $-defects is
the linear category which underlies~$\Phi $, so a defect is labeled (locally)
by an object of~$\Phi $.  The fusion product is the tensor product in~$\Phi
$.  We can also have nontrivial stratified $\rho $-defects, such as
illustrated in Figure~\ref{fig:27}.  In the figure the~$x_i$ are objects
of~$\Phi $ and the label at the central point is a vector in $\Hom_{\Phi
}(1,x_1\otimes \cdots\otimes x_5)$.

  \subsection{Higher group symmetries: composition of defects}\label{subsec:4.4}

 The two examples in this section demonstrate that in general there is no
sensible composition law on classical labels in finite homotopy theories.  (See
\autoref{thm:115} for a definition of \emph{classical labels}.)  We illustrate
that nonzero $k$-invariants manifest as higher multiplicative structures in the
categories obtained by quantization.  We will also emphasize both at the
semiclassical and quantum levels how automorphisms of local defects play a role
in their globalization by means of the topology of the normal bundle.  (See
\autoref{thm:41} for an example of this phenomenon.)

   \subsubsection{A 2-group example}\label{subsubsec:4.4.1}
 Let $G$~be a finite group, let $A$~be a finite abelian group, and fix a
cocycle~$k$ for a cohomology class $[k]\in H^3(G;A)$.  (We leave the reader
to include a nontrivial $G$-action on~$A$ in what follows.)  Realize~$k$ as a
map $k\:BG\to \BnA3$, and form the $\pi $-finite space~$\sX$ as a pullback:
  \begin{equation}\label{eq:38}
     \begin{gathered} \xymatrix@C+1.5pc{\BnA2\ar@{=}[r] \ar[d] & \BnA2\ar[d]\\
     \sX\ar@{-->}[r] \ar[d] & \ast\ar[d]\\ BG\ar[r]^{k} & \BnA3} \end{gathered} 
  \end{equation}
Then $\sX$~is the classifying space of a 2-group~$\Omega \sX$ which is an
extension of~$G$ by~$BA$.  Note that $BA$~is the \emph{sub} and $G$~is the
\emph{quotient}.  The extension class is $[k]\in H^3(G;A)\cong H^2(G;BA)$.  A
nonzero~$[k]$ means the extension is not split, in which case $\Omega \sX$~is
not the product 2-group.  We do not use the 2-group directly, but rather use
its classifying space~$\sX$.  Maps into~$\sX$ are ``background gauge fields''
for the 2-group symmetry.

Set~$n=2$ and let $\sigma =\sXd3{\sX}$ be the finite homotopy theory built
from~$\sX$.  It has a regular boundary theory~$\rho $ as the quantization of a
basepoint $*\to \sX$.  Let the codomain of~$\sigma $ be $\Alg(\Cat)$, a
3-category of tensor categories; the domain is the bordism category~$\Bord_3$
of unoriented manifolds.  The tensor category~$\sigma (\pt)$ is obtained by
quantizing~$\sX$.  One way to compute it,\footnote{See~\cite[\S8.1]{FHLT} for
an alternative approach.} based on Remark~\ref{thm:37}(2), is to quantize~$\OX$
to a linear category and induce the monoidal structure from multiplication
on~$\OX$.  First assume $G=1$, so that $\sX=\BnA2$ and $\OX$~is the homotopical
group~$BA$.  The category of (flat) vector bundles on $\OX=BA$ is~$\Rep(A)$,
the category of linear representations of~$A$.  The monoidal structure induced
by multiplication on~$BA$ is \emph{not} the usual tensor product of
representations.  Rather, identify $\Rep(A)\simeq \Vect(A\dual)$, where
$\Vect(A\dual)$ is the category of vector bundles on the Pontrjagin dual group.
The monoidal structure on~$\Vect(A\dual)$ is pointwise tensor product.  Denote
this tensor category as\footnote{$c$ for convolution} `$\RepcA$'.  Observe that
the tensor unit is the trivial line bundle on~$\Vect(A\dual)$ with fiber~$\CC$,
which corresponds to the regular representation of~$A$ in~$\RepcA$.  For
general finite~$G$, but zero $k$-invariant, the quantization is the group ring
of~$G$ with coefficients in $\RepcA$, which we denote $\RepcA[G]$.  The objects
of the underlying linear category~$\sL$ are vector bundles on~$G$ whose fibers
are representations of~$A$, or equivalently the fibers are vector bundles
over~$A\dual$.  A $k$-invariant $[k]\in H^3(G;A)\cong H^2(G;BA)$ can be
represented as a principal $A$-bundle $K\to G\times G$---compare
\cite[\S4.1]{FHLT}---together with further cocycle data/conditions.  Observe
that an $A$-torsor~$K_{g_1,g_2}$ produces a complex line bundle $L_{g_1,g_2}\to
A\dual$.  So there is a twisted convolution product: if $W_i\to G$,
$i=1,2$, are bundles over~$G$ whose fibers are vector bundles over~$A\dual$,
then define
  \begin{equation}\label{eq:46}
     (W_1*W_2)_g = \bigoplus\limits_{g_1g_2=g} L_{g_1,g_2}\otimes
     (W_1)_{g_1}\otimes (W_2)_{g_2},\qquad g\in G. 
  \end{equation}
This produces a tensor category $\scrT=\RepcA_k[G]$; it is the desired
quantization of~$\sX$.  The tensor unit~1 is the vector bundle over~$G$
supported at~$e\in G$ with fiber the regular representation of~$A$.  Notice
that $\Hom_{\scrT}(1,1)$ is the vector space underlying this representation,
the vector space~$\Fun(A)$ of functions $A\to \CC$, and it carries the
algebra structure of the group algebra.\footnote{So $\scrT$~is a fusion
category which does not have a simple unit, hence $\scrT$~ is sometimes
called a `multifusion category', as in~\cite{EGNO}.}

  \begin{figure}[ht]
  \centering
  \includegraphics[scale=.25]{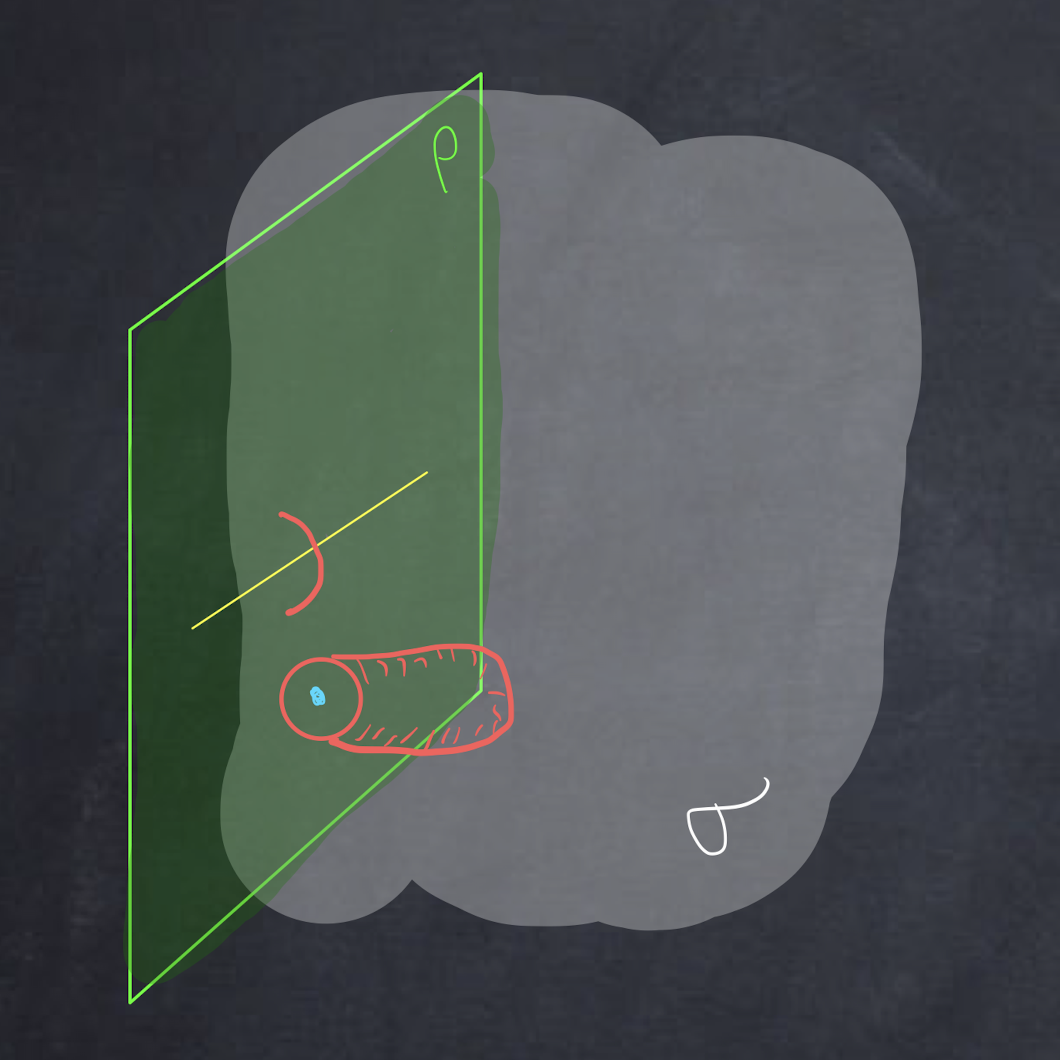}
  \vskip -.5pc
  \caption{A point defect and a line defect on the $\rho $-colored boundary}\label{fig:22}
  \end{figure}

Now consider point and line $\rho $-defects, as illustrated together with
their links in Figure~\ref{fig:22}.  The mapping spaces of these links
into~$\sX$ are---respectively for the point defect and line defect---the
(iterated) based loop spaces
  \begin{align}
      \Omega ^2\sX&\simeq A \label{eq:39}\\ 
      \Omega \sX&\simeq G\times BA \label{eq:40}
  \end{align}
(See~\eqref{eq:34} for more detail.)  The classical labels for invertible
defects are---respectively for the point defects and line defects---elements of
the abelian group~$A$ and of the group~$G$.  The quantizations are the vector
space $\Hom_{\scrT}(1,1)$ and the linear category~$\sL$, respectively:
  \begin{gather}
      \Fun(A) \label{eq:41}\\ 
      \sL=\Vect(G)\times \Rep(A) \label{eq:42}
  \end{gather}

  \begin{figure}[ht]
  \centering
  \includegraphics[scale=.4]{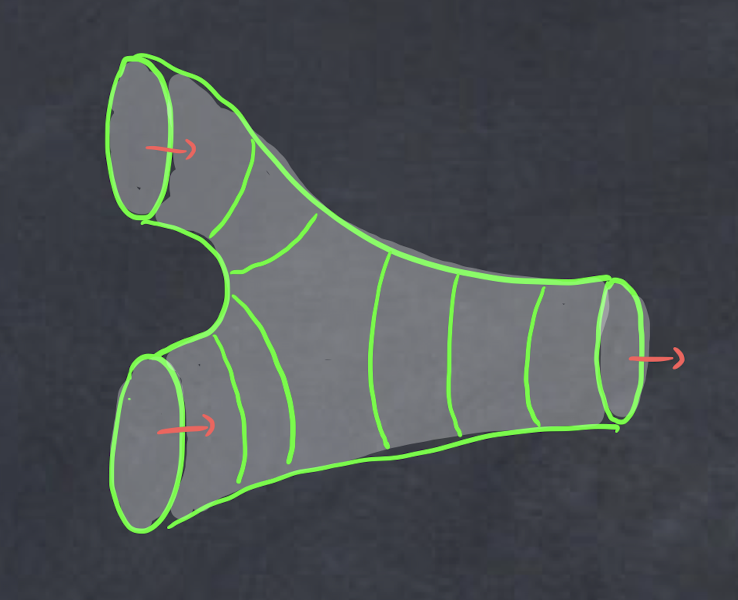}
  \vskip -.5pc
  \caption{Computation of the fusion rule for point defects}\label{fig:23}
  \end{figure}

The (quantum) composition law on point defects is computed by quantization of
a 3-dimensional pair of chaps~$S$, which can also be described as a
solid pair of pants: see Figure~\ref{fig:23}.  From the correspondence
  \begin{equation}\label{eq:43}
     \begin{gathered} \xymatrix{&\Map(S,\sX)\ar[dl]\ar[dr]\\ \Omega
     ^2\sX\times \Omega ^2\sX &&\Omega ^2\sX} \end{gathered} 
  \end{equation}
which is essentially the diagram for the group law on~$\pi _2\sX$, we deduce
the commutative algebra structure of convolution on $\Fun(A)$, which is then
the group algebra~$\CC[A]$.  Therefore, for point defects the composition law
for quantum defects, labeled by elements of the vector space~$\Fun(A)$,
specializes to multiplication on~$A$, which is the natural composition law on
classical labels.
 
  \begin{figure}[ht]
  \centering
  \includegraphics[scale=1.5]{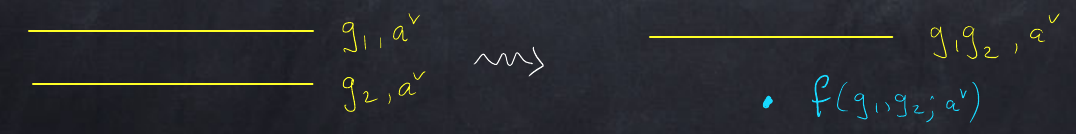}
  \vskip -.5pc
  \caption{A putative composition law on labels for line defects}\label{fig:24}
  \end{figure} 

Now consider line defects.  Recall that classical labels---path components
in~\eqref{eq:40}---are elements of~$G$ and their composition law is
multiplication in~$G$.  By contrast, the set of isomorphism classes of simple
objects in the category~$\sL$ in~\eqref{eq:42} is $G\times A\dual$.  Under the
tensor structure on~$\sL$, given in~\eqref{eq:46}, the product of two simples
is simple; the induced composition law on $G\times A\dual$ is the group law
of~$G$ on $G\times \{a\dual\}$ for all $a\dual\in A\dual$.

If the $k$-invariant vanishes, so the symmetry group splits as $G\times BA$,
then simple line defects compose as just described; see~\S\ref{subsec:4.3}.
The general composition law, which is based on \emph{quantum} labels, is
computed from the pair of chaps~$C$, as in Figure~\ref{fig:17}.  It leads to
the correspondence
  \begin{equation}\label{eq:44}
     \begin{gathered} \xymatrix{&\Map(C,\sX)\ar[dl]\ar[dr]\\ \Omega \sX\times
     \Omega \sX&&\Omega \sX} 
     \end{gathered} 
  \end{equation}
which induces a tensor structure on the quantization of $\Omega \sX\simeq
G\times BA$.  By Remark~\ref{thm:37}(2) this recovers~$\sigma (\pt)$, namely
the tensor category $\scrT=\RepcA_k[G]$ with its monoidal
structure~\eqref{eq:46}.  This is the correct composition law on line defects.
The important point is that the cocycle~$k$ is part of the tensor structure; if
$[k]\neq 0$ then whereas the \emph{space}~$\Omega \sX$ is a Cartesian product,
the \emph{group}~$\sX$ is not a direct product: it is a nonsplit 2-group.  A
choice of splitting $\Omega \sX\simeq G\times BA$ as a space does not lead to
the group law on~$G$, which is the quotient of the 2-group~$\Omega \sX$ by the
subgroup~$BA$.  (This is apparent in the model $K\to G\times G$ for the
$k$-invariant given before~\eqref{eq:46}.)

  \begin{remark}[]\label{thm:114}
 If $k\neq 0$, then one might be tempted to make an ansatz that the fusion of
two line defects is the union of a line defect and a point defect, as in
Figure~\ref{fig:24}, for some putative function $f(g_1,g_2;a\dual)$.  The
problem appears when fusing three line defects.  Suppose the labels are
$g_1,g_2,g_3\in G$ and the same $a\dual\in A\dual$.  Because of the associator
in the category~$\scrT$, the compositions $(\ell _{g_1}*\ell _{g_2})*\ell
_{g_3}$ and $\ell _{g_1}*(\ell _{g_2}*\ell _{g_3})$ differ by the contraction
$\langle a\dual,k \rangle$.  On the other hand, the ansatz implies that these
compositions differ by a point defect with label $\delta
f(g_1,g_2,g_3;a\dual)$, having viewed $f\in C^2(G;A)$.  However, if $[k]\in
H^3(G;A)$ is nonzero, then no such~$f$ exists.
  \end{remark}

  \begin{figure}[ht]
  \centering
  \includegraphics[scale=.3]{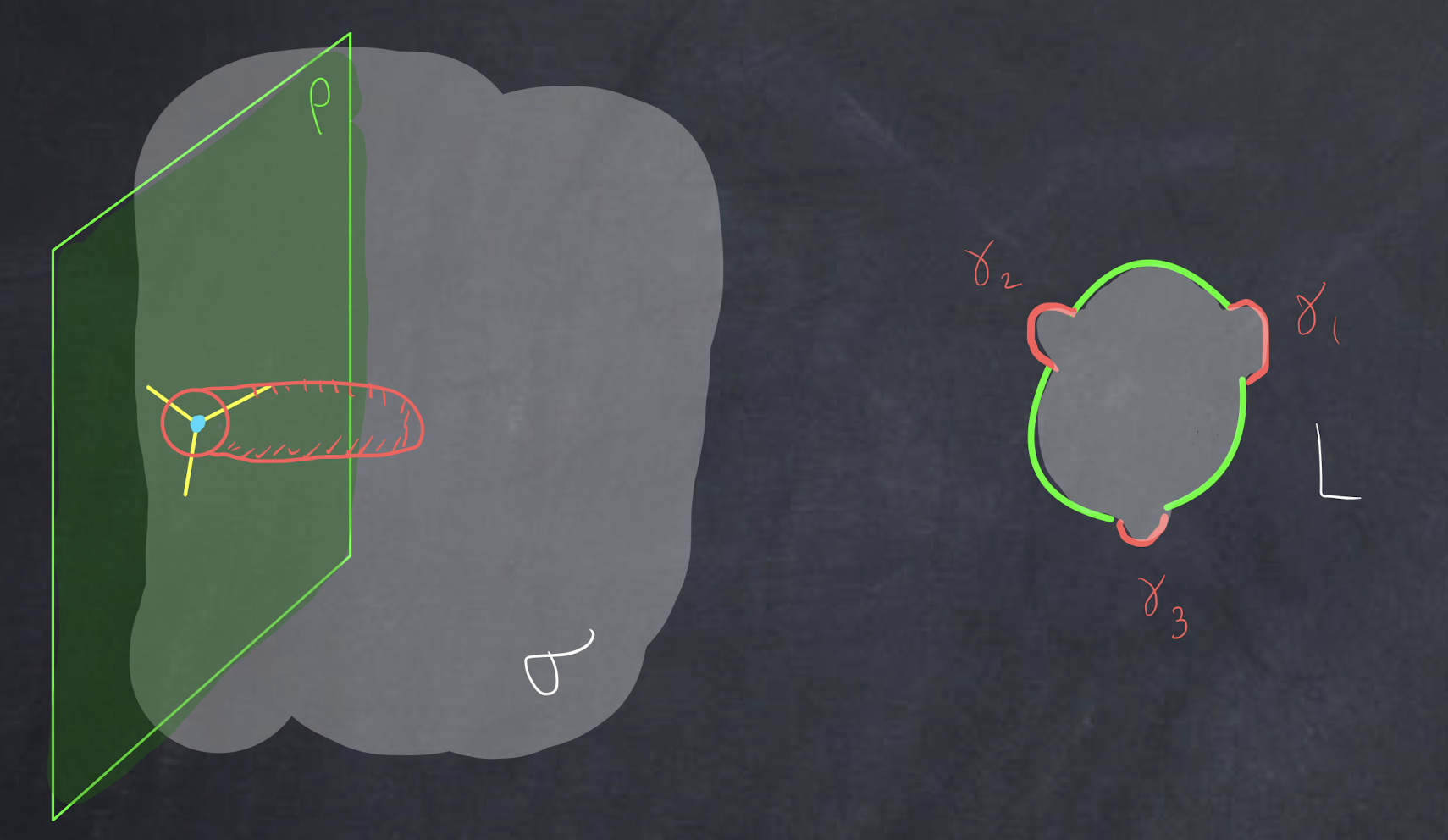}
  \vskip -.5pc
  \caption{A defect supported on a stratified submanifold and its link~$L$}\label{fig:25}
  \end{figure}

As a further example, consider a $\rho $-defect whose support is a graph, as
depicted in Figure~\ref{fig:25}.  This is a special case of the defect in
Figure~\ref{fig:27}.  The three line defects are labeled by an object in the
category~$\sL$.  We must supply a label for the point defect at the
intersection of the line defects.  Let $L$~be the link of the point.  Observe
that some portions of~$\partial L$ are $\rho $-colored.  By restriction we
obtain a map
  \begin{equation}\label{eq:45}
     \Map(L,\sX)\longrightarrow \OX\times \OX\times \OX 
  \end{equation}
The image consists of triples $(\gamma _1,\gamma _2,\gamma _3)$ of based
loops in~$\sX$ whose product~$\gamma _1\gamma _2\gamma _3$ is null homotopic;
the fiber over such a triple consists of null homotopies which, up to
homotopy, form an $A$-torsor, since $\pi _2\sX=A$.  Therefore, in the
quantization we expect that for fixed labels on the line defects, the
possible labels on the point defect form a $\Fun(A)$-module.  Indeed, if the
lines are labeled by objects $\ell _1,\ell _2,\ell _3\in \sL$, then $\ell
_1*\ell _2*\ell _3$ must be isomorphic to~$1$.  The label on the point defect
is a vector in $\Hom_{\sL}(1,\ell _1*\ell _2*\ell _3)$, and as expected this
vector space is a module over~$\Hom_{\sL}(1,1)=\Fun(A)$.

   \subsubsection{An example with higher homotopy groups}\label{subsubsec:4.4.2}
 Let $\sX$ be the $\pi $-finite space whose Postnikov tower is 
  \begin{equation}\label{eq:47}
     B^3\zt\longrightarrow \sX\longrightarrow B^2\zt 
  \end{equation}
with $k$-invariant $\Sq^2\:B^2\zt\longrightarrow B^4\zt$.  Since the Steenrod
square is a stable cohomology operation, $\sX$~is an infinite loop space.  We
grade so that it is the 2-space in a spectrum~$h$ which is an extension 
  \begin{equation}\label{eq:48}
     \Sigma H\zt\xrightarrow{\;\;i\;\;}h\xrightarrow{\;\;j\;\;} H\zt 
  \end{equation}
The spectrum~$h$ is similar to the spectrum~$e$ in \cite[Proposition~4.4]{F2}
with which it shares the following properties:

 \begin{enumerate}[label=\textnormal{(\arabic*)}]

 \item $h$~is a module over~$ko$, the connective real $K$-theory spectrum,

 \item $h$~is oriented for spin bundles, and

 \item for any space~$Z$ there is a natural identification\footnote{This can be
sharpened to an identification of Picard groupoids if we use the Koszul sign
rule for $\zt$-graded double covers.  (For the cohomology theory~$e$
in~\cite{F2}, the double covers are $\ZZ$-graded rather than $\zt$-graded.)}
of~$h^0(Z)$ with the group of isomorphism classes of $\zt$-graded double covers
of~$Z$.
 \end{enumerate}

\noindent 
 We will exploit these properties to facilitate some computations.

Fix a positive integer~$n$ and let $\sigma =\sXd{n+1}{\sX}$ be the indicated
finite homotopy theory.  We emphasize that $\sigma $~is an unoriented theory,
that is, there are no nontrivial background fields.  Below we use a spin
structure to derive some formulas for the quantum invariants, but the spin
structure is not necessary to define the theory.  Fix a basepoint of~$\sX$ to
construct a regular boundary theory~$\rho $.  Let $X$~be an $n$-manifold and
consider $\sigma $~on $[0,1)\times X$ with $\rho $-colored boundary at
$\{0\}\times X$.  We consider $\rho $-defects in this theory.

   \subsubsection*{Codimension two defects: semiclassical and quantum}\label{subsubsec:4.4.3}
 Let $Z\subset X$ be a codimension~2 submanifold with normal bundle $\pi \:\nu
\to Z$.   
 
\medskip\noindent
 {\emph 1}. \emph{Semiclassical local defects}.  The link of $\{0\}\times
Z\subset [0,1)\times X$ at~$p\in Z$ can be identified with the unit disk~$D(\nu
_p)$ in the fiber of the normal bundle.  As in~\S\ref{subsubsec:4.4.1}, the
mapping space of semiclassical local defects is the space of pointed maps
$D(\nu _p)\!\bigm/\!\partial D(\nu _p)\to \sX$.  A framing
$\RR^2\xrightarrow{\;\cong \;}\nu _p$ identifies this mapping space as
  \begin{equation}\label{eq:123}
     \Omega ^2\sX\simeq \zt\times B\zt. 
  \end{equation} 
 
\medskip\noindent
 {\emph 2}. \emph{Automorphisms}.  The space of oriented framings
$\RR^2\xrightarrow{\;\cong \;}\nu _p$ is homotopy equivalent to a circle.  Over
that circle we have two fiber bundles and an isomorphism between them.  One has
fiber the space of pointed maps $D(\nu _p)\!\bigm/\!\partial D(\nu _p)\to \sX$,
and the other has fiber the space of pointed maps $D(\RR^2)\!\bigm/\!\partial
D(\RR^2)\to \sX$.  The monodromy of the isomorphism between them is an
automorphism of the identity of~\eqref{eq:123}.  On the component labeled
by~$0\in \zt$ it is the identity automorphism.  On the component labeled
by~$1\in \zt$ it is the nontrivial automorphism of the identity functor
of~$B\zt$.

\medskip\noindent
 {\emph 3}. \emph{Semiclassical global defects}.  Globally over~$Z$, consider
the space $\Map(Z^\nu ,\sX)$ of pointed maps out of the Thom space~$Z^\nu $ of
the normal bundle.  Assume that the normal bundle admits a spin structure.
Homotopy classes of maps $Z^\nu \to \sX$ form the cohomology group~$h^2(Z^\nu
)$, and the existence of the spin structure on the normal bundle implies that
$h^2(Z^\nu )$~sits in an exact sequence (compare~\eqref{eq:108} below)
  \begin{equation}\label{eq:49}
     0\longrightarrow H^1(Z;\zt)\longrightarrow  h^2(Z^\nu )\longrightarrow
     H^0(Z;\zt) \longrightarrow 0,
 \end{equation}
where we have used the Thom isomorphism in cohomology with $\zt$~coefficients.
The composition law on defects is the standard abelian group structure
on~$h^2(Z^\nu )$.  

\medskip\noindent
 {\emph 4}. \emph{Splitting the sequence}.  Now choose a spin structure on the
normal bundle $\pi \:\nu \to Z$.  Then the Thom isomorphism for the cohomology
theory~$h$ identifies~$h^2(Z^\nu )$ as the abelian group~$h^0(Z)$, which is
isomorphic to a direct product of the quotient and sub in~\eqref{eq:49}.  Its
elements are pairs~$(g,\rho )$ consisting of a locally constant function
$g\:Z\to \zt$ and a double cover $\rho \:\tZ\to Z$.  Now shift the spin
structure on $\pi \:\nu \to Z$ by a double cover of~$Z$.  It follows from
essentially the same argument as in \cite[Proposition~4.4]{F2} that the shift
in the Thom isomorphism $h^2(Z^\nu )\xrightarrow{\;\cong \;}h^0(Z)$ is the
shearing
  \begin{equation}\label{eq:122}
      (g,\rho )\mapsto (g,\rho +g\delta ) 
  \end{equation}
on~$h^0(Z)$, where $\delta \in H^1(Z;\zt)$ classifies the difference of spin
structures.  
 
\medskip\noindent
 {\emph 5}. \emph{Dehn twists}.  This dependence on spin structures can be
manifested as follows.  Cut $Z$~along a codimension one hypersurface Poincar\'e
dual to $\delta \in H^1(Z;\zt)$ and reglue using a full twist on the link.
This gives a ``Dehn twist'' automorphism of~$Z^\nu $.  It shifts the spin
structure by~$\delta $, and so it acts on the space of global labels by the
shearing~\eqref{eq:122}.

\medskip\noindent
 {\emph 6}. \emph{The $w_2$ obstruction}.  Dropping the assumption that the
normal bundle~$\pi $ has a spin structure, \eqref{eq:49}~is replaced by the
exact sequence
  \begin{equation}\label{eq:108}
     0\longrightarrow H^1(Z;\zt)\longrightarrow h^2(Z^\nu )\longrightarrow
     H^0(Z;\zt)\xrightarrow{\;\;w_2(\nu )\;\;} H^2(Z;\zt) .
  \end{equation}
Hence the group~$h^2(Z^\nu )$ of isomorphism classes of defects does not
allow a nonzero $H^0$~``part'' on components of~$Z$ over which the normal
bundle is not spinable.   
 
The link of a codimension~2 defect is a circle, and the value of any
topological field theory on a circle has an $E_2$-structure.  Semiclassically,
that $E_2$-structure on the double loop space~$\Omega ^2\sX$ in~\eqref{eq:123}
recovers the space~$\sX$.  

This concludes the semiclassical discussion.

\medskip\noindent
 {\emph 7}. \emph{Local quantum defects}.  For definiteness take~$n=3$.  Then
the quantization of~\eqref{eq:123} is the linear category of $\zt$-graded
representations of~$\zt$.  As an $E_2$-category, i.e., a braided tensor
category, this is $\Vect\langle \zt \rangle\oplus s\!\Vect$, the sum of the
braided tensor categories of $\zt$-graded vector spaces and super vector
spaces.  This comes about as the Karoubi completion of the span of two objects
($\pi _2\sX$), each with endomorphism algebra~$\CC[\pi_3\sX]=\CC[\zt]$.  This
is the linear category of local topological line $\rho $-defects.

 As tensor categories, $\Vect\langle \zt  \rangle$ is equivalent
to~$s\!\Vect$.  However, the braiding is different due to the Koszul sign rule
in the latter.  It is induced from the $k$-invariant between $\pi _2\sX$
and~$\pi_3\sX$ and is the action of the nonidentity element in the latter
group.  Said differently, the identification in~\eqref{eq:123} preserves the
group structure but not the commutativity at the $E_2$~level.  This manifests
in the braiding of global line defects in the 3-manifold~$X$.

\medskip\noindent
 {\emph 8}. \emph{Global quantum defects}.  The global quantum defects
supported on~$Z\subset X$ form the vector space of functions on~$h^2(Z^\nu )$;
see point~(3) above.  The integration of local quantum defects to global
quantum defects assigns a vector in this vector space to each object in the
category $\Vect\langle \zt \rangle\oplus s\!\Vect$.  This map depends on a
trivialization of $\nu \to Z$; a change of trivialization acts by the
shearing~\eqref{eq:122}. 

If~$Z'$ is a parallel copy of~$Z$, then we can identify the vector spaces of
global defects on~$Z$ and~$Z'$ and define a composition law; see
\autoref{thm:108}.  In this case the result is the group algebra~$\CC[h^2(Z^\nu
)]$.

\medskip
   \subsubsection*{Codimension one defects}\label{subsubsec:4.4.4}
 We now undertake an analogous study of codimension one defects.

\medskip\noindent
 {\emph 1}. \emph{Semiclassical global defects}.  Consider a codimension~1
submanifold~$W\subset X$.  Then the analog of~\eqref{eq:108} is the exact
sequence 
  \begin{equation}\label{eq:121}
     0\longrightarrow H^2(W;\zt)\longrightarrow h^2(W^\nu )\longrightarrow
     H^1(W;\zt) \xrightarrow{\;\;w_1(\nu )\,\smile\, \Sq^1\;\;} H^3(W;\zt)
  \end{equation}
Assume the normal bundle $\nu \to W$ to~$W\subset X$ is trivialized.  As
before, a choice of spin structure on $\nu \to W$ induces a Thom isomorphism
on~$h^{\bullet }$ and a splitting
  \begin{equation}\label{eq:109}
     h^2(W^\nu )\cong h^1(W)\cong H^1(W;\zt) \oplus H^2(W;\zt). 
  \end{equation}
Different spin structures shear the splitting, as in~\eqref{eq:122}.  However,
\eqref{eq:109}~ is only an isomorphism of \emph{sets}, not of \emph{abelian
groups}.  The nonzero $k$-invariant $\Sq^2$ in~\eqref{eq:47} implies that the
abelian group law on~$h^2(W^\nu )$ transports to the group law
  \begin{equation}\label{eq:110}
     (a_1,b_1) + (a_2,b_2) = (a_1+a_2,b_1+b_2+a_1\smile a_2) 
  \end{equation}
on $H^1(W;\zt) \oplus H^2(W;\zt)$.

\medskip\noindent
 {\emph 2}. \emph{Composition law}.  One consequence of the group
law~\eqref{eq:110} is that the square of a defect whose class in~\eqref{eq:109}
is~$(a,0)$ has equivalence class~$(0,a\smile a)$.  In particular, such a defect
need not be of order two.

When the normal bundle $\nu \to W$ is not trivial we can still have a self
composition law of double covering defects of~$W$.  Let $\tW$~be the boundary
of a tubular neighborhood of~$W$, and denote by $p\:\tW\to W$ the double cover.
Write $\delta = w_1(\nu ) \in H^1(W;\zt)$.  Consider the following process:
begin with a defect supported on~$W$, pull back to a defect supported on~$\tW$,
and then compose the defects on the two sheets to obtain a defect supported
on~$W$.  On isomorphism classes of defects this process induces the map
  \begin{equation}\label{eq:124}
     p_*\circ  p^*\:h^2(W^\nu )\longrightarrow h^2(W^\nu ) 
  \end{equation}
The map $p_*\circ p^*$ on ordinary cohomology is multiplication by the order of
the cover, which is~2, and so it vanishes on mod~2 cohomology.  Apply this
twice to the short exact sequences~\eqref{eq:121} for~$W$ and~$\tW$ to
conclude: (1)~the value of~\eqref{eq:124} on a class in~$h^2(W^\nu )$ only
depends on its image in the quotient $H^1(W;\zt)$, and (2)~the result lies in
the subgroup $H^2(W;\zt)$.  We claim that \eqref{eq:124}~is the restriction of
the quadratic map
  \begin{equation}\label{eq:125}
     \begin{aligned} H^1(W;\zt)&\longrightarrow H^2(W;\zt) \\ a\qquad &\longmapsto
      \quad\; a^2 + \delta a\end{aligned} 
  \end{equation}
to the subgroup of~$H^1(W;\zt)$ cut out by the last map in~\eqref{eq:121}.
Namely, \eqref{eq:124}~is a natural quadratic map for all spaces~$W$ equipped
with a double cover, hence is a linear combination of~$a^2,\delta a,\delta ^2$.
Furthermore, it vanishes when~$a=0$ or~$a=\delta $, and it reduces to~$a^2$
when~$\delta =0$ by the previous paragraph.  It follows that~\eqref{eq:125} is
the only possibility.
 
\medskip\noindent
 {\emph 3}. \emph{Local quantum defects}.  Specialize to~$n=3$.  For local
defects, we quantize~$\Omega \sX$ to a tensor category.  Since $\Omega \sX$~is
a based loop space the tensor category has an additional $E_1$-structure.
Proceeding as in point~ (7) above for codimension two defects, we first
quantize $\Omega ^2\sX$ as a linear category, use one loop to derive the tensor
structure, and use the second loop to derive the $E_1$-structure.  The result is
a braided tensor category, namely the same braided tensor category
$\Vect\langle \zt \rangle\oplus s\!\Vect$ as in point~ (7) above.
 
\medskip\noindent
 {\emph 4}. \emph{Global quantum defects}.  On~$W\subset X$, the quantization
of the semiclassical global defects of point~(1) above is the vector space of
functions on the group~$h^2(W^\nu )$ in~\eqref{eq:121}.  Integration of an
unobstructed object in the category $\Vect\langle \zt \rangle\oplus s\!\Vect$
gives a vector in this vector space, but the result depends on a choice of spin
structure on the normal bundle.

If the normal bundle $\nu \to W$ is trivialized, then the composition of
defects is encoded in an algebra structure on this vector space, namely the
group algebra of the group law~\eqref{eq:110}.

   \section{Quotient and duality defects}\label{sec:6}

In this section we take up two types of defects which have been discussed in
the literature recently: \emph{quotient}\footnote{The word `condensation' is
sometimes used in place of `quotient', but we refrain from doing so.}
\emph{defects} and \emph{duality defects}.  Recall from~\S\ref{subsec:3.4} that
the quotient~$F\bs$ of a field theory~$F$ by a symmetry~$\sigma $ is defined
using an augmentation.  Now, in~\S\ref{subsec:4.5}, we use an augmentation to
define a defect on a positive codimensional submanifold which, in effect, takes
the quotient on that submanifold.  Returning to the quotient theory, there are
special situations in which there exists an isomorphism $F\bs\to F$.
In~\S\ref{subsec:4.6} we use such an isomorphism to define a self-domain wall
of~$F$ called a \emph{duality defect}, and we give some applications.

  \subsection{Quotient defects: quotienting on a submanifold}\label{subsec:4.5}

Fix a positive integer~$n$ and an $n$-dimensional quiche~$\sr$.  Suppose
$\epsilon $~is an augmentation of~$\sigma $, as in Definition~\ref{thm:21}.  As
explained in Definition~\ref{thm:25}, if $\tFt$ is a $\sr$-module structure on
an $n$-dimensional quantum field theory~$F$, then dimensional reduction
of~$\sigma $ along the closed interval depicted in Figure~\ref{fig:9}, which is
the sandwich $\epsilon \otimes _\sigma \tF$, is the quotient~$F\bs$ of~$F$ by
the symmetry.  This can be interpreted as the theory~$F$ with the topological
space-filling defect~$\epsilon $.
 
There is a generalization which places the defect on a submanifold;
see~\cite{RSS} and the references therein.  For this, recall the
Dirichlet-to-Neumann and Neumann-to-Dirichlet domain walls~$\delta ,\delta ^*$
introduced in Definition~\ref{thm:86}.  Suppose $M$~is a bordism on which we
evaluate~$F$, and suppose $Z\subset M$ is a submanifold of codimension~$\ell $
on which we place the defect.  (We do not make background fields explicit here;
see~\S\ref{subsec:2.5}.)  Form the sandwich $\zo\times M$ with $\{0\}\times M$
colored with~$\rho $ and $\{1\}\times M$ colored with~$\tF$.  Let $\nu \subset
M$ be an open tubular neighborhood of $Z\subset M$ with projection $\pi \:\nu
\to Z$, and arrange that the closure $\bn$ of~$\nu $ is the total space of a
disk bundle $\bn\to Z$.

  \begin{figure}[ht]
  \centering
  \includegraphics[scale=.4]{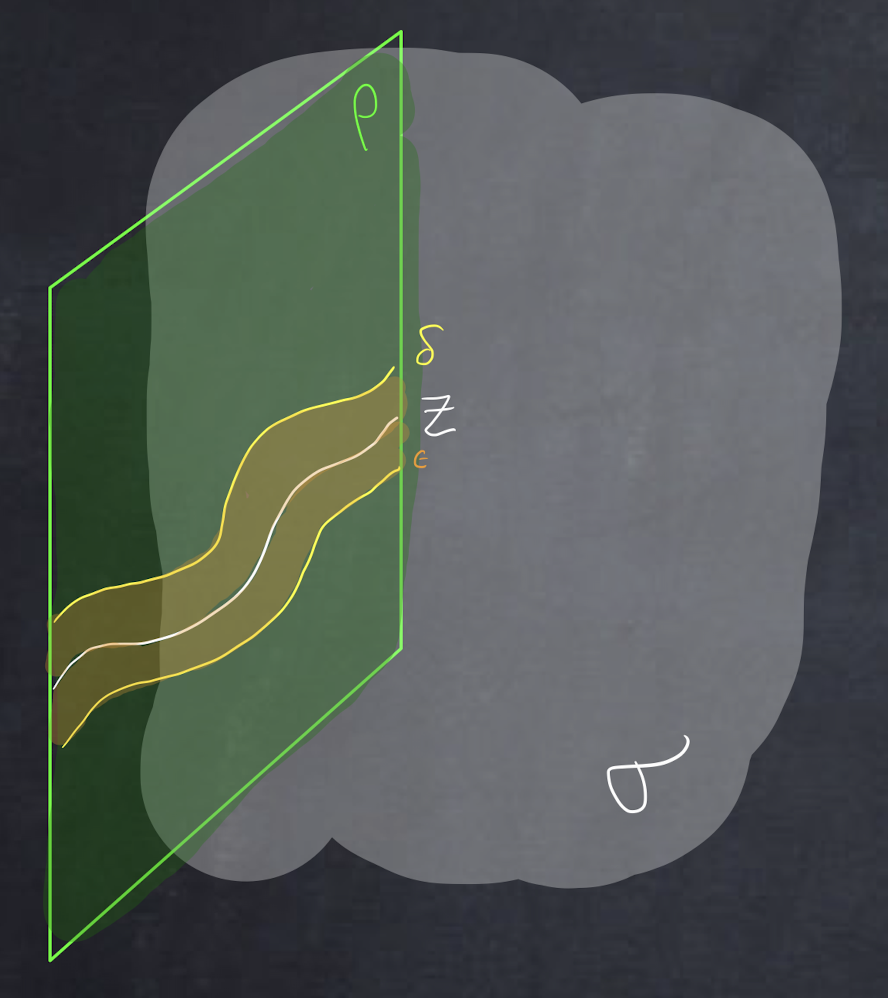}
  \vskip -.5pc
  \caption{The quotient defect~$\epsilon (Z)$}\label{fig:28}
  \end{figure}

  \begin{definition}[]\label{thm:45}
 The \emph{quotient defect}~$\epsilon (Z)$ is the $\rho $-defect supported on
$\{0\}\times \bn$ with $\{0\}\times \nu $ colored with~$\epsilon $ and
$\{0\}\times \partial \bn$ colored with~$\delta $.
  \end{definition}

\noindent
 This defect is depicted in Figure~\ref{fig:28}.  The label~$\delta $ is for
the domain wall from~$\rho $ to~$\epsilon $; if we read in the other
direction from~$\epsilon $ to~$\rho $, then the label is~$\delta ^*$.

  \begin{figure}[ht]
  \centering
  \includegraphics[scale=1.6]{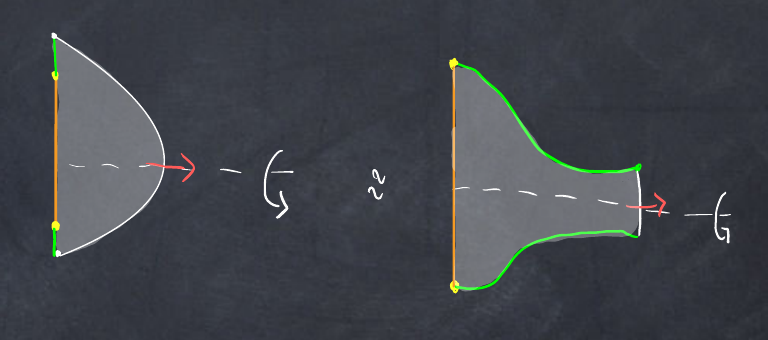}
  \vskip -.5pc
  \caption{The local label of~$\epsilon (Z)$ in codimension~1}\label{fig:29}
  \end{figure} 
 
Next, we compute the local label of the quotient defect~$\epsilon (Z)$, as in
Definition~\ref{thm:7}(1), and so express~$\epsilon (Z)$ as a defect supported
on~$Z$.  Consider a somewhat larger tubular neighborhood, now of $\{0\}\times
Z\subset \zoh\times M$.  Recall $\ell =\codim\mstrut _{\hneg M}\!Z$.  The
tubular neighborhood for~$\ell =1$ is depicted in Figure~\ref{fig:29}.  Its
value in the topological theory~$\sigma $---with boundaries and defects $\rho
,\epsilon ,\delta $---is an object in $\Hom\bigl(1,\sigma (D^1,S^0_\delta )
\bigr)$.  (If $\sC=\Alg(\sC')$ is the codomain of~$\sigma$, and $\sigma
(\pt)=A$ is an algebra object in~$\sC'$, then $\sigma (D^1,S^0_\delta )=A$ as
an object of~$\sC'$.)

  \begin{figure}[ht]
  \centering
  \includegraphics[scale=.5]{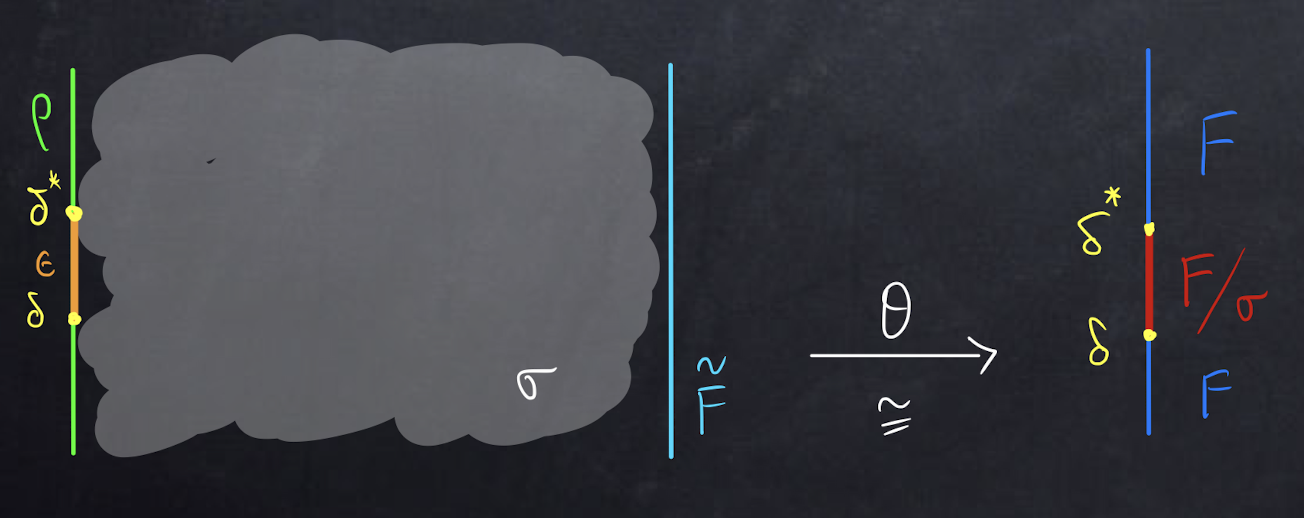}
  \vskip -.5pc
  \caption{The codimension~1 quotient defect as a composition of domain walls}\label{fig:30}
  \end{figure}

  \begin{remark}[]\label{thm:47}
 The defect~$\epsilon (Z)$ for~$\ell =1$ can be interpreted as follows,
assuming $Z\subset M$ has trivialized normal bundle.  Let $Z_1,Z_2$ be parallel
normal translates of~$Z$, color the region in between $\{0\}\times Z_1$ and
$\{0\}\times Z_2$ with~$\epsilon $, color the remainder of~$\{0\}\times M$
with~$\rho $, and use the domain wall~$\delta $ at $\{0\}\times Z_1$ and
$\{0\}\times Z_2$; see Figure~\ref{fig:30}.  Then $\epsilon (Z)$~is the
composition $\delta^* (Z_2)*\delta (Z_1)$.  If a quantum field theory~$F$ has a
$\sr$-module structure, then $\delta (Z_1)$~is a domain wall from~$F$ to~$F\bs$
and $\delta ^*(Z_2)$ is a domain wall from~$F\bs$ to~$F$; the
composition~$\epsilon (Z)$ is a self domain wall of~$F$.
  \end{remark}

  \begin{figure}[ht]
  \centering
  \includegraphics[scale=1.4]{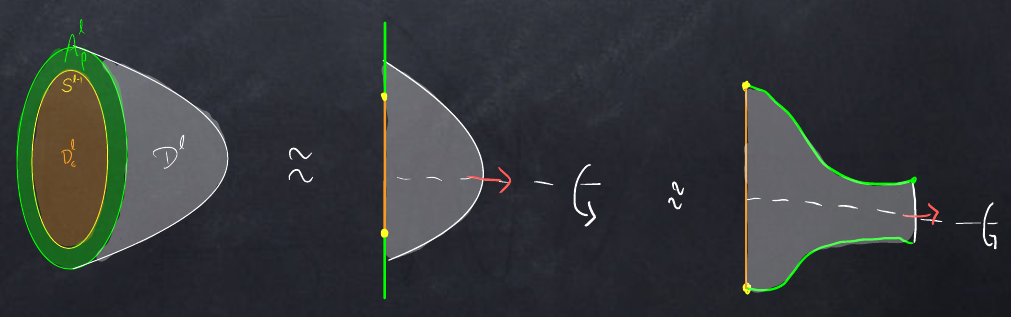}
  \vskip -.5pc
  \caption{The local label of~$\epsilon (Z)$ in codimension~2}\label{fig:31}
  \end{figure}

The tubular neighborhood of~$\{0\}\times Z\subset \zoh\times M$ for
codimension~$\ell =2$ is the 3-dimensional bordism obtained from
\autoref{fig:29} by revolution in 3-space, as illustrated in
Figure~\ref{fig:31}.  For general~$\ell >1$, the bordism is the $(\ell
+1)$-disk~$D^{\ell +1}$ with boundary~$S^\ell $ partitioned as
  \begin{equation}\label{eq:52}
     \partial D^{\ell +1} = D^\ell _\epsilon \cup A^\ell _\rho \cup D^\ell
     _{\relax} 
 \end{equation}
into a pair of disks~$D^\ell $ and an annulus~$A^\ell $; the domain
wall~$\delta $ is placed at the intersection of the~$\epsilon $ and~$\rho
$-colored regions.

  \begin{remark}[]\label{thm:48}
These are the local defects.  As always, the global defects are a
section of a bundle (local system) of local defects over the
submanifold~$Z\subset M$.  
  \end{remark}

  \begin{example}[finite homotopy theory: local label]\label{thm:49}
 Let $\sigma =\sXd{n+1}\sX$ be the finite homotopy theory built from a $\pi
$-finite space~$\sX$.  Then we can use the calculus described in
Appendix~\ref{sec:5} to compute semiclassical spaces of defects.   Suppose $\rho
$~is specified by a basepoint $*\to \sX$ and $\epsilon $~is specified by the
identity map $\sX\xrightarrow{\;\id\;}\sX$.  Then $\delta $, which is a domain
wall between boundary theories (\autoref{thm:109}), is specified by the
homotopy fiber product
  \begin{equation}\label{eq:53}
     \begin{gathered} \xymatrix{&\ast\ar@{-->}[dl]\ar@{-->}[dr] \\
     \ast\ar[dr]&&\sX\ar[dl]^>>>>>>>{\id}\\ &\sX} \end{gathered} 
  \end{equation}
That the homotopy fiber product is a point~$*$ is the manifestation of the
uniqueness of~$\delta $.
 
The space of maps from the link of a point of~$Z$ into~$\sX$ is
  \begin{equation}\label{eq:54}
     \Map\bigl((D^\ell ,S^{\ell -1}),(\sX,*) \bigr)=\Omega ^\ell \sX. 
  \end{equation}
Set 
  \begin{equation}\label{eq:55}
     N^\ell =\bigl(D^{\ell +1},A^\ell  \bigr), 
  \end{equation}
where $A^\ell \subset \partial D^{\ell +1}$; see~\eqref{eq:52} and
Figure~\ref{fig:31}.  The semiclassical local defect~$\epsilon (Z)$, in the
sense of \autoref{thm:68}, is the map induced by restriction to~$\OlX$:
  \begin{equation}\label{eq:56}
     \Map(N^\ell ,\sX)\longrightarrow \OlX .
  \end{equation}
Here the basepoint in~$\sX$ is implicit.  This map is a homotopy equivalence,
as can be proved using the technique in~\cite[Example~0.8]{H}.  So we can
replace~\eqref{eq:56} with the identity map on~$\OlX$.

Now the map $\id\:\OlX\to \OlX$, viewed as a correspondence from a point~$*$
to~$\OlX$, quantizes to an object in the quantization of~$\OlX$, and it is
typically noninvertible.  For example, if we are at the level in which the
quantization of~$\OlX$ is a vector space, then the vector space is\footnote{The
homotopy group $\pi _{\ell }\sX=\pi _\ell (\sX,*)$ uses the basepoint~$*\in
\sX$.} $\Fun(\pi _0\OlX)=\Fun(\pi _\ell \sX)$, which typically has
dimension~$>1$.  The local label we compute is the constant function~1.  If the
quantization is a linear category, then it is the category $\Vect(\OlX)$ of
flat vector bundles over~$\OlX$, i.e., vector bundles on the fundamental
groupoid $\pi _{\le1}\OlX$, and the local label is the trivial bundle with
fiber~$\CC$.
  \end{example}

  \begin{example}[finite homotopy theory: global label]\label{thm:94}
 We continue with \autoref{thm:49}, but now compute the global label of the
defect~$\epsilon (Z)$.  As in~\S\ref{subsubsec:4.4.2}, we must quantize 
  \begin{equation}\label{eq:59}
     \id\:\Map(Z^\nu ,\sX)\longrightarrow \Map(Z^\nu ,\sX), 
  \end{equation}
where $Z^\nu $~is the Thom space of the normal bundle.  As an example,
suppose~$\ell =1$ and assume that the normal bundle $\nu \to Z$ has been
trivialized.  (This amounts to a coorientation of the codimension~1
submanifold~$Z\subset M$---a direction for the domain wall.)  Then 
  \begin{equation}\label{eq:60}
     \Map(Z^\nu ,\sX)\simeq \Map(Z,\Omega \sX). 
  \end{equation}
For example, if $A$~is a finite abelian group and $\sX=\BtA$---so $\sigma
$~encodes a $BA$-symmetry---then $\Map(Z^\nu ,\BtA)\simeq \Map(Z,BA)$ is the
``space'' of principal $A$-bundles $P\to Z$.  One should, rather, treat it as a
groupoid, the groupoid $\Bun_A(Z)$ of principal $A$-bundles over~$Z$ and
isomorphisms between them.  A point $*\to \Bun_A(Z)$ is a principal $A$-bundle
$P\to Z$, and this map quantizes to a global defect~$\eta (P)$ supported
on~$Z$.  The quantization of $\id\:\Bun_A(Z)\to \Bun_A(Z)$ is a sum of the
quantizations of $*\!\!\bigm/\!\!\Aut P\to \Bun_A(Z)$ over isomorphism classes
of principal $A$-bundles $P\to Z$.  Informally, we might write this as a sum
over isomorphism classes of~$P\to Z$ of
  \begin{equation}\label{eq:61}
     \frac{1}{\#\Aut P}\,\eta (P) = \frac{1}{\#H^0(Z;A)}\,\eta (P). 
  \end{equation}
This sort of expression appears in~\cite[(1.3)]{CCHLS}, for example. 
 
The $\rho $-defect~$\eta (P)$ has a geometric semiclassical interpretation.
Without the defect one is summing over $A$-gerbes on $\zoh\times M$ which are
trivialized on~$\{0\}\times M$.  Putting the defect~$\eta (P)$ on~$\{0\}\times
Z$ amounts to the instruction to trivialize the $A$-gerbe only on
$\bigl(\{0\}\times M\bigr)\;\setminus \;\bigl(\{0\}\times Z\bigr)$ and to
demand---relative to the coorientation of~$Z$---that the trivialization jump by
the $A$-bundle $P\to Z$.  (Compare Remark~\ref{thm:36}(1).)
  \end{example}

  \begin{remark}[]\label{thm:51}
 \ 
 \begin{enumerate}[label=\textnormal{(\arabic*)}]

 \item As stated in Remark~\ref{thm:47}, the defect~$\epsilon (Z)$ is the
composition of the Dirichlet$\to $Neumann and Neumann$\to $Dirichlet domain
walls in case~$\ell =1$.  In terms of $\pi $-finite spaces, that computation
is the homotopy fiber product 
  \begin{equation}\label{eq:62}
     \begin{gathered} \xymatrix@R-.5pc@C-.5pc{&&\OX\ar@{-->}[dl]\ar@{-->}[dr]\\
     &\ast\ar[dl]\ar[dr]&&\ast\ar[dl]\ar[dr]\\
     \ast\ar[ddrr]&&\sX\ar[dd]&&\ast\ar[ddll]\\ \\ &&\sX} \end{gathered} 
  \end{equation}
which is then the domain wall 
  \begin{equation}\label{eq:p14}
     \begin{gathered} \xymatrix{&\OX\ar[dl]\ar[dr] \\
     \ast\ar[dr]&&\ast\ar[dl] \\&\sX }
     \end{gathered}
  \end{equation}
This recovers the description of~$\epsilon (Z)$ in~\eqref{eq:56}.

 \item If the $\pi $-finite space~$\sX$ is equipped with a cocycle~$\lambda $
which represents a cohomology class $[\lambda ]\in h^n(\sX)$ for some
cohomology theory~$h$, then a codimension~$\ell $ quotient defect has
semiclassical label space~$\OlX$ with transgressed cocycle and its cohomology
class $[\tau ^\ell \lambda ]\in h^{n-\ell }(\OlX)$.  A nonzero cohomology
class obstructs the quotient.  However, as observed in~\cite{RSS} it is
possible that $[\lambda ]\neq 0$ but $[\tau ^\ell \lambda ]=0$ for some~$\ell
$, which means that the quotient by~$\sigma $ does not exist but quotient
defects of sufficiently high codimension do exist.
 \end{enumerate}
  \end{remark}

  \begin{example}[Turaev-Viro symmetry]\label{thm:52}
 Suppose~$n=2$ and the 3-dimensional theory~$\sigma $ is of Turaev-Viro type
with $\sigma (\pt)=\Phi $ a fusion category.  Assume $\rho $~is given by the
right regular module~$\Phi _\Phi $ and $\epsilon $~is given by a fiber
functor $\eP\:\Phi \to \Vect$.  Then the codimension~1
quotient defect has local label the object $\xreg\in \Phi $ defined as 
  \begin{equation}\label{eq:63}
     \xreg=\sum\limits_{x}\eP (x)^*\otimes x, 
  \end{equation}
where the sum is over a representative set of simple objects~$x$.  See
\cite[Proposition~8.9]{FT1} for a very similar computation. 
  \end{example}

  \subsection{Duality defects}\label{subsec:4.6}

This section is inspired by~\cite{CCHLS,KOZ}.  Our approach separates out a
topological sector of these quantum field theories and uses the calculus of
defects we have developed.

Resume the general setup: $\sr$~is an $n$-dimensional quiche, with $\rho $~a
regular right $\sigma $-module, and $\sigma $~is equipped with an augmentation.
Suppose $F$~is a quantum field theory equipped with a $\sr$-module structure
$\srtF$.  Assume further that $F$~is equipped with an isomorphism
  \begin{equation}\label{eq:p59}
     \phi \:F\bs\xrightarrow{\;\;\cong \;\;} F.
  \end{equation}

  \begin{example}[]\label{thm:81}
 The existence of \eqref{eq:p59} is a special feature of~$F$ and~$\sigma $.
An example with~$n=2$ is the Ising model (for the group~$\bmut$) at the
critical temperature; see Example~\ref{thm:89}.  In this case $\sigma $~is
the 3-dimensional $\bmut$-gauge theory.  For the five-state Potts model, also
discussed in Example~\ref{thm:89}, $\sigma $~is the 3-dimensional
$\bmu5$-gauge theory.  There are examples in~$n=4$ discussed
in~\cite{CCHLS,KOZ}.  In these cases $\sigma $~is the 5-dimensional
$\bmut$-gerbe theory~$\sXd5{B^2\!\bmut}$.  These include $\U_1$~
Yang-Mills theory with coupling constant $\tau =2\sqmo$ and
$N=4$~supersymmetric $\SU_2$ Yang-Mills theory with $\tau =\sqmo$.
  \end{example}

Recall the domain walls $\delta \:F\to F\bs$ and $\delta^*\:F\bs\to F$
introduced in Definition~\ref{thm:86}.

  \begin{definition}[]\label{thm:p51}
 The \emph{duality defect} is the self-domain wall
  \begin{equation}\label{eq:p60}
     \Delta =\phi \circ \delta \:F\longrightarrow F. 
  \end{equation}
  \end{definition}

\noindent 
 Since $\delta $~is a topological defect, and $\phi $~is an isomorphism of
theories, the composition~$\Delta $ is also a topological defect.

View~$\phi $ as a domain wall from~$F$ to~$F\bs$, and furthermore imagine that
there is a 2-category of theories, domain walls, and domain walls between
domain walls.  Then we can consider the adjoint~$\phi ^*$.  It is a general
fact in 2-categories that if the 1-morphism~ $\phi $ is invertible, then its
adjoint equals its inverse: $\phi ^*=\phi \inv $.  Accepting all this, we
compute
  \begin{equation}\label{eq:p61}
     \Delta ^*\circ \Delta  =(\phi \delta )^*(\phi \delta )=\delta ^*\phi
     ^*\phi \delta =\delta ^*\phi\inv 
     \phi \delta =\delta ^*\circ \delta 
  \end{equation}
The composition~$\delta ^*\circ \delta $ is the quotient defect; see
Remark~\ref{thm:47}.   
 
The following example illustrates a situation in which there is a larger quiche
$(\hat\sigma ,\hat\rho )$ in which we can interpret $\hat\sigma $ as~$\sigma $
with $\Delta $~adjoined.  In this situation $F$~has a $(\hat\sigma ,\hat\rho
)$-module structure.  We do not attempt a general construction beyond the
example.

  \begin{example}[]\label{thm:82}
 In the case of the $n=2$~Ising model introduced in Example~\ref{thm:81}, the
three dimensional $\bmut$-gauge theory~$\sigma $ has $\sigma (\pt)$~the fusion
category whose set of isomorphism classes of simple objects is identified with
$\bmut=\{1,\psi \}$.  Then $\hat\sigma (\pt)$~is the fusion category whose set
of isomorphism classes of simple objects is $\{1,\psi ,\Delta \}$, where the
fusion rules are
  \begin{equation}\label{eq:97}
     \begin{aligned} \psi ^2&=1 \\ \psi \Delta &=\Delta  \\ \Delta
     ^2&=1+\psi \end{aligned}  
  \end{equation}
For the last equation, combine~\eqref{eq:p61} with~\eqref{eq:60}
or~\eqref{eq:63}.  This is an example of a Tambara--Yamagami fusion
category~\cite{TY}.  
  \end{example}

One can use this enhanced symmetry to draw dynamical conclusions.  Namely,
assume that $F$~is gapped and furthermore its infrared behavior is modeled by
an \emph{invertible} field theory~$\lambda $.  Furthermore, suppose that
$\lambda $~carry a $\sr$-module structure as well as a self-defect~$\Delta $
which satisfies~\eqref{eq:p61}.  (If we construct the larger symmetry
$(\hat\sigma ,\hat\rho )$, then we posit that $\lambda $~carry a $(\hat\sigma
,\hat\rho )$-module structure.)  Now because $\lambda $~is invertible,
self-domain walls of~$\lambda $ do not couple to~$\lambda $; they are
independent field theories that act as an endomorphisms on~$\lambda $.
(Compare: an endomorphism of a line is multiplication by a complex number.
More formally, since $\lambda $~is invertible, it follows that $\End(\lambda
)\cong \End(\bone)$.)  So $\delta ^*\circ \delta $~acts as multiplication by an
$(n-1)$-dimensional \emph{topological} field theory, and so too does $\Delta
$~act as multiplication by a topological field theory.  Those theories
satisfy~\eqref{eq:p61}: $\Delta $~is a kind of square root of~$\delta ^*\circ
\delta $.  But in some situations no such square root exists, as we can prove
using the well-developed principles of topological field theory.  If so, this
rules out the possibility of an invertible field theory in the infrared, i.e.,
the possibility that $F$~be ``trivially gapped''.  (We find the term `infrared
invertible' more suitable.)

The following are examples of this phenomenon.

  \begin{example}[]\label{thm:87}
  Take $\sigma =\sXd5{B^2\!\bmut}$ to be the $\bmut$-gerbe theory in
5~dimensions; it acts on 4-dimensional theories with $B\!\bmut$-symmetry.
Assume that $\Delta $~is adjoined and that \eqref{eq:p61}~is satisfied.  The
composition~$\delta ^*\circ \delta $ is computed in Remark~\ref{thm:51}(1);
from~\eqref{eq:p14} we see that it acts on an invertible 4-dimensional theory
as multiplication by 3-dimensional $\bmut$-gauge theory~$\Gamma
=\sXd3{B\!\bmut}$.
  \end{example}

  \begin{example}[]\label{thm:88}
 Continue with Example~\ref{thm:87}.  We claim there is no 3-dimensional
topological field theory~$T$ such that $T^*\circ T=\Gamma $.  If so, evaluate
on a point to obtain fusion categories~$T(\pt)$ and $\Gamma
(\pt)=\Vect[\bmut]$.  The number of simple objects in~$\Vect[\bmut]$ is~2,
which is not a perfect square.  The number of simple objects in
$T^*(\pt)\otimes T(\pt)$ is a perfect square.  This contradiction proves that
there is no invertible left $\sigma $-module on which $\Delta $~acts.
  \end{example} 

  \begin{example}[]\label{thm:90}
 Consider pure $U_1$-gauge theory~$F$ in $n=4$ dimensions.  Such a
theory has a coupling constant~$\tau $ which lies in the upper half plane.
The quotient~$F\bs$ by~ $B\!\bmut$ is another $U_1$-gauge theory but with
coupling constant~$\tau /4$.  The transformation $\tau \mapsto -1/\tau $
lifts to an isomorphism~$\phi $ of the corresponding gauge theories.  Hence
for~$\tau =2\sqmo$ the isomorphism~$\phi $ maps as in~\eqref{eq:p59}.  The
previous arguments show that $F$~is not infrared invertible.  (We know from
other arguments that $F$~is not even gapped, much less infrared invertible.)
  \end{example}

  \begin{example}[]\label{thm:116}
 Another example with $B\!\bmut$-symmetry is $N=4$ supersymmetric $\SU_2$-gauge
theory, which also has a coupling constant~$\tau $.  The quotient is $N=4$~
supersymmetric $\SO_3$-gauge theory, and now S-duality can be used to supply
the isomorphism~$\phi $.  (See~\cite[\S2.4]{AST} for a discussion of S-duality
in these theories.)  It turns out that one must take~$\tau =\sqmo$.  So we
learn that this theory is not infrared invertible.  (Again, it is not even
gapped.)
  \end{example}

  \begin{example}[]\label{thm:83}
 Continue with the $n=2$~Ising model at the critical temperature
(Example~\ref{thm:81} and Example~\ref{thm:82}).  In this case $\delta
^*\circ \delta $~ acts on an invertible 2-dimensional theory as
multiplication by the 1-dimensional topological field theory which is the
$\sigma $-model into~$\bmut$; see Remark~\ref{thm:51}(1) or
Example~\ref{thm:52}.  In particular, the vector space attached to a point
has dimension~2.  Hence there is no square root: $\Delta $~acts as
multiplication by a 1-dimensional topological field theory, as does~$\Delta
^*$, and the vector space attached to a point has the same dimension in both.
Since $\sqrt2$~is not an integer, this cannot happen.  

The Ising model at the critical temperature is not gapped, but the argument
has more power applied instead to the five-state Potts model.  There again
the argument shows there cannot be a unique vacuum at the critical value of
the parameter (the fixed point of the Kramers-Wannier involution of
Example~\ref{thm:89}).  Now, as opposed to in the previous examples, the
theory is gapped at this critical parameter.
  \end{example} 

  \begin{remark}[]\label{thm:91}
 In these examples the theory~$F$ sits in a 1-parameter family~$F_s,\;s\in \RR$,
of theories in which the duality defect~\eqref{eq:p60} extends to an involution
$F_s\leftrightarrow F_{s^*}$ where $s\leftrightarrow s^*$ is an involution of
the parameter space with unique fixed point~$s_c$ and $F=F_{s_c}$.  If we
assume that $F_s$, $s\neq s_c$, is infrared invertible, \emph{and} we also
assume that there is a phase transition at~ $s_c$, then we easily conclude that
$F=F_{s_c}$~cannot be infrared invertible.  (Either $F_{s_c}$~is gapless or if
it is gapped the phase transition is of first-order and there is more than one
vacuum.)  But without the assumption that there is a phase transition, we
cannot rule out the possibility that all $F_s$~are infrared invertible.
  \end{remark}

\appendix
   \section{Finite homotopy theories}\label{sec:5}

The class of topological field theories described here was introduced by
Kontsevich~\cite{Ko} in~1988 and was picked up by Quinn~\cite{Q} a few years
later.  They are also the subject of a series of papers by Turaev~\cite{Tu}
in the early 2000's.  These \emph{finite homotopy theories} lend themselves
to explicit computation using topological techniques.  Not only do they arise
in examples, but they also form a useful playground for the general study of
quantum field theory.

Quantization proceeds via finite algebraic processes, as opposed to the
infinite dimensional analysis required for typical quantum field theories.  The
``finite path integral'' quantization in \emph{fully local} field theory was
introduced in~\cite{F1} with further development in~\cite[\S3,\S8]{FHLT}; see
also \cite[\S9]{FT1}.  The modern approach uses \emph{ambidexterity} or
\emph{higher semiadditivity}, as introduced by Hopkins--Lurie~\cite{HL}; see
also \cite{HeL,Ha,CSY,RS}.  Nonetheless, as far as we know a definitive
treatment is still missing.  Here we summarize a bit about quantization of
theories with an illustrative example.  Then we indicate how to use mapping
spaces to encode semiclassical defects, and how to quantize them via a finite
path integral.

We can drop the $\pi $-finiteness assumption at the cost of only being able
to carry out quantization below the top dimension; the sum which leads to a
complex number is no longer finite and there is no topology to control
convergence.  Just below the top dimension we obtain functions on a possibly
infinite discrete set, which is a well-defined vector space albeit not finite
dimensional in general.  Put differently, in the absence of $\pi $-finiteness
we construct a once-categorified topological field theory
(see Remark~\ref{thm:4}(1)).

  \subsection{$\pi $-finiteness}\label{subsec:5.1}

  \begin{definition}[]\label{thm:53}
 \ 
 \begin{enumerate}[label=\textnormal{(\arabic*)}]

 \item A topological space~$\sX$ is \emph{$\pi $-finite} if (i)~$\pi
_0\sX$~is a finite set, (ii)~for all~$x\in \sX$, the homotopy group $\pi
_q(\sX,x)$, $q\ge1$, is finite, and (iii)~there exists $Q\in \ZZ^{>0}$ such
that $\pi _q(\sX,x)=0$ for all $q> Q$, $x\in \sX$.  (For a fixed bound~$Q$
we say that $\sX$~is $Q$-finite.)

 \item A continuous map $f\:\sY\to \sZ$ of topological spaces is \emph{$\pi
$-finite} if for all~$z\in \sZ$ the homotopy fiber\footnote{The homotopy
fiber over~$z\in \sZ$ consists of pairs~$(y,\gamma )$ of a point~$y\in \sY$
and a path~$\gamma $ in~$\sZ$ from~$z$ to~$f(y)$.} over~$z$ is a $\pi
$-finite space.

 \item A spectrum~$E$ is \emph{$\pi $-finite} if each space in the spectrum
is a $\pi $-finite space.

 \end{enumerate} 
  \end{definition}

  \begin{example}[]\label{thm:54}
 An Eilenberg-MacLane space~$K(\pi ,q)$ is $\pi $-finite if $\pi $~is a
finite group.  We use notation which emphasizes the role of~$\sX$ as a
classifying space: if~$q=1$ we denote $K(\pi ,1)$ by~$B\pi $, and if~$q\ge 1$
and $A$~is a finite abelian group, we denote~$K(A,q)$ by~$\BnA q$.  In the
text, for example in~\S\ref{subsec:4.4}, we encounter $\pi $-finite spaces
with two nonzero homotopy groups.
  \end{example}

  \begin{remark}[]\label{thm:55}
 If $\sX$~is a path connected topological space with basepoint~$x\in \sX$, then
$\sX$~is the classifying space\footnote{The space~$\scP_*\sX$ of continuous
paths $\gamma \:[0,1]\to \sX$ with $\gamma (0)=*$ is contractible, and there is
a continuous map $\scP_*\sX\to \sX$ by evaluation at~$1$.  The fiber over~$*$
is the based loop space~$\OX$.  The path-loop fibration $\scP_*\sX\to \sX$
exhibits~ $\sX$ as the classifying space of~$\OX$.} of its based loop
space~$\OX$, where the latter is a higher finite group (\autoref{thm:1}) by
composition of based loops.
  \end{remark}

  \begin{remark}[]\label{thm:56}
 \ 

 \begin{enumerate}[label=\textnormal{(\arabic*)}]

 \item A topological space~$\sX$ gives rise to a sequence of higher
fundamental groupoids $\pi _0\sX$, $\pi _{\le1}\sX$, $\pi _{\le2}\sX$,\dots ,
or indeed to an $\infty $-groupoid.  There is a classifying space
construction which passes in the opposite direction from higher groupoids to
topological spaces.  An $\infty $-groupoid is $\pi $-finite iff the
corresponding topological space is $\pi $-finite.

 \item In a similar way, one can define $\pi $-finiteness for a simplicial
set. 

 \item A simplicial sheaf is $\pi $-finite if its values are $\pi $-finite
simplicial sets.

 \item These variations pertain to the relative cases of maps, as in
Definition~\ref{thm:53}(2).  

 \end{enumerate} 
  \end{remark}

  \begin{example}[]\label{thm:57}
 Fix $m\in \ZZ^{\ge1}$ and consider the simplicial sheaves of fields (as in
footnote~\footref{field}) which assign to an $m$-manifold~$W$: 
  \begin{equation}\label{eq:64}
     \begin{aligned} \tsF(W) &= \{\textnormal{Riemannian metrics,
      $\SU_2$-connections on~$W$}\} \\ \sF(W) &= \{\textnormal{Riemannian metrics,
      $\SO_3$-connections on~$W$}\} \\ \end{aligned} 
  \end{equation}
There is a map $p\:\tsF\to \sF$ which takes an $\SU_2$-connection to the
associated $\SO_3$-connection.  The map~$p$ is a fiber bundle of simplicial
sheaves.  Neither $\tsF$~nor $\sF$~is $\pi $-finite, but the map~$p$ is $\pi
$-finite.  The fiber over a principal $\SO_3$-bundle $\overline{P}\to W$ is the
groupoid of lifts to a principal $\SU_2$-bundle $P\to W$.  These lifts, if they
exist, form\footnote{The fiber product of a $\bmut$-bundle and a principal
$\SU_2$-bundle is a principal $(\bmut\times \SU_2)$-bundle, and multiplication
$\bmut\times \SU_2\to \SU_2$ is a group homomorphism, so there is an associated
principal $\SU_2$-bundle.} a torsor over the groupoid of double covers of~$W$.
The groupoid of double covers is the fundamental groupoid of the mapping space
$\Map(W,B\!\bmut)$, where $\bmut=\{\pm1\}$~is the center of~$\SU_2$.  Observe
that $B\!\bmut\simeq \RP^{\infty}$ is a $\pi $-finite space, in fact a $\pi
$-finite infinite loop space.
  \end{example}

  \subsection{Field theories from $\pi $-finite spaces and
  maps}\label{subsec:5.2} 

Fix~$m\in \ZZ^{\ge1}$ and suppose $p\:\tsF\to \sF$ is a $\pi $-finite fiber
bundle of simplicial sheaves $\Man_m\op\to \Set_\Delta $, as in
Example~\ref{thm:57}.  The basic idea is that there is a finite process which
takes an $m$-dimensional field theory~$\ts$ over~$\tsF$ as input and produces
an $m$-dimensional field theory~$\sigma $ over~$\sF$ as output.  One
obtains~$\sigma$ from~$\ts$ by summing over the (fluctuating) fields in the
fibers of~$p$.  Since $p$~is $\pi $-finite, this is a finite sum---a finite
version of the Feynman path integral.  In this generality the
theories~$\ts,\sigma $ need not be topological.

  \begin{remark}[]\label{thm:58}
 \ 

 \begin{enumerate}[label=\textnormal{(\arabic*)}]

 \item It often happens that the theory~$\ts$ is ``classical'', in which case
it is an invertible field theory.  Then $\sigma$~is its quantization.
 
 \item If $\sX$~is a $\pi $-finite space, $\sF_{\sX}$~is the simplicial sheaf
of maps into~$\sX$, and $p\:\sF'\times \sF_{\sX}\to \sF'$~is projection for
some  simplicial sheaf~$\sF'$, then we use the notation $\sigma =\sXd m{\sX}$.

 \item The framework is most developed for \emph{topological} field theories,
in which case we can work in \emph{fully local} field theory.

 \end{enumerate} 
  \end{remark}

The basic idea of quantization---see \cite[\S3]{FHLT} for details---is as
follows.  Let $\sX$~be a $\pi $-finite space and fix a dimension~$m$ for the
quantized theory~$\sigma $.  If $Y$~is a closed $(m-1)$-manifold, then we
construct the vector space~$\sigma (Y)$ in two steps.  First, consider the
mapping space $\Map(Y,\sX)$; this is the space of ``classical fields'' on~$Y$.
To each point we attach the trivial line~$\CC$, and the second step of the
quantization is to form the space of sections of this trivial line bundle over
$\Map(Y,\sX)$.  However, we must take sections in a homotopical sense---so
``flat'' sections---which here simply means locally constant sections.  Now a
locally constant function $\Map(Y,\sX)\to\CC$ factors through the set $\pi
_0\Map(Y,\sX)$ of path components, and so $\sigma (Y)$~can be identified with
the space of functions on this set.  There is a similar, but more algebraically
more complicated procedure in other dimensions.  Put together, the finite path
integral is a composition of functors
  \begin{equation}\label{eq:111}
      \Bord_m(\sF)\xrightarrow{\;\;\pi _{\le
      m}\Map(-,\sX)\;\;}\Fam_m(\sC)\xrightarrow{\;\;\Sum_m\;\;}\sC   
  \end{equation}
in which $\sC$~is an $m$-category and $\Fam_m(\sC)$~is an $m$-category of
$m$-groupoids equipped with local systems valued in morphisms at the
appropriate level in~$\sC$; morphisms in $\Fam_m(\sC)$ are correspondences.
The first map in~\eqref{eq:111} takes a bordism to the mapping space
into~$\sX$, viewed as a (higher) correspondence via restrictions to boundaries
and corners.  A cocycle on~$\sX$ is used to construct an invertible
$\sC$-valued local system.  The second map is a finite sum, constructed as a
limit or colimit, as developed in the theory of higher semiadditivity referred
to at the beginning of this appendix.  It is this map $\Sum_m$ that is called
`quantization'.
 
The above abstract reasoning leads to the following concrete formulae for the
values of the $m$-dimensional field theory $\sigma_{\sX}$ associated to a
$\pi$-finite space $\sX$. The domain of $\sigma_{\sX}$ is the bordism category
with fields given by maps to $\sX$.  Here we take the codomain of
$\sigma_{\sX}$ to be a choice of $m$-category $\sC$ such that $\Omega^m \sC =
\CC$, $\Omega^{m-1}(\sC)$ is the category of finite-dimensional complex vector
spaces, and $\Omega^{m-2}(\sC)$ is a 2-category of complex linear categories.
Often in this paper our codomain~$\sC$ has $\Omega^{m-2}(\sC)$~equal to the
2-category of complex algebras.  For such codomains the quantization is
described in~\cite[\S8]{FHLT}.

As we have stated above the state space associated a compact closed
$(m-1)$-dimensional manifold $M_{m-1}$ is the vector space of complex-valued
locally constant functions on $\sX^{M_{m-1}}=\Map(M_{m-1},\sX)$, so
  \begin{equation}\label{eq:CX-theory-cod1}
\sigma_{\sX}(M_{m-1}) \cong \Fun(\pi_0(\sX^{M_{m-1}})) ~ .
  \end{equation}
We next write the amplitudes between state spaces associated with
bordisms; see \cite[\S8]{FHLT} for a fuller treatment. If $M_m: N^0_{m-1} \to N^1_{m-1}$ is a bordism and $\iota_a:
N^a_{m-1}\hookrightarrow M_{m-1}$ is the embedding into the appropriate
boundary of $M_m$ then we have a correspondence of spaces
  \begin{equation}\label{eq:CX-theory-CorrDiag}
\xymatrix{   &  \sX^{M_m}\ar[dl]_{p_0}\ar[dr]^{p_1}  &   \\
\sX^{N^0_{m-1}} &   &   \sX^{N^1_{m-1}} \\
}
  \end{equation}
where $p_a = \iota_a^*$. The amplitude is the linear map given by the
``push-pull formula'': $\sigma_{\sX}(M_m):= p_{1,*}\circ  p_0^*$ applied to locally
constant functions.  Here $p_0^*(\Psi)$, for a function $\Psi$ on the mapping
space from $N^0_{m-1}$ to $\sX$ is simply the function on a mapping space from
$M_m$ to $\sX$ given by restriction of the mapping to the boundary $N_{m-1}^0$.
On the other hand, defining,
  \begin{equation}
p_{1,*}: \Fun(\pi_0(\sX^{M_{m}})) \rightarrow \Fun(\pi_0(\sX^{N^1_{m-1}}))
  \end{equation}
requires some more care. If $\Psi \in \Fun(\pi_0(\sX^{M_{m}}))$ and
$h:N^1_{m-1} \to \sX$ then 
  \begin{equation}\label{eq:CX-theory-amp}
p_{1,*}(\Psi)(h):= \sum_{[\phi] \in \pi_0(p_1^{-1}(h))}  \left( \prod_{i=1}^\infty \left( \# \pi\mstrut _i(p_1^{-1}(h), \phi ) \right)^{(-1)^i} \right)   \Psi(\phi)
  \end{equation}
Note that for each connected component of the fiber above the mapping
$h:N^1_{m-1} \to \sX$ we choose a mapping $\phi: M_m \to \sX$ in that
component. Since $\Psi$ is locally constant on $\sX^{M_m}$ the
\underline{choice} of $\phi$ does not affect the right hand side. Thanks to the
$\pi$-finiteness condition $p^{-1}(h)$ is a $\pi$-finite space and hence the
sum is finite and the infinite product is well-defined. Note that the function
$p_{1,*}(\Psi)(h)$ is locally constant in $h$. Note too that $p_{1,*}$ is not
simply ``sum along fibers.''  Rather, it is the formula for \emph{homotopy
cardinality}; see~\cite{BaDo}, for example.
A pleasant 
property of the homotopy fiber product is that the amplitudes compose 
properly for composition of bordisms, as is necessary when defining a 
functor from $\Bord_n$.
An immediate consequence of
\eqref{eq:CX-theory-amp} is the formula for the partition function on
$m$-manifolds without boundary: 
  \begin{equation}\label{eq:CX-theory-pf}
\sigma_{\sX}(M_m) = \sum_{[\phi] \in \pi_0(\sX^{M_m})}  \left(
\prod_{i=1}^\infty \left( \# \pi_i(\sX^{M_m}, \phi ) \right)^{(-1)^i} \right) ~
.  
  \end{equation}

Now, as explained in \cite{FHLT} there is an inductive procedure for
determining the value of $\sigma_{\sX}(M_{m-\ell})$ for closed manifolds of
smaller dimension. An important idea in that procedure is that the algebra
objects in a symmetric monoidal $j$-category form a $(j+1)$-category. The
result of the discussion in \cite{FHLT} is that $\sigma_{\sX}(M_{m-2})$ is the
$1$-category of ``locally constant vector bundles'' over
$\sX^{M_{m-2}}$. ``Locally constant vector bundles'' should be interpreted as
flat vector bundles, a.k.a. local systems. A local system on $\sX^{M_{m-2}}$ is
the same thing as a vector bundle over the groupoid $\pi_{\leq
1}(\sX^{M_{m-2}})$ and hence we have:\footnote{A vector bundle over a
topological groupoid is a vector bundle over the space of objects and an
isomorphism of the pullback bundles over the space of morphisms given by the
source and target maps. This isomorphism must furthermore satisfy a cocycle
condition for composable morphisms. See \cite[Appendix~A]{FHT2} for more
details.}
  \begin{equation}\label{eq:120}
     \sigma_{\sX}(M_{m-2}) = \Vect(\pi_{\leq 1}( \sX^{M_{m-2}})) ~ . 
  \end{equation}
Next, a bordism $M_{m-1}: N^0_{m-2} \to N^1_{m-2}$ produces a functor
$\sigma_{\sX}(M_{m-1})$ again given by a push-pull formula associated with a
correspondence diagram analogous to \eqref{eq:CX-theory-CorrDiag}. 

The theories $\sigma_{\sX}$ can be enhanced in interesting ways if one provides
the extra data of a $\CC^*$-valued $m$-cocycle $\lambda$ on $\sX$. In general
this will require an extension of the fields in the domain of $\sigma_{\sX}$ to
include orientations.
 Equation \eqref{eq:CX-theory-cod1}  for state spaces is now modified to be the vector space of flat sections of a distinguished flat complex line bundle over $\sX^{M_{m-1}}$ determined by $\lambda$.
The line bundle is determined by integrating $ev^*(\lambda)$ over $M_{m-1}$, where $ev: M_{m-1} \times \sX^{M_{m-1}} \to \sX$ is the evaluation map, to obtain a $1$-cocycle on $\sX^{M_{m-1}}$. The $1$-cocycle determines a flat line bundle over $\sX^{M_{m-1}}$.  The amplitudes are modified by including a factor $\langle \phi^*\lambda, [M_m] \rangle$
in sums such as \eqref{eq:CX-theory-pf},
 the category of vector bundles over $\pi_{\leq 1}(\sX^{M_{m-2}})$ is replaced by a category of twisted vector bundles, and so forth.

We now give an explicit example in which the target is an Eilenberg-MacLane
space.  This example recurs in~\cite{F4}.

  \begin{example}[]\label{thm:59}
 Let $A$~be a finite abelian group and set~$\sX=\BnA2$.  For definiteness fix
dimension~$m=5$.  Our aim is to construct a 5-dimensional topological field
theory\footnote{Sometimes this is called the theory of a ``B-field'', which is
the background field for a ``1-form symmetry~$A$''.}~$\sigma=\sXd5{\BnA2}$.  In
the terms above: $\tsF$~is the simplicial sheaf on~$\Man_5$ which assigns to a
5-manifold~$W$ the 2-groupoid $\pi _{\le2}\Map(W,\BnA2)$, $\ts$~is the tensor
unit theory, and $\sF$~is the trivial simplicial sheaf which assigns a point to
each 5-manifold~$W$.  (The triviality of~$\sF$ is the statement that $\sigma$~
is an ``unoriented theory''---there are no background fields.)  We have not
specified the codomain~$\sC$ of the theory, and one has latitude in this
choice.  For our purposes we assume standard choices at the top three levels:
$\Omega ^3\sC=\Cat$ is a linear 2-category of complex linear categories, from
which it follows that $\Omega ^4\sC=\Vect$ is a linear 1-category of complex
vector spaces and $\Omega ^5\sC=\CC$.
 
Let $M$~be a closed manifold.  Then $\sigma(M)$~is the quantization of the
mapping space
  \begin{equation}\label{eq:65}
     \Xm M=\Map(M,\sX) 
  \end{equation}
The nature of that quantization depends on~$\dim M$.  Here we simply report
the results.
 
\medskip
 $\dim M=5$: \parbox[t]{36em}{The quantization is a (rational) number, a
weighted sum over homotopy classes of maps $M\to \sX$:}
  \begin{equation}\label{eq:66}
     \sigma(M) = \sum\limits_{[\phi ]\in \pi _0(\Xm M)}\frac{\#\pi _2(\Xm M,\phi
     )}{\#\pi _1(\Xm M,\phi )} \;\;=\;\;
     \frac{\#H^0(M;A)}{\#H^1(M;A)}\;\#H^2(M;A).  
  \end{equation}

\medskip
 $\dim M=4$: \parbox[t]{36em}{The quantization is the vector space of locally
constant complex-valued functions on~$\Xm M$:} 
  \begin{equation}\label{eq:67}
     \sigma(M) = \Fun\bigl(\pi _0(\Xm M)\bigr) = \Fun\bigl(H^2(M;A)
     \bigr). \phantom{MMMMMMMN} 
  \end{equation}

\medskip
 $\dim M=3$: \parbox[t]{36em}{The quantization is the linear category of flat
vector bundles  (local systems) over~$\Xm M$:}  
  \begin{equation}\label{eq:68}
     \phantom{M}\begin{aligned} \sigma(M) = \Vect\bigl(\pi _{\le1}(\Xm M)
     \bigr) &= 
      \Vect\bigl(H^2(M;A) \bigr)\times \Rep\bigl(H^1(M;A) \bigr) \\ &\simeq
      \Vect\bigl(H^2(M;A)\times H^1(M;A)\dual \bigr),\end{aligned} 
  \end{equation}
 \hspace{6.5em}\parbox[t]{36em}{where $\sA\dual$~is the Pontrjagin dual
group of characters of the finite abelian group~$\sA$.  (If $M$~is oriented,
there is an isomorphism $H^1(M;A)\dual\cong H^2(M;A\dual)$.)}
  \end{example}

  \begin{remark}[]\label{thm:60}
 \ 
 \begin{enumerate}[label=\textnormal{(\arabic*)}]

 \item In this example $\sX$~is an Eilenberg-MacLane space, which explains the
cohomological translations in \eqref{eq:66}--\eqref{eq:68}.

 \item All homotopy groups of the mapping space are used in the top dimension
of the theory---see~\eqref{eq:66}---whereas in codimension~$\ell $ only $\pi
_{\le(\ell -1)}$~enters.  

 \item See \cite[\S8]{FHLT} for a fuller treatment, including higher
codimension; in that reference the codomain is a higher category of algebras.

 \end{enumerate}
  \end{remark}

In terms of the paradigm at the beginning of this subsection, the example so
far has trivial~ $\ts$.  We now give an example in which $\ts$~is a
nontrivial invertible theory.  The data which defines it is a
pair~$(\sX,\lambda )$ consisting of a $\pi $-finite space~$\sX$ and a
``cocycle''~$\lambda $ on~$\sX$.  Typically we need a generalized orientation
to integrate~$\lambda $, depending on the generalized cohomology theory in
which $\lambda $~is a cocycle.  (The model for a geometric representative of
a cohomology class---a ``cocycle''---may vary.)

  \begin{example}[twisted $\sX=\BnA2$]\label{thm:63}
 We continue with $\sX=\BnA2$, and we illustrate with $A=\zt$ the cyclic group
of order~2.  Then\footnote{Let $\iota \in H^2(B^2\zt;\zt)$ be the tautological
class.  Then $\iota \smile \Sq^1\!\iota \in H^5(B^2\zt;\zt)$ becomes the
nonzero class after extending coefficients $\zt\to \Cx$.}  $H^5(\sX;\Cx)\cong
H^6(\sX;\ZZ)$ is cyclic of order~2.  Let $\lambda $~be a cocycle which
represents this class.  The quantizations in Example~\ref{thm:59} are altered
as follows.  For $\dim M=5$ weight the sum in~\eqref{eq:66} by $\langle \phi
^*\lambda ,[M] \rangle$, where $[M]$~is the fundamental class.\footnote{Since
$\lambda $~is induced from a mod~2 class, orientations are not necessary---we
can proceed in mod~2 cohomology.}  For $\dim M=4$ the transgression of~$\lambda
$ to~$\Xm M$ induces a flat complex line bundle (of order~2) $L\to \Xm M$; now
\eqref{eq:67}~becomes the space of flat sections of $L\to \Xm M$.  Note that a
flat section vanishes on a component of~$\Xm M$ on which the automorphisms act
by a nonidentity character.  Similarly, for $\dim M=3$ the cocycle~$\lambda $
transgresses to a twisting of $K$-theory, and the quantization is a category of
twisted vector bundles.
  \end{example}

Some finite homotopy theories are constructed from an invertible theory~$\ts$
based on a cocycle that uses the intrinsic geometry, possibly mixed with the
extrinsic geometry that we have been using heretofore, as illustrated in the
next example.  See~\cite{De} for one situation in which such a theory arises
from a lattice model.

  \begin{example}[twisted $\sX=\BnA2$ mixed with intrinsic geometry]\label{thm:64}
 Continue with $\sX=B^2(\zt)$ and $m=5$.  We construct a topological field
theory~$\sigma$ of oriented manifolds ($\sF=\{\textnormal{orientation}\}$) from
an invertible topological field theory~$\ts$ of oriented manifolds equipped
with a $\zt$-gerbe ($\tsF=\{\textnormal{orientation, $\zt$-gerbe}\}$).  The
latter are classified by maps into the spectrum
  \begin{equation}\label{eq:71}
     \Sigma ^5\MTSO_5\wedge B^2(\zt)_+; 
  \end{equation}
see \cite{FHT1,FH2} for the notation and for more on invertible field theories
and homotopy theory.  Let $\iota \in H^2(B\zt;\zt)$ be the tautological class,
and let $w_3=\Sq^1\!w_2\in H^3(\Sigma ^5\MTSO_5;\zt)$ be the third
Stiefel-Whitney class.  Use the cup product $w_3\smile\iota $ to define an
invertible field theory~$\ts$, and then the finite path integral to define a
topological field theory~$\sigma$ whose partition function on a closed oriented
5-manifold~$M$ is
  \begin{equation}\label{eq:72}
     \sigma(M) = \sum\limits_{[\phi ]\in \pi _0(\Xm M)}\frac{\#\pi _2(\Xm M,\phi
     )}{\#\pi _1(\Xm M,\phi )} \;(-1)^{\langle w_3(M)\smile \phi ^*\iota
     ,\,[M] \rangle}.  
  \end{equation} 
The weighting factor $(-1)^{\langle w_3(M)\smile \phi ^*\iota ,\,[M] \rangle}$
reflects the mixing with the intrinsic geometry.
  \end{example}

  \begin{remark}[]\label{thm:67}
 There is a composition law---tensor product---on field theories with fixed
domain and codomain.  This is sometimes called ``stacking'' of quantum
systems.  The tensor product of finite homotopy theories based on
$(\sX_1,\lambda _1)$ and $(\sX_2,\lambda _2)$ is the finite homotopy theory
based on $(\sX_1\times \sX_2,\lambda _1+\lambda _2)$.  In the relative
setting of Definition~\ref{thm:53}(2), the Cartesian product is generalized
to the fiber product over the base.
  \end{remark}

  \subsection{Defects in finite homotopy theories}\label{subsec:5.3}
 
Our account here implicitly assumes framings.  One could combine with the ideas
in~\S\ref{subsec:2.5} to generalize to arbitrary tangential structures.

Consider a finite homotopy theory~$\sigma$ based on a $\pi $-finite
space~$\sX$.  For a defect on a submanifold of codimension~$\ell \in
\ZZ^{\ge1}$, the link is~$S^{\ell -1}$---canonically if the normal bundle is
framed---and so the mapping space on the link is
  \begin{equation}\label{eq:73}
     \LlX:=\Map(S^{\ell -1},\sX).
  \end{equation}
Note that if $\sX$~is equipped with a cocycle~$\lambda $, then $\lambda
$~transgresses\footnote{Use the diagram 
  \begin{equation}\label{eq:119}
     \begin{gathered} \xymatrix{\LlX\times S^{\ell
     -1}\ar[r]^<<<<<{e}\ar[d]_<<<<<{\pi } 
     & \sX\\\LlX} \end{gathered} 
  \end{equation}
to form the map $(\pi )_*\circ e^*$ on cohomology; this is transgression of the
sort that does not require a trip to confession.} to a cocycle $\tau ^{\ell
-1}\lambda $ on~$\LlX$ with a drop of degree by~$\ell -1$.  Recall the
definition of a local defect in Definition~\ref{thm:7}(1) .

  \begin{definition}[]\label{thm:68}
 Fix $m,\ell \in \ZZ^{\ge1}$ with $\ell \le m$.  Let $\sX$~be a $\pi $-finite
space and suppose $\lambda $~is a cocycle of degree~$m$ on~$\sX$.  A
\emph{semiclassical local defect} of codimension~$\ell $ for~$\Xl$ is a $\pi
$-finite map
  \begin{equation}\label{eq:74}
     \delta \:\sY\longrightarrow \LlX 
  \end{equation}
and a trivialization~$\mu $ of~$\delta ^*(\tau ^{\ell -1}\lambda )$. 
  \end{definition}

\noindent
 Since $\LlX$~is $\pi $-finite, \eqref{eq:74}~ amounts to a $\pi $-finite
space~$\sY$ and a continuous map~$\delta $.  Intuitively, $\sY$~takes into
account the degrees of freedom on the defect.  The local quantum defect in
$\Hom\bigl(1,\sigma(S^{\ell -1}) \bigr)$ is the quantization of the
map~\eqref{eq:74}, viewed as a correspondence 
  \begin{equation}\label{eq:118}
     \begin{gathered} \xymatrix{&\sY\ar[dl]\ar[dr]^{\delta }\\ \ast && \LlX}
     \end{gathered} 
  \end{equation}

  \begin{remark}[]\label{thm:115}
 The term `classical label' is used in the main text in the context of finite
homotopy theories; the set of classical labels of codimension~$\ell $ is~$\pi
_0(\LlX)$.  (For $\rho $-defects it is $\pi _0(\Omega ^{\ell -1}\sX)$.)  We
illustrate in~\S\ref{subsec:4.4} that classical labels do not adequately label
quantum defects. 
  \end{remark}

We now turn to semiclassical \emph{global} defects.  As an example, based on
\autoref{thm:97}, if $M$~is a closed manifold and $Z\subset M$ is a
\emph{normally framed} codimension~$\ell $ submanifold on which the
defect~\eqref{eq:74} is placed, the value of the theory~$\sigma$ on~$M$ with
the defect on~$Z$ is the quantization of the mapping space
  \begin{equation}\label{eq:77}
     \Map\bigl((M,Z),(\sX,\sY) \bigr) 
  \end{equation}
consisting of  pairs of maps  $\phi \:M\to \sX$  and $\psi \:Z\to  \sY$ which
satisfy a compatibility condition:  if $i\:Z\times S^{\ell -1}\hookrightarrow M$
is the inclusion  of the boundary of a tubular  neighborhood of~$Z\subset M$,
and $\phi '\:Z\to \LlX$ is the adjoint of the composition
  \begin{equation}\label{eq:75}
     Z\times S^{\ell -1}\longhookrightarrow
     ^{\!\!\!\!\!\!\!\!i\mstrut }\;\;M\xrightarrow{\;\;\phi 
     \;\;}\sX, 
  \end{equation}
then the diagram 
  \begin{equation}\label{eq:76}
     \begin{gathered} \xymatrix@C+1pc@R+1pc{&\sY\ar[d]^{\delta }\\ Z\ar[r]^{\phi
     '}\ar[ur]^{\psi }&\LlX} \end{gathered}\qquad \qquad 
  \end{equation}
is required to commute.

  \begin{remark}[]\label{thm:66}
 \ 
 \begin{enumerate}[label=\textnormal{(\arabic*)}]

 \item Strict commutation is unnatural in this context.  One can use instead a
mapping space of triples $(\phi ,\psi ,\gamma )$ with a specified homotopy
$\gamma \:\delta \circ \psi \to \phi '$.  However, the homotopy can be
incorporated into a tubular neighborhood of~$Z$, so nothing is lost by using
the strict mapping space.

 \item There are many variations of this basic scenario.  The defect may have
support on a manifold with boundary or corners, or more generally on a
stratified manifold.  Such is the case for the $\rho $-defects in
Definition~\ref{thm:33}; a further example is in Figure~\ref{fig:25}.

 \item Also, to include background fields and more complicated cocycles, we use
a relative version with $\pi $-finite maps; see~\S\ref{subsec:2.5} for the
quantum picture.  We leave the general development to the reader or to the
future.

 \end{enumerate}
  \end{remark}

Our thesis is that there is a calculus of semiclassical mapping spaces which
encodes defects and their fusion laws.  Rather than pursue general theory, we
indicate some general classes of defects and their composition laws,
beginning with boundaries and domain walls.  (Observe that boundaries are
naturally normally framed, and we will assume a normal framing on domain
walls, though see Remark~\ref{thm:41} for a non-coframable domain wall.)

   \subsubsection{Domain walls}\label{subsubsec:5.3.1}
 Fix $m\in \ZZ^{\ge1}$ and let $\Xln1$, $\Xln2$ be pairs of $\pi $-finite
spaces and degree~$m$ cocycles.  The following is a variation of
Definition~\ref{thm:68}.  

  \begin{definition}[]\label{thm:69}
 A \emph{semiclassical \textnormal{(}local\textnormal{)} domain wall} from
$\Xln1$ to~$\Xln2$ is a pair~$(\sY,\mu )$ consisting of a $\pi $-finite
space~$\sY$ equipped with a correspondence
  \begin{equation}\label{eq:78}
     \begin{gathered} \xymatrix{&(\sY,\mu )\ar[dl]_{f_1} \ar[dr]^{f_2} \\
     (\sX_1,\lambda _1)&&(\sX_2,\lambda _2)} \end{gathered}
  \end{equation}
where $\mu $~is a trivialization of $f_2^*\lambda _2 - f_1^*\lambda _1$.
  \end{definition}

  \begin{remark}[]\label{thm:70}
 \ 
 \begin{enumerate}[label=\textnormal{(\arabic*)}]

 \item We have written~\eqref{eq:78} to conform to standard practice for a
correspondence from~$\sX_1$ to~$\sX_2$, but to fit our right/left
conventions, as illustrated in Figure~\ref{fig:2}, we could swap~$\sX_1$
and~$\sX_2$.

 \item The link of a domain wall is~$S^0$, and \eqref{eq:78}~is the analog
of~\eqref{eq:74} for $\ell =1$.

 \item If $\sY'$ is a $\pi $-finite space equipped with a degree~$m-1$
cocycle~$\mu '$, then there is a new semiclassical domain wall
  \begin{equation}\label{eq:79}
     \begin{gathered} \xymatrix@C-1pc{&(\sY\times \sY',\mu +\mu
     ')\ar[dl]\ar[dr]\\      \Xln1&&\Xln2} \end{gathered} 
  \end{equation}
This corresponds to tensoring with the $(m-1)$-dimensional theory~$(\sY',\mu
')$ on the domain wall; see Remark~\ref{thm:67}.

 \end{enumerate}
  \end{remark}

To quantize a semiclassical domain wall, we use~\eqref{eq:78} to construct a
mapping space.  Let $M$~be a closed manifold of dimension~$\le m$ separated by
a cooriented hypersurface~$Z$:
  \begin{equation}\label{eq:80}
     M=M_1\cup \mstrut _ZM_2 .
  \end{equation}
Form the mapping space
  \begin{equation}\label{eq:81}
     \sM=\bigl\{ (\phi _1,\phi _2,\psi ):\phi _i\:M_i\to \sX_i,\; \psi \:Z\to
     \sY,\; f_i\circ \psi =\phi _i\res Z \bigr\}. 
  \end{equation}
This is essentially a special case of~\eqref{eq:77}.  Now quantize~$\sM$ as
illustrated around~\eqref{eq:CX-theory-CorrDiag}.
 
Suppose $\Xln1$, $\Xln2$, $\Xln3$ are $\pi $-finite spaces and degree~$m$
cocycles, and let
  \begin{equation}\label{eq:82}
     \begin{aligned} (\sY',\mu ')\:\Xln1\longrightarrow \Xln2 \\ (\sY'',\mu
      '')\:\Xln2\longrightarrow \Xln3 \\ \end{aligned} 
  \end{equation}
be semiclassical domain walls.  Their composition 
  \begin{equation}\label{eq:83}
     (\sY,\mu )\:\Xln1\longrightarrow \Xln3 
  \end{equation}
is the homotopy fiber product\footnote{More properly, it is the \emph{homotopy
limit} of the diagram~\eqref{eq:84} with dashed arrows omitted, but that
reduces to the indicated homotopy fiber product.} over the maps to~$\sX_2$
  \begin{equation}\label{eq:84}
     \begin{gathered} \xymatrix{&&\sY\ar@{-->}[dl]\ar@{-->}[dr]\\
     &\sY'\ar[dl]\ar[dr] &&\sY''\ar[dl]\ar[dr] \\ \sX_1&&\sX_2&&\sX_3}
     \end{gathered} 
  \end{equation}
This is the composition of correspondence diagrams (in the homotopy
category); the trivialization~$\mu $ of~$\lambda _3-\lambda _1$ is the sum
$\mu _1+\mu _2$.  (For ease of reading, we omitted pullbacks in the previous
clause.)  We write~\eqref{eq:84} with cocycles and trivializations as
follows:
  \begin{equation}\label{eq:85}
     \begin{gathered} \xymatrix@C-1pc{&&(\sY,\mu '+\mu
     '')\ar@{-->}[dl]\ar@{-->}[dr]\\ 
     &(\sY',\mu ')\ar[dl]\ar[dr] &&(\sY'',\mu '')\ar[dl]\ar[dr] \\
     \Xln1&&\Xln2&&\Xln3}  \end{gathered}
  \end{equation}

  \begin{remark}[]\label{thm:71}
 This prescription for composition is a special case of~\eqref{eq:93} below. 
  \end{remark}

   \subsubsection{Boundaries}\label{subsubsec:5.3.2}
 As in~\S\ref{subsec:2.3} we specialize domain walls to boundary theories. 

  \begin{definition}[]\label{thm:72}
 Let $\sX$~be a $\pi $-finite space and suppose $\lambda $~is a cocycle of
degree~$m$ on~$\sX$.

 \begin{enumerate}[label=\textnormal{(\arabic*)}]

 \item A \emph{right semiclassical boundary theory} of~$\Xl$ is a
pair~$(\sY,\mu )$ consisting of a $\pi $-finite space~$\sY$, a map $f\:\sY\to
\sX$, and a trivialization~$\mu $ of~$-f^*\lambda $. 

 \item A \emph{left semiclassical boundary theory} of~$\Xl$ is a
pair~$(\sY,\mu )$ consisting of a $\pi $-finite space~$\sY$, a map $f\:\sY\to
\sX$, and a trivialization~$\mu $ of~$f^*\lambda $. 

 \end{enumerate} 
  \end{definition}

\noindent
 The mapping spaces used for quantization specialize~\eqref{eq:81}. 
 
In this finite homotopy context there are special forms for Dirichlet and
Neumann boundary theories, which we call regular and augmentation,
respectively.  

  \begin{definition}[]\label{thm:73}
 Let $\sX$~be a connected $\pi $-finite space and suppose $\lambda $~is a
cocycle of degree~$m$ on~$\sX$.

 \begin{enumerate}[label=\textnormal{(\arabic*)}]

 \item A \emph{semiclassical right regular boundary theory} of~$\Xl$ is a
basepoint $f\:*\to \sX$ and a trivialization~$\mu $ of~$-f^*\lambda $.

 \item A \emph{semiclassical right augmentation} of~$\Xl$ is a
trivialization~$\mu $ of~$-\lambda $; the map~$f$ in Definition~\ref{thm:72}
is the identity $\id_{\sX}\:\sX\to \sX$.

 \end{enumerate} 
  \end{definition}

  \begin{remark}[]\label{thm:78}
 Note that the regular boundary condition amounts to an extra semiclassical
(fluctuating) field on the boundary which is a trivialization of the bulk
field (map to~$\sX$).
  \end{remark}

  \begin{example}[]\label{thm:74}
 Let~$m=2$.  Fix a finite group~$G$ and let $\sX=BG$ with basepoint $*\to BG$.
The value of the theory on the interval depicted in Figure~\ref{fig:5} is the
quantization of the restriction map to the right endpoint
  \begin{equation}\label{eq:86}
     \Map\bigl(([0,1],\{0\}),(BG,*) \bigr)\longrightarrow \Map\bigl(\{1\},BG
     \bigr), 
  \end{equation}
which up to homotopy is the map $*\to BG$.  Choose the codomain $\sC=\Cat$ so
that, as in~\eqref{eq:68}, the quantization of~$\Map(*,BG)$ is the category
$\Vect(\pi _{\le1}BG)\simeq \Rep(G)$.  Then the quantization of the map $*\to
BG$, or better of the correspondence
  \begin{equation}\label{eq:87}
     \begin{gathered} \xymatrix{&\ast\ar[dl]\ar[dr] \\\ast&&BG}
     \end{gathered} 
  \end{equation}
of mapping spaces derived from Figure~\ref{fig:5}, is the pushforward of the
trivial bundle over~$*$ with fiber~$\CC$ (the tensor unit).  This is the
regular representation of~$G$ in~$\Rep(G)$.  If, instead, we choose
$\sC=\Alg(\Vect)$, then the prescription~\eqref{eq:120} is altered so that
$\Map(*,BG)$~quantizes to the group algebra~$\GA$ and $*\to BG$ quantizes to
the right regular module: see \cite[Example~3.6]{FHLT}.  We leave the reader to
incorporate a nonzero cocycle in the form of a central extension
  \begin{equation}\label{eq:88}
     1\longrightarrow \Cx\longrightarrow G^\tau \longrightarrow
     G\longrightarrow 1 
  \end{equation}
as in~\S\ref{subsec:1.4}. 
  \end{example}

Let $\Xl$ be given and suppose $(\sY',\mu ')$ and $(\sY'',\mu '')$ are right
and left semiclassical boundary theories for~$\Xl$.  Then, as a special case
of the composition~\eqref{eq:85}, the $(m-1)$-dimensional semiclassical
sandwich of~$\Xl$ between $(\sY',\mu ')$ and $(\sY'',\mu '')$ has as its
semiclassical data the pair $(\sY'\overset h{\times}\mstrut _{\sX}\sY'',\mu
'+\mu '')$, where $\sY'\overset h{\times}\mstrut _{\sX}\sY''$ is the
homotopy fiber product; observe that $\mu '+\mu ''$ is a \emph{cocycle} of
degree~$m-1$.  This is the data that defines an $(m-1)$-dimensional theory.

  \begin{remark}[]\label{thm:109}
 One could go on to define semiclassical defects within defects using a
variation of the setup in~\S\ref{subsubsec:5.3.1}.  In particular, we will
encounter semiclassical domain walls between semiclassical boundary theories in
\autoref{thm:49}.
  \end{remark}

   \subsubsection{Composition laws in higher codimension}\label{subsubsec:5.3.3}
 The general composition law on local defects is constructed using the higher
dimensional pair of pants---see the end of~\S\ref{sec:2}---or, in the case of
$\rho $-defects as in Figure~\ref{fig:17}, using the higher dimensional pair
of chaps.  Here we state the semiclassical version of the first of these. 
 
Resume the setup of Definition~\ref{thm:68}: $m,\ell \in \ZZ^{\ge1}$ are
integers with $\ell \le m$, and $\Xl$~is the finite homotopy data for an
$m$-dimensional theory~$\sigma$.  Let $P$~be the $\ell $-dimensional pair of
pants: as a manifold with boundary,
  \begin{equation}\label{eq:89}
     P=D^\ell \;\setminus\; B^\ell \amalg B^\ell , 
  \end{equation}
where $B^\ell \amalg B^\ell $ are embedded balls in the interior of~$D^\ell
$.  As a bordism, 
  \begin{equation}\label{eq:90}
     P\:S^{\ell -1}\amalg S^{\ell -1}\longrightarrow S^{\ell -1}, 
  \end{equation}
where the domain spheres are the inner boundaries of~$P$ and the codomain
sphere is the outer boundary.  By integration over~$P$, the cocycle~$\lambda $
on~$\sX$ transgresses to an isomorphism
  \begin{equation}\label{eq:91}
     \mu \:r_0^*\pi _1^*(\tll) + r_0^*\pi _2^*(\tll)\longrightarrow r_1^*(\tll) 
  \end{equation}
of cocycles on the mapping space~$\sX^P$.  Here $\pi _i\:\LlX\times \LlX\to
\LlX$ is projection onto the $i^{\textnormal{th}}$~factor.  Then the
composition law on~$\sigma(S^{\ell -1})$ is the quantization of the
correspondence
  \begin{equation}\label{eq:92}
     \begin{gathered} \xymatrix@!@R-13pc@C-11pc{&\bigl(\sX^P\!,\,\mu
     \bigr)\ar[dl]_{r_0}\ar[dr]^{r_1} \\ 
     \bigl(\LlX\times \LlX,\,\pi _1^*(\tll) + \pi _2^*(\tll)\bigr) &&
     \bigl(\LlX,\,\tll\bigr)} \end{gathered} 
  \end{equation}
The composition law on~$\sigma(S^{\ell -1})$ induces the composition law---the
fusion product---on $\Hom\bigl(1,\sigma(S^{\ell -1}) \bigr)$, the higher
category of local codimension~$\ell $ defects.  Suppose given $(\sY_1,\mu _1)$
and $(\sY_2,\mu _2)$ semiclassical local defects of codimension~$\ell $, as in
Definition~\ref{thm:68}.  Then the product of their quantizations in
$\Hom\bigl(1,\sigma(S^{\ell -1}) \bigr)$ is the quantization of the
composition~$r_1\circ g$ in the homotopy fiber product~$\sY$ of $\delta
_1\times \delta _2$ and~$r_0$ in the diagram
  \begin{equation}\label{eq:93}
     \begin{gathered} \xymatrix@!@R-12pc@C-9pc{&\bigl(\sY,\,\pi _1^*\mu _1 +
     \pi _2^*\mu _2+\mu \bigr) \ar@{-->}[dl]\ar@{-->}[d]^{g}\\
     \bigl(\sY_1\times \sY_2,\,\pi ^*_1\mu _1 + \pi _2^*\mu _2
     \bigr)\ar[d]_{\delta _1\times \delta _2}&\bigl(\sX^P\!,\,\mu
     \bigr)\ar[dl]_{r_0}\ar[dr]^{r_1} \\ 
     \bigl(\LlX\times \LlX,\,\pi _1^*(\tll) + \pi _2^*(\tll)\bigr) &&
     \bigl(\LlX,\,\tll\bigr)} \end{gathered} 
  \end{equation}
This diagram is the general composition law on semiclassical local defects.
By quantizing we obtain the general composition law on local defect theories.

  \begin{remark}[]\label{thm:76}
 The identity object---the tensor unit or transparent defect---in
$\Hom(1,\sigma(S^{\ell -1}))$ is the quantization of the semiclassical defect
  \begin{equation}\label{eq:94}
     \sX^{D^\ell }\longrightarrow \LlX 
  \end{equation}
given by the restriction from maps out of~$D^\ell $ to maps out of its
boundary~$S^{\ell -1}$.  
  \end{remark}

  \subsection{An example: finite gauge theories}\label{subsec:5.4}

Recall that to a finite group~$G$ and a positive integer~$m$ is associated
the finite gauge theory $\sigma =\sXd m{BG}$ built from the $\pi $-finite
space $\sX=BG$.  Furthermore, if $\lambda $~is a cocycle for a class in
$H^m(BG;\Cx)$, then there is a twisted version of this theory on oriented
manifolds: the Dijkgraaf-Witten theory $\sigma =\sXd m{BG,\lambda }$.  Here
we give examples of boundaries and domain walls in these theories. 
 
Fix $m\in \ZZ^{\ge1}$.  Suppose $f\:H\to G$ is a homomorphism of finite
groups, let $\lambda \in Z^m(BG;\Cx)$ be a cocycle, and let $\mu \in
C^{m-1}(BH;\Cx)$ be a cochain which satisfies 
  \begin{equation}\label{eq:98}
     \delta \mu =(Bf)^*\lambda , 
  \end{equation}
where $Bf\:BH\to BG$ is the induced map on classifying spaces.  Then
(Definition~\ref{thm:72}) the pair $(BH,\pm\mu )$ is a left/right
semiclassical boundary theory of~$(BG,\lambda )$.   
 
For a space~$M$, the groupoid $\pi _{\le1}(BG^M)$ is equivalent to the
groupoid of principal $G$-bundles over~$M$.  For a pair of spaces $(M,N)$ in
which $N\subset M$~is a subspace, let $\Map\bigl((M,N),(BG,BH) \bigr)$ be the
mapping space of pairs $\phi \:M\to BG$ and $\psi \:N\to BH$ such that 
  \begin{equation}\label{eq:99}
     \begin{gathered} \xymatrix@C+1pc@R+1pc{&BH\ar[d]^{Bf }\\
     N\ar[r]^{\phi |\mstrut _ N}\ar[ur]^{\psi }&BG} \end{gathered} 
  \end{equation}
commutes.  Then $\pi _{\le1}\Map\bigl((M,N),(BG,BH) \bigr)$ is equivalent to
the groupoid of principal $G$-bundles $P\to M$ equipped with a reduction
along~$f$ to a principal $H$-bundle~$P'\to N$.  (That reduction, which is~$\psi
$ in~\eqref{eq:99}, is a fluctuating field in the semiclassical model for this
boundary theory.)  For such data the cochains~$\lambda ,\mu $ determine a
relative characteristic class $(\lambda ,\mu )(P,P')\in H^m(M,N;\Cx)$.
 
Now if $M$~is a compact $m$-manifold with boundary, and we color the boundary
with the right boundary theory given by $(H,f,\mu )$, then the partition
function on~$M$ is (compare to~\eqref{eq:66})
  \begin{equation}\label{eq:100}
     \sum\limits_{[P,P']}\frac{\bigl\langle(\lambda ,\mu )(P,P'),[M,\partial
     M]\bigr\rangle\inv }{\#\Aut(P,P')},  
  \end{equation}
where the sum is over equivalence classes of principal $G$-bundles $P\to M$
equipped with a reduction along~$f$ to a principal $H$-bundle $P'\to \partial
M$.  The inverse is due to the minus sign in \autoref{thm:73}(1).

  \begin{example}[$m=3$]\label{thm:84}
 In lieu of a cocycle~$\lambda $, represent a class in~$H^3(BG;\Cx)$ as a line
bundle $L\to G\times G$, where $H^3(BG;\Cx)$ is interpreted as a 2-cocycle
on~$G$ with values in lines; see the text before~\eqref{eq:46} and
\cite[\S4.1]{FHLT}.  The quantum theory $\sigma =\sXd3{BG,L }\:\Bord_3\to
\Alg(\Cat)$ with values in the 3-category of tensor categories has $\sigma
(\pt)=\Vect^L [G]$, the fusion category of vector bundles over~$G$ under
convolution with a twist from the line bundle~$L \to G\times G$.  The
quantization of the bordism in Figure~\ref{fig:5} is the right
$\Vect[G]$-module $\Vect(G\gpd H)$ whose objects are vector bundles over the
stack~$G\gpd H$.  If $W\to G$ and $V\to G\gpd H$ are vector bundles, then
$V*W\to G\gpd H$ is the vector bundle
  \begin{equation}\label{eq:101}
     (V*W)_{K}=\bigoplus \limits_{g'\in G} 
     L_{K',g'}\otimes V _{K'}\otimes W_{g'}, \qquad
     K=g'K', \quad K\in G/f(H).
  \end{equation} 
The cochain~$\mu $ in~\eqref{eq:98} is implemented as data to descend $L\to
G\times G$ to a line bundle over $G\gpd H\times G$.  In~\eqref{eq:101} the
vector spaces $(V*W)_{K}$, $V_{K'}$ and $L_{K',g'}$ carry representations of
stabilizer groups in the stack~$G\gpd H$.
  \end{example}

We sketch a similar example of a domain wall (\S\ref{subsubsec:5.3.1}).  Let
$G_1,G_2$ be finite groups and suppose $\lambda _1,\lambda _2$ are degree~$m$
cocycles with values in~$\Cx$ on $BG_1,BG_2$, respectively.  Then a
correspondence 
  \begin{equation}\label{eq:102}
     \begin{gathered} \xymatrix{&H_{12}\ar[dl]_{f_1}\ar[dr]^{f_2}\\ G_1&&G_2}
     \end{gathered} 
  \end{equation}
of finite groups together with a cochain $\mu _{12}\in C^{m-1}(BH_{12};\Cx)$
that satisfies 
  \begin{equation}\label{eq:104}
     \delta \mu _{12} = (Bf_2)^*\lambda _2 - (Bf_1)^*\lambda _1 
  \end{equation}
determines a domain wall from $\sXd m{BG_1,\lambda _1}$ to $\sXd m{BG_2,\lambda
_2}$; see Definition~\ref{thm:69}.

To illustrate the composition of semiclassical domain walls, suppose 
  \begin{equation}\label{eq:103}
     \begin{gathered} \xymatrix{&H_{12}\ar[dl]\ar[dr]^{f_{12}}
     &&H_{23}\ar[dl]_{f_{23}}\ar[dr]\\G_1&&G_2&&G_3} \end{gathered} 
  \end{equation}
is a diagram of finite groups and homomorphisms.  Furthermore, assume $\lambda
_1,\lambda _2,\lambda _3$ are cocycles on $BG_1,BG_2,BG_3$ and $\mu _{12},\mu
_{23}$ are cochains on $BH_{12},BH_{23}$ which satisfy analogs
of~\eqref{eq:104}.  The composition~\eqref{eq:85} is computed as the fiber
product of the interior maps in the diagram of classifying spaces:
  \begin{equation}\label{eq:128}
     \begin{gathered}
     \xymatrix{&&\sY\ar@{-->}[dl]\ar@{-->}[dr]\\&BH_{12}\ar[dl]\ar[dr] 
     &&BH_{23}\ar[dl]\ar[dr]\\BG_1&&BG_2&&BG_3} \end{gathered} 
  \end{equation}
We leave it as a pleasant exercise for the reader to identify 
  \begin{equation}\label{eq:129}
     \sY\simeq \bigsqcup\limits_{K\in
     f_{12}(H_{12})\backslash G_2 / f_{23}(H_{23})} BZ(g_K), 
  \end{equation}
where $g_K\in K$ is a representative element of the double coset, and 
  \begin{equation}\label{eq:130}
     Z(g_K) = \bigl\{ (h_{12},h_{23})\in H_{12}\times H_{23}:
     f_{12}(h_{12})\,g_K\,f_{23}(h_{23})\inv  = g_K \bigr\} . 
  \end{equation}
Upon quantization on a particular manifold~$M$ with domain walls supported on a
codimension one submanifold~$N\subset M$, this composition takes the shape
  \begin{equation}\label{eq:139}
     \sD_{12}*\sD_{23} = \sum\limits_{K\in f_{12}(H_{12})\backslash G_2 /
     f_{23}(H_{23})} \sD_{K}, 
  \end{equation}
where the right hand side is a sum of quantum domain walls supported on~$N$.

 \bigskip\bigskip
\newcommand{\etalchar}[1]{$^{#1}$}
\providecommand{\bysame}{\leavevmode\hbox to3em{\hrulefill}\thinspace}
\providecommand{\MR}{\relax\ifhmode\unskip\space\fi MR }
\providecommand{\MRhref}[2]{%
  \href{http://www.ams.org/mathscinet-getitem?mr=#1}{#2}
}
\providecommand{\href}[2]{#2}

\end{document}